\documentstyle[12pt]{article}
\begin{document}

\onecolumn

\begin{titlepage}
\begin{center}
{\LARGE \bf Exact Solutions of Relativistic Two-Body Motion in Lineal Gravity} 
\\ \vspace{2cm}
R.B. Mann \footnotemark\footnotetext{email: 
mann@avatar.uwaterloo.ca} 
and D. Robbins\footnotemark\footnotetext{email: 
dgr@gpu.srv.ualberta.ca} 
\\
\vspace{1cm}
Dept. of Physics,
University of Waterloo
Waterloo, ONT N2L 3G1, Canada\\
and \\
\vspace{1cm}
T. Ohta \footnotemark\footnotetext{email:
t-oo1@ipc.miyakyo-u.ac.jp}\\
\vspace{1cm} 
Department of Physics, Miyagi University of Education,
Aoba-Aramaki, Sendai 980, Japan\\
\vspace{2cm}
\vspace{2cm}
\today\\
\end{center}

\begin{abstract}
We develop the canonical formalism for a system of $N$ bodies 
in lineal gravity and obtain exact solutions to the equations of motion for $N=2$. 
The determining equation of the Hamiltonian is derived in the form of a 
transcendental equation, which leads to the exact Hamiltonian to infinite 
order of the gravitational coupling constant.  In the equal mass case explicit expressions 
of the trajectories of the 
particles are given as the functions of the proper time, which show 
characteristic features of the motion depending on the strength of gravity 
(mass) and the magnitude and sign of the cosmological constant.
As expected, we find that  a positive cosmological constant has 
a repulsive effect on the motion, while a negative one has an attractive 
effect. However, some surprising features emerge that are absent for vanishing
cosmological constant.
For a certain range of the negative cosmological constant the motion shows 
a double maximum behavior as a combined result of an induced  
momentum-dependent cosmological potential and the gravitational attraction between
the particles. For a positive cosmological constant, not only 
bounded motions but also unbounded ones are realized. 
The change of the metric along the movement of the particles is also exactly 
derived. 
\end{abstract}
\end{titlepage}
\onecolumn

\section{INTRODUCTION}

Solving the motion of a system of $N$ particles interacting through their
mutual gravitational forces has long been one of the important themes in
physics \cite{EIH}. Though an exact solution is known in the $N=2$ case in
Newtonian theory, in the context of the general theory of relativity the
motion of the $N$ bodies cannot be solved exactly due to dissipation of
energy in the form of gravitational radiation, even when $N=2$. Hence
analysis of a two body system in general relativity (e.g. binary pulsars)
necessarily involves resorting to approximation methods such as a
post-Newtonian expansion \cite{300yrs,ookh}.

However in the past decade lower dimensional versions of general relativity
( both in (1+1) and (2+1) dimensions) have been extensively investigated
from both classical and quantum perspectives. Here the reduced
dimensionality yields an absence of gravitational radiation. Although this
desirable physical feature is lost in such models (at least in the vacuum),
most (if not all) of the remaining conceptual features of relativistic
gravity are retained. Hence their mathematical simplificity offers the hope
of obtaining a deep understanding of the nature of gravitation in a wide
variety of physical situations. It is with this motivation that we consider
the $N$-body problem in lower dimensional gravity.

Specifically, we consider the gravitational $N$-body problem in two
spacetime dimensions. Such lineal theories of gravity have found widespread
use in other problems in physics. The simplest such theory (sometimes
referred to as Jackiw-Teitelboim (JT) theory \cite{JT}) sets the Ricci
scalar equal to a constant, with other matter fields evolving in this
constant-curvature two-dimensional spacetime. Another such theory (sometimes
referred to as $R=T$ theory) sets the Ricci scalar equal to the trace of the
stress-energy of the prescribed matter fields and sources -- in this manner,
matter governs the evolution of spacetime curvature which reciprocally
governs the evolution of matter \cite{r3}. This theory has a consistent
Newtonian limit \cite{r3} (a problematic issue for a generic $(1+1)$%
-dimensional gravity theory \cite{jchan}), and reduces to JT theory if the
stress-energy is that of a cosmological constant.

The $N$-body problem, then, can be formulated in relativistic gravity by
taking the matter action to be that of $N$ point-particles minimally coupled
to gravity. In previous work we developed the canonical formalism for this
action in $R=T$ lineal gravity \cite{OR} and derived the exact Hamiltonian
for $N=2$ as a solution to a transcendental equation which is valid to
infinite order in the gravitational coupling constant \cite{2bd}. In the
slow motion, weak field limit this Hamiltonian coincides with that of
Newtonian gravity in $(1+1)$ dimensions, and in the limit where all bodies
are massless, spacetime is flat.

More recently we have extended this case to include a cosmological constant $%
\Lambda $, so that in the limit where all bodies are massless, spacetime has
constant curvature (ie the JT theory is obtained), and when $\Lambda $
vanishes the situation described in the previous paragraph is recovered \cite
{2bdcossh}. For $N=2$, we derived an exact solution for the Hamiltonian as a
function of the proper separation and the centre-of-inertia momentum of the
bodies. In the equal mass case an exact solution to the equations of motion
for the proper separation of the two point masses as a function of their
mutual proper time was also obtained. The trajectories showed characteristic
structures depending on the values of a cosmological constant $\Lambda $.
The purpose of this paper is to more fully describe these results and to
expand upon them. Specifically, we generalize our previous formalism with $%
\Lambda =0$ \cite{OR} to a system of $N$ particles in (1+1) dimensional
gravity with cosmological constant. When $N=2$ we obtain exact solutions for
the motion of two bodies of unequal (and equal) mass.

Since the Einstein action is a topological invariant in (1+1) dimensions, we
must incor\-porate a scalar (dilaton) field into the action \cite{BanksMann}%
. By a canonical reduction of the action, the Hamiltonian is defined as a
spatial integral of the second derivative of the dilaton field, which is a
function of the canonical variables of the particles (coordinates and
momenta) and is determined from the constraint equations. For a system of
two particles an equation which determines the Hamiltonian in terms of the
remaining degrees of freedom of the system is derived from the matching
conditions of the solution to the constraint equations. We refer to this
equation as the determining equation; it is a transcendental equation which
allows one to determine the Hamiltonian in terms of the centre of inertia
momentum and proper separation of the bodies. The canonical equations of
motion are derived from the Hamiltonian. For the equal mass case they can be
transformed so that the separation and momentum are given by differential
equations in terms of the proper time. In this latter form they can be
solved exactly in terms of hyperbolic and/or trigonometric functions.

Corresponding to the values of the magnitudes (and signs) of the energy and
other parameters (e.g. gravitational coupling constant, masses, cosmological
constant) several different types of motion are expected in the 2 body case.
Broadly speaking, the two particles could remain either bounded or
unbounded, or else balanced between these two conditions. We shall analyze
these various states of motion, and discuss the transitions which occur
between them. We find several surprising situations, including the onset of
mutual repulsion for a range of values of negative $\Lambda$ and the masses,
and the diverging separation of the two bodies at finite proper time for a
range of values of positive $\Lambda$.

We shall also consider the unequal mass case. In this situation the proper
time is no longer the same for the two particles, and so a description of
the motion requires a more careful analysis. We find that we are able to
obtain phase space trajectories from the determining equation. We also can
obtain explicit solutions for the proper separation in terms of a
transformed time coordinate which reduces to the mutual proper time in the
case of equal mass.

In Sec.II we describe the canonical reduction of the theory and define the
Hamiltonian for the $N$-body system. In Sec.III we solve the constraint
equations for the two-body case and get the determining equation of the
Hamiltonian, from which the canonical equations of motion are explicitly
derived. We investigate the motion for $\Lambda =0$ in Sec.IV, by using the
exact solutions to the canonical equations. The motion of equal masses for $%
\Lambda \neq 0$ are analyzed in Sec.V where a general discussion on the
structure of the determining equation, the plots of phase space
trajectories, the analysis of the explicit solutions in terms of the proper
time are developed. We treat the unequal mass case in Sec.VI. Sec.VII
contains concluding remarks and directions for further work. The solution of
the metric tensor, a test particle approximation in the small mass limit of
one of the particles and the causal relationships between particles in
unbounded motion are given in Appendices.

\section{CANONICALLY REDUCED HAMILTONIAN of $N$ PARTICLES}

The action integral for the gravitational field coupled to $N$ point masses
is 
\begin{eqnarray}  \label{act1}
I&=&\int dx^{2}\left[ \frac{1}{2\kappa}\sqrt{-g}g^{\mu\nu} \left\{\Psi
R_{\mu\nu}+\frac{1}{2}\nabla_{\mu}\Psi\nabla_{\nu}\Psi +\frac{1}{2}%
g_{\mu\nu}\Lambda \right\} \right.  \nonumber \\
&&\makebox[2em]{}\left.+\sum_{a}\int d\tau_{a}\left\{
-m_{a}\left(-g_{\mu\nu}(x)\frac{dz^{\mu}_{a}}{d\tau_{a}} \frac{dz^{\nu}_{a}}{%
d\tau_{a}}\right)^{1/2}\right\} \delta^{2}(x-z_{a}(\tau_{a})) \right]\;,
\end{eqnarray}
where $\Psi$ is the dilaton field, $\Lambda$ is the cosmological constant, $%
g_{\mu\nu}$ and $g$ are the metric and its determinant, $R$ is the Ricci
scalar, and $\tau_{a}$ is the proper time of $a$-th particle, respectively,
with $\kappa=8\pi G/c^4$. The symbol $\nabla_{\mu}$ denotes the covariant
derivative associated with $g_{\mu\nu}$.

The field equations derived from the action (\ref{act1}) are 
\begin{eqnarray}
&&\makebox[2em]{}R-g^{\mu \nu }\nabla _{\mu }\nabla _{\nu }\Psi =0\;,
\label{eq-R} \\
&&\frac{1}{2}\nabla _{\mu }\Psi \nabla _{\nu }\Psi -\frac{1}{4}g_{\mu \nu
}\nabla ^{\lambda }\Psi \nabla _{\lambda }\Psi +g_{\mu \nu }\nabla ^{\lambda
}\nabla _{\lambda }\Psi -\nabla _{\mu }\nabla _{\nu }\Psi =\kappa T_{\mu \nu
}+\frac{1}{2}g_{\mu \nu }\Lambda \;,  \label{eq-Psi} \\
&&\frac{d}{d\tau _{a}}\left\{ g_{\mu \nu }(z_{a})\frac{dz_{a}^{\nu }}{d\tau
_{a}}\right\} -\frac{1}{2}g_{\nu \lambda ,\mu }(z_{a})\frac{dz_{a}^{\nu }}{%
d\tau _{a}}\frac{dz_{a}^{\lambda }}{d\tau _{a}}=0\;,  \label{eq-z}
\end{eqnarray}
where the stress-energy due to the point masses is 
\begin{equation}
T_{\mu \nu }=\sum_{a}m_{a}\int d\tau _{a}\frac{1}{\sqrt{-g}}g_{\mu \sigma
}g_{\nu \rho }\frac{dz_{a}^{\sigma }}{d\tau _{a}}\frac{dz_{a}^{\rho }}{d\tau
_{a}}\delta ^{2}(x-z_{a}(\tau _{a}))\;.  \nonumber  \label{stress1} \\
\end{equation}
Eq.(\ref{eq-Psi}) guarantees the conservation of $T_{\mu \nu }$. Inserting
the trace of Eq.(\ref{eq-Psi}) into Eq.(\ref{eq-R}) yields 
\begin{equation}
R-\Lambda =\kappa T_{\;\;\mu }^{\mu }\;\;.  \label{RT}
\end{equation}
Eqs. (\ref{eq-z}) and (\ref{RT}) form a closed sytem of equations for the
matter-gravity system. The evolution of the dilaton then follows from
inserting the solutions to these equations into(\ref{eq-R}), and then
solving for its motion, the traceless part of \ (\ref{eq-Psi}) being
identities once the other equations are satisfied. Alternatively, one can
solve the independent parts of equations (\ref{eq-R}), (\ref{eq-Psi}), and (%
\ref{eq-z}) for the metric, dilaton and matter degrees of freedom, which is
the approach we shall take in this paper. If the masses of all particles are
taken to be zero then the field equations reduce to those of constant
curvature lineal gravity, or JT theory \cite{JT,r3}. 

Consider next the transformation of the action (\ref{act1}) to canonical
form. We decompose the scalar curvature in terms of the extrinsic curvature $%
K$ via 
\begin{equation}
\sqrt{-g}R=-2\partial _{0}(\sqrt{\gamma }K)+2\partial _{1}(\sqrt{\gamma }%
N^{1}K-\gamma ^{-1}\partial _{1}N_{0})\;,  \label{extK}
\end{equation}
where $K=(2N_{0}\gamma )^{-1}(2\partial _{1}N_{1}-\gamma ^{-1}N_{1}\partial
_{1}\gamma -\partial _{0}\gamma )$, and the metric is 
\begin{equation}
ds^{2}=-N_{0}^{2}dt^{2}+\gamma \left( dx+\frac{N_{1}}{\gamma }dt\right)
^{2}\;,  \label{lineel}
\end{equation}
so that $\gamma =g_{11},N_{0}=(-g^{00})^{-1/2}$ and $N_{1}=g_{10}$, and then
rewrite the particle action in first-order form. After some manipulation we
find that the action (\ref{act1}) may be rewritten in the form 
\begin{equation}
I=\int dx^{2}\left\{ \sum_{a}p_{a}\dot{z}_{a}\delta (x-z_{a}(t))+\pi \dot{%
\gamma}+\Pi \dot{\Psi}+N_{0}R^{0}+N_{1}R^{1}\right\} \;,  \label{act2}
\end{equation}
where $\pi $ and $\Pi $ are conjugate momenta to $\gamma $ and $\Psi $,
respectively, and 
\begin{eqnarray}
R^{0} &=&-\kappa \sqrt{\gamma }\gamma \pi ^{2}+2\kappa \sqrt{\gamma }\pi \Pi
+\frac{1}{4\kappa \sqrt{\gamma }}(\Psi ^{\prime })^{2}-\frac{1}{\kappa }%
\left( \frac{\Psi ^{\prime }}{\sqrt{\gamma }}\right) ^{\prime }-\frac{%
\Lambda }{2\kappa }\sqrt{\gamma }  \nonumber \\
&&-\sum_{a}\sqrt{\frac{p_{a}^{2}}{\gamma }+m_{a}^{2}}\;\delta (x-z_{a}(t))\;,
\nonumber  \label{R0} \\
&& \\
R^{1} &=&\frac{\gamma ^{\prime }}{\gamma }\pi -\frac{1}{\gamma }\Pi \Psi
^{\prime }+2\pi ^{\prime }+\sum_{a}\frac{p_{a}}{\gamma }\delta
(x-z_{a}(t))\;,  \label{R1}
\end{eqnarray}
with the symbols $(\;\dot{}\;)$ and $(\;^{\prime }\;)$ denoting $\partial
_{0}$ and $\partial _{1}$, respectively.

The action (\ref{act2}) leads to the system of equations 
\begin{eqnarray}  \label{e-pi}
\dot{\pi}&+&N_{0}\left\{\frac{3\kappa}{2}\sqrt{\gamma}\pi^{2} -\frac{\kappa}{%
\sqrt{\gamma}}\pi\Pi +\frac{1}{8\kappa\sqrt{\gamma}\gamma}%
(\Psi^{\prime})^{2} -\frac{1}{4\sqrt{\gamma}}\frac{\Lambda}{\kappa} \right. 
\nonumber \\
&&\makebox[10em]{}\left. -\sum_{a}\frac{p^{2}_{a}}{2\gamma^{2}\sqrt{\frac{%
p^{2}_{a}}{\gamma} +m^{2}_{a}}}\;\delta(x-z_{a}(t))\right\}  \nonumber \\
&+&N_{1}\left\{-\frac{1}{\gamma^{2}}\Pi\Psi^{\prime} +\frac{\pi^{\prime}}{%
\gamma} +\sum_{a}\frac{p_{a}}{\gamma^{2}}\;\delta(x-z_{a}(t))\right\}
+N^{\prime}_{0}\frac{1}{2\kappa\sqrt{\gamma}\gamma}\Psi^{\prime}
+N^{\prime}_{1}\frac{\pi}{\gamma}=0\;,
\end{eqnarray}
\begin{eqnarray}
&& \dot{\gamma}-N_{0}(2\kappa\sqrt{\gamma}\gamma\pi-2\kappa\sqrt{\gamma}\Pi)
+N_{1}\frac{\gamma^{\prime}}{\gamma}-2N^{\prime}_{1}=0\;,  \label{e-gamma} \\
&& R^{0}=0\;,  \label{e-R0} \\
&& R^{1}=0\;,  \label{e-R1} \\
&& \dot{\Pi}+\partial_{1}(-\frac{1}{\gamma}N_{1}\Pi +\frac{1}{2\kappa\sqrt{%
\gamma}}N_{0}\Psi^{\prime} +\frac{1}{\kappa\sqrt{\gamma}}N^{\prime}_{0})=0\;,
\label{e-Pi} \\
&& \dot{\Psi}+N_{0}(2\kappa\sqrt{\gamma}\pi)-N_{1}(\frac{1}{\gamma}
\Psi^{\prime})=0\;,  \label{e-Psi} \\
&& \dot{p}_{a}+\frac{\partial N_{0}}{\partial z_{a}}\sqrt{\frac{p^{2}_{a}} {%
\gamma}+m^{2}_{a}}-\frac{N_{0}}{2\sqrt{\frac{p^{2}_{a}}{\gamma}+m^{2}_{a}}} 
\frac{p^{2}_{a}}{\gamma^{2}}\frac{\partial\gamma}{\partial z_{a}} -\frac{%
\partial N_{1}}{\partial z_{a}}\frac{p_{a}}{\gamma}  \nonumber \\
&&\makebox[2em]{}+N_{1}\frac{p_{a}}{\gamma^{2}}\frac{\partial\gamma}{%
\partial z_{a}}=0\;,  \label{e-p} \\
&& \dot{z_{a}}-N_{0}\frac{\frac{p_{a}}{\gamma}}{\sqrt{\frac{p^{2}_{a}}{\gamma%
} +m^{2}_{a}}} +\frac{N_{1}}{\gamma}=0 \;.  \label{e-z}
\end{eqnarray}
In the equations (\ref{e-p}) and (\ref{e-z}), all metric components ($N_{0}$%
, $N_{1}$, $\gamma$) are evaluated at the point $x=z_{a}$ and 
\[
\frac{\partial f}{\partial z_{a}}\equiv \left.\frac{\partial f(x)}{\partial x%
}\right|_{x=z_{a}}\;. 
\]
This system of equations can be shown to be equivalent to the set of
equations (\ref{eq-R}), (\ref{eq-Psi}) and (\ref{eq-z}).

Since $N_{0}$ and $N_{1}$ are Lagrange multipliers, equations (\ref{e-R0})
and (\ref{e-R1}) are constraints; specifically they are the energy and
momentum constraints of the $(1+1)$ dimensional gravitational system we
consider. We may solve them for $(\Psi^{\prime}/\sqrt{\gamma})^{\prime}$ and 
$\pi^{\prime}$ in terms of the dynamical and gauge ({\it i.e.} co-ordinate)
degrees of freedom, since these are the only linear terms in these
constraints. We identify these coordinate degrees of freedom by writing the
generator arising from the variation of the action at the boundaries in
terms of $(\Psi^{\prime}/\sqrt{\gamma})^{\prime}$ and $\pi^{\prime}$, and
then finding which quantities serve to fix the frame of the physical
space-time coordinates in a manner similar to the $(3+1)$-dimensional case.

Carrying out the same procedure as in the $\Lambda=0$ case \cite{2bd} we
find that we can consistently choose the coordinate conditions 
\begin{equation}  \label{cc}
\gamma=1 \makebox[2em]{} \mbox{and} \makebox[2em]{} \Pi=0 \;.
\end{equation}
Eliminating the constraints, the action reduces to 
\begin{equation}
I=\int dx^{2}\left\{\sum_{a}p_{a}\dot{z}_{a}\delta(x-z_{a}) -{\cal H\mit}%
\right\}\;,
\end{equation}
and the reduced Hamiltonian for the $N$-body system is 
\begin{equation}  \label{ham1}
H=\int dx {\cal H\mit =-\frac{1}{\kappa}\int dx \triangle\Psi \;,}
\end{equation}
where $\Psi$ is a function of $z_{a}$ and $p_{a}$, determined by solving the
constraints which are under the coordinate conditions (\ref{cc}) 
\begin{equation}  \label{Psi}
\triangle\Psi-\frac{1}{4}(\Psi^{\prime})^{2} +\kappa^{2}\pi^{2}-\frac{%
\Lambda }{2} +\kappa\sum_{a}\sqrt{p^{2}_{a}+m^{2}_{a}}\delta(x-z_{a})=0 \;,
\end{equation}
\begin{equation}  \label{pi}
2\pi^{\prime}+\sum_{a}p_{a}\delta(x-z_{a})=0 \;.
\end{equation}

The expression for the Hamiltonian (\ref{ham1}) is analogous to the reduced
Hamiltonian in $(3+1)$ dimensional general relativity. In $(1+1)$ dimensions
it is determined by the dilaton field at spatial infinity. The consistency
of this canonical reduction may be demonstrated in a manner analogous to
that employed in the $\Lambda=0$ case: namely the canonical equations of
motion derived from the reduced Hamiltonian (\ref{ham1}) are identical with
the equations (\ref{e-p}) and (\ref{e-z}).

The methodology at this point is then as follows. First we must solve (\ref
{Psi}) and (\ref{pi}) for $\Psi$ and $\pi$ in terms of the $p_a$ and the $%
z_a $, consistently matching solutions across the boundaries of the
particles. Then we compute from (\ref{ham1}) the Hamiltonian in terms of the
independent momenta and coordinates of the particles. This expression is
sufficient to obtain the phase-space trajectories for a given set of initial
conditions. Finally we solve equations (\ref{e-pi} -- \ref{e-z}) to obtain a
complete solution for the $N$ body system. Throughout the remainder of this
paper we shall consider only $N=2$, i.e. 2-body dynamics.

\section{SOLUTION TO THE CONSTRAINT EQUATIONS AND THE HAMILTONIAN FOR A
SYSTEM OF TWO PARTICLES}

The standard approach for investigating the dynamics of particles is to
derive an explicit expression of the Hamiltonian, from which the equations
of motion and the solution of trajectories are obtained. In this section we
explain how to derive the Hamiltonian from the solution to the constraint
equations (\ref{Psi}) and (\ref{pi}) and get the explicit Hamiltonian for
two particles in a spacetime with a cosmological constant.

We first express the equations (\ref{Psi}) and (\ref{pi}) as 
\begin{equation}  \label{Psi1}
\triangle\Psi=\frac{1}{4}\left(\Psi^{\prime}\right)^{2}
-\kappa^{2}\left(\chi^{\prime}\right)^{2} +\frac{1}{2}\Lambda -\kappa\sum_{a}%
\sqrt{p^{2}_{a}+m^{2}_{a}}\delta(x-z_{a})\;,
\end{equation}
\begin{equation}  \label{chi1}
\triangle\chi= -\frac{1}{2}\sum_{a}p_{a}\delta(x-z_{a})\;,
\end{equation}
where we set $\chi^{\prime}=\pi$. Rewriting (\ref{Psi1}) as 
\begin{equation}  \label{Psi1a}
(1+\frac{\Psi}{4})\triangle\Psi= \frac{1}{8}\triangle(\Psi^2-4\kappa^2%
\chi^2) +\frac{1}{2}\Lambda + \kappa^2 \chi\triangle\chi -\kappa\sum_{a}%
\sqrt{p^{2}_{a}+m^{2}_{a}}\delta(x-z_{a})\;,
\end{equation}
and using (\ref{chi1}), we can rewrite (\ref{ham1}) as 
\begin{eqnarray}  \label{ham2}
H&=&\sum_{a}\frac{\sqrt{p^{2}_{a}+m^{2}_{a}}}{1+\frac{1}{4}\Psi(z_{a})} +%
\frac{\kappa}{2}\sum_{a}\frac{p_{a}\chi(z_{a})}{1+\frac{1}{4}\Psi(z_{a})} 
\nonumber \\
&&-\frac{1}{8\kappa}\int dx\frac{1}{1+\frac{1}{4}\Psi(x)} \triangle\left(%
\Psi^{2}-4\kappa^{2}\chi^{2}+2\Lambda x^{2}\right)\;,
\end{eqnarray}
which can also be obtained by inserting (\ref{Psi1}) and (\ref{chi1}) into (%
\ref{ham1}) and iterating by partial integration (assuming convergence). We
shall refer to this formula later when we consider boundary conditions.

Defining $\phi$ by 
\begin{equation}
\Psi=-4\mbox{log}|\phi| \;,
\end{equation}
the constraints (\ref{Psi1}) and (\ref{chi1}) for a two-particle system
become 
\begin{eqnarray}
\triangle\phi-\frac{1}{4}\left\{ \kappa^{2}\left(\chi^{\prime}\right)^{2} -%
\frac{1}{2}\Lambda\right\}\phi &=&\frac{\kappa}{4} \left\{\sqrt{%
p^{2}_{1}+m^{2}_{1}}\;\phi(z_{1})\delta(x-z_{1})\right.  \nonumber \\
&&\left.\quad +\sqrt{p^{2}_{2}+m^{2}_{2}}\;\phi(z_{2})\delta(x-z_{2})\right%
\}\;,  \label{phi-eq}
\end{eqnarray}
\begin{equation}
\triangle\chi =-\frac{1}{2}\left\{p_{1}\delta(x-z_{1})
+p_{2}\delta(x-z_{2})\right\}\;.  \label{chi-eq}
\end{equation}
The general solution to (\ref{chi-eq}) is 
\begin{equation}  \label{chi-sol}
\chi=-\frac{1}{4}\left\{p_{1}\mid x-z_{1}\mid+p_{2}\mid x-z_{2}\mid\right\}
-\epsilon Xx+\epsilon C_{\chi} \;.
\end{equation}
The factor $\epsilon$ ($\epsilon^{2}=1$) has been introduced in the
constants $X$ and $C_{\chi}$ so that the T-inversion (time reversal)
properties of $\chi$ are explicitly manifest. By definition, $\epsilon$
changes sign under time reversal and so, therefore, does $\chi$.

Consider first the case $z_{2}<z_{1}$, for which we may divide space into
three regions: $z_{1}<x$ ((+) region), $z_{2}<x<z_{1}$ ((0) region) and $%
x<z_{2}$ ((-) region). In each region $\chi^{\prime}$ is constant: 
\begin{equation}
\chi^{\prime}=\left\{ 
\begin{array}{ll}
-\epsilon X-\frac{1}{4}(p_{1}+p_{2}) & \makebox[3em]{}\mbox{(+) region}, \\ 
-\epsilon X+\frac{1}{4}(p_{1}-p_{2}) & \makebox[3em]{}\mbox{(0) region}, \\ 
-\epsilon X+\frac{1}{4}(p_{1}+p_{2}) & \makebox[3em]{}\mbox{(-) region}\;.
\end{array}
\right.
\end{equation}
General solutions to the homogeneous equation $\triangle\phi-\frac{1}{4}%
\left\{\kappa^{2}\left(\chi^{\prime}\right)^{2} -\frac{1}{2}%
\Lambda\right\}\phi=0$ in each region are 
\begin{equation}  \label{phi2}
\left\{ 
\begin{array}{l}
\phi_{+}(x)=A_{+}e^{\frac{1}{2}K_{+}x}+B_{+}e^{-\frac{1}{2}K_{+}x}\;, \\ 
\phi_{0}(x)=A_{0}e^{\frac{1}{2}K_{0}x}+B_{0}e^{-\frac{1}{2}K_{0}x}\;, \\ 
\phi_{-}(x)=A_{-}e^{\frac{1}{2}K_{-}x}+B_{-}e^{-\frac{1}{2}K_{-}x}\;,
\end{array}
\right.
\end{equation}
where 
\begin{equation}  \label{K+-0}
\left\{ 
\begin{array}{lll}
K_{+} & = \sqrt{\kappa^{2}\left(X+\frac{\epsilon}{4}(p_{1}+
p_{2})\right)^{2}-\frac{1}{2}\Lambda} & \qquad \mbox{(+) region}\;, \\ 
K_{0} & = \sqrt{\kappa^{2}\left(X-\frac{\epsilon}{4}(p_{1}
-p_{2})\right)^{2} -\frac{1}{2}\Lambda} & \qquad \mbox{(0) region}\label{K}%
\;, \\ 
K_{-} & = \sqrt{\kappa^{2}\left(X-\frac{\epsilon}{4}(p_{1}+
p_{2})\right)^{2}-\frac{1}{2}\Lambda} & \qquad \mbox{(-) region}\;.
\end{array}
\right.
\end{equation}
For these solutions to be the actual solutions to Eq.(\ref{phi-eq}) with
delta function source terms, they must satisfy the following matching
conditions at $x=z_{1}, z_{2}$: {\ \setcounter{enumi}{\value{equation}} %
\addtocounter{enumi}{1} \setcounter{equation}{0} \renewcommand{%
\theequation}{\theenumi\alph{equation}} 
\begin{eqnarray}
&&\phi_{+}(z_{1})=\phi_{0}(z_{1})=\phi(z_{1})\;,  \label{match1} \\
&&\phi_{-}(z_{2})=\phi_{0}(z_{2})=\phi(z_{2})\;,  \label{match2} \\
&&\phi^{\prime}_{+}(z_{1})-\phi^{\prime}_{0}(z_{1}) =\frac{\kappa}{4}\sqrt{%
p^{2}_{1}+m^{2}_{1}}\phi(z_{1})\;,  \label{match3} \\
&&\phi^{\prime}_{0}(z_{2})-\phi^{\prime}_{-}(z_{2}) =\frac{\kappa}{4}\sqrt{%
p^{2}_{2}+m^{2}_{2}}\phi(z_{2}) \;.  \label{match4}
\end{eqnarray}
\setcounter{equation}{\value{enumi}} } The conditions (\ref{match1}) and (%
\ref{match3}) lead to 
\begin{equation}
e^{\frac{1}{2}K_{+}z_{1}}A_{+}+e^{-\frac{1}{2}K_{+}z_{1}}B_{+} =e^{\frac{1}{2%
}K_{0}z_{1}}A_{0} +e^{-\frac{1}{2}K_{0}z_{1}}B_{0}\;,
\end{equation}
\begin{eqnarray}
e^{\frac{1}{2}K_{+}z_{1}}A_{+}-e^{-\frac{1}{2}K_{+}z_{1}}B_{+} &=&\frac{%
\kappa\sqrt{p^{2}_{1}+m^{2}_{1}}+2K_{0}}{2K_{+}} e^{\frac{1}{2}%
K_{0}z_{1}}A_{0}  \nonumber \\
&&+\frac{\kappa\sqrt{p^{2}_{1}+m^{2}_{1}}-2K_{0}}{2K_{+}} e^{-\frac{1}{2}%
K_{0}z_{1}}B_{0}\;,
\end{eqnarray}
yielding {\ \setcounter{enumi}{\value{equation}} \addtocounter{enumi}{1} %
\setcounter{equation}{0} \renewcommand{\theequation}{\theenumi%
\alph{equation}} 
\begin{eqnarray}
A_{+}&=&\frac{\kappa\sqrt{p^{2}_{1}+m^{2}_{1}}+2K_{0}+2K_{+}}{4K_{+}} e^{%
\frac{1}{2}(K_{0}-K_{+})z_{1}}A_{0}  \nonumber \\
&&\makebox[10em]{} +\frac{\kappa\sqrt{p^{2}_{1}+m^{2}_{1}}-2K_{0}+2K_{+}}{%
4K_{+}} e^{-\frac{1}{2}(K_{0}+K_{+})z_{1}}B_{0}\;,  \label{a+} \\
B_{+}&=&-\frac{\kappa\sqrt{p^{2}_{1}+m^{2}_{1}}+2K_{0}-2K_{+}}{4K_{+}} e^{%
\frac{1}{2}(K_{0}+K_{+})z_{1}}A_{0}  \nonumber \\
&&\makebox[10em]{} -\frac{\kappa\sqrt{p^{2}_{1}+m^{2}_{1}}-2K_{0}-2K_{+}}{%
4K_{+}} e^{-\frac{1}{2}(K_{0}-K_{+})z_{1}}B_{0} \;.  \label{b+}
\end{eqnarray}
\setcounter{equation}{\value{enumi}} } Similarly from (\ref{match2}) and (%
\ref{match4}) we have 
\begin{equation}
e^{\frac{1}{2}K_{-}z_{2}}A_{-}+e^{-\frac{1}{2}K_{-}z_{2}}B_{-} =e^{\frac{1}{2%
}K_{0}z_{2}}A_{0} +e^{-\frac{1}{2}K_{0}z_{2}}B_{0}\;,
\end{equation}
\begin{eqnarray}
-e^{\frac{1}{2}K_{-}z_{2}}A_{-}+e^{-\frac{1}{2}K_{-}z_{2}}B_{-} &=&\frac{%
\kappa\sqrt{p^{2}_{2}+m^{2}_{2}}-2K_{0}}{2K_{-}}e^{\frac{1}{2}
K_{0}z_{2}}A_{0}  \nonumber \\
&&+\frac{\kappa\sqrt{p^{2}_{2}+m^{2}_{2}}+2K_{0}}{2K_{-}}e^{-\frac{1}{2}
K_{0}z_{2}}B_{0}\;,
\end{eqnarray}
and then {\ \setcounter{enumi}{\value{equation}} \addtocounter{enumi}{1} %
\setcounter{equation}{0} \renewcommand{\theequation}{\theenumi%
\alph{equation}} 
\begin{eqnarray}
A_{-}&=&-\frac{\kappa\sqrt{p^{2}_{2}+m^{2}_{2}}-2K_{0}-2K_{-}}{4K_{-}} e^{%
\frac{1}{2}(K_{0}-K_{-})z_{2}}A_{0}  \nonumber \\
&&\makebox[10em]{} -\frac{\kappa\sqrt{p^{2}_{2}+m^{2}_{2}}+2K_{0}-2K_{-}}{%
4K_{-}} e^{-\frac{1}{2}(K_{0}+K_{-})z_{2}}B_{0}\;,  \label{a-} \\
B_{-}&=&\frac{\kappa\sqrt{p^{2}_{2}+m^{2}_{2}}-2K_{0}+2K_{-}}{4K_{-}} e^{%
\frac{1}{2}(K_{0}+K_{-})z_{2}}A_{0}  \nonumber \\
&&\makebox[10em]{} +\frac{\kappa\sqrt{p^{2}_{2}+m^{2}_{2}}+2K_{0}+2K_{-}}{%
4K_{-}} e^{-\frac{1}{2}(K_{0}-K_{-})z_{2}}B_{0}\;.  \label{b-}
\end{eqnarray}
\setcounter{equation}{\value{enumi}} }

Since the magnitudes of both $\phi$ and $\chi$ increase with increasing $|x|$%
, it is necessary to impose a boundary condition which guarantees that the
surface terms which arise in transforming the action vanish and
simultaneously preserves the finiteness of the Hamiltonian. A consideration
of (\ref{ham2}) implies that we may choose the boundary condition 
\begin{equation}  \label{bound}
\Psi^{2}-4\kappa^{2}\chi^{2}+2\Lambda x^{2}=C_{\pm}x \qquad 
\mbox{for (+)
and (-) regions}
\end{equation}
with $C_{\pm}$ being constants to be determined. This boundary condition
means 
\begin{eqnarray}
&&A_{-}=B_{+}=0\;,  \label{cond1} \\
&&\left\{2K_{+}x+4\mbox{log}|A_{+}|\right\}^{2} -4\kappa^{2}\left\{-\left[%
\epsilon X+\frac{1}{4}(p_{1}+p_{2})\right]x +\epsilon C_{\chi}+\frac{1}{4}%
(p_{1}z_{1}+p_{2}z_{2})\right\}^{2}  \nonumber \\
&&\makebox[1em]{}+2\Lambda x^{2}=C_{+}x \;,  \label{cond2} \\
&&\left\{2K_{-}x-4\mbox{log}|B_{-}|\right\}^{2} -4\kappa^{2}\left\{-\left[%
\epsilon X-\frac{1}{4}(p_{1}+p_{2})\right]x +\epsilon C_{\chi}-\frac{1}{4}%
(p_{1}z_{1}+p_{2}z_{2})\right\}^{2}  \nonumber \\
&&\makebox[1em]{}+2\Lambda x^{2}=C_{-}x \;.  \label{cond3}
\end{eqnarray}
It may seem that instead of (\ref{cond1}) we could have made the alternate
choices $(A_+=0, B_-=0), (A_+=0, A_-=0)$ or $(B_+=0, B_-=0)$. However the
definitions (\ref{K+-0}) imply that $K_{\pm}$ are positive quantities, which
in turn leads to a negative Hamiltonian for the choice $(A_+=0, B_-=0)$. For
the choices $(A_+=0, A_-=0)$ and $(B_+=0, B_-=0)$ the Hamiltonian
identically vanishes.

The terms quadratic in $x$ from (\ref{cond2}) and (\ref{cond3}) merely
recapitulate the definitions of $K_{\pm}$. Equating terms linear in $x$
yield the relations 
\begin{eqnarray}
16K_{+}\mbox{log}|A_{+}|+8\kappa^{2} \left[\epsilon X+\frac{1}{4}%
(p_{1}+p_{2})\right] \left[\epsilon C_{\chi}+\frac{1}{4}%
(p_{1}z_{1}+p_{2}z_{2})\right]&=&C_{+}\;,  \label{cond4} \\
-16K_{-}\mbox{log}|B_{-}|+8\kappa^{2} \left[\epsilon X-\frac{1}{4}%
(p_{1}+p_{2})\right] \left[\epsilon C_{\chi}-\frac{1}{4}%
(p_{1}z_{1}+p_{2}z_{2})\right]&=& C_{-} \;.  \label{cond5}
\end{eqnarray}
Equating the constant terms of (\ref{cond2}) and (\ref{cond3}) leads to 
\begin{eqnarray}
&&16\left(\mbox{log}|A_{+}|\right)^{2} -4\kappa^{2}\left[\epsilon C_{\chi}+%
\frac{1}{4}(p_{1}z_{1}+p_{2}z_{2})\right] ^{2}=0\;,  \label{cond6} \\
&&16\left(\mbox{log}|B_{-}|\right)^{2} -4\kappa^{2}\left[\epsilon C_{\chi}-%
\frac{1}{4}(p_{1}z_{1}+p_{2}z_{2})\right] ^{2}=0 \;.  \label{cond7}
\end{eqnarray}
We choose the solutions 
\begin{eqnarray}  \label{a+b-1}
\mbox{log}|A_{+}|&=&-\frac{\kappa}{2}\left[C_{\chi} +\frac{\epsilon}{4}%
(p_{1}z_{1}+p_{2}z_{2})\right]\;,  \nonumber \\
\\
\mbox{log}|B_{-}|&=&\frac{\kappa}{2}\left[C_{\chi} -\frac{\epsilon}{4}%
(p_{1}z_{1}+p_{2}z_{2})\right]\;.  \nonumber
\end{eqnarray}
Before proceeding, we add a remark to (\ref{a+b-1}). In solving (\ref{cond6}%
) and (\ref{cond7}), there are actually four combinations $(\mp, \pm)$ of
sign choices for $\mbox{log}|A_{+}|$ and $\mbox{log}|B_{-}|$. However the
choices $(+, +)$ and $(-, -)$ do not lead to any relations among the
coefficients and gives us an unphysical Hamiltonian. The choices $(+, -)$
and $(- , +)$ lead to identical physical results once the signs of the
momenta $p_{i}$ and the coefficient $C_{\chi}$ are reversed.

The condition (\ref{cond1}) leads to 
\begin{equation}  \label{a/b-1}
\frac{A_{0}}{B_{0}}=-\;\frac{\kappa\sqrt{p^{2}_{1}+m^{2}_{1}}-2K_{0}-2K_{+}%
} {\kappa\sqrt{p^{2}_{1}+m^{2}_{1}}+2K_{0}-2K_{+}}e^{-K_{0}z_{1}}\;,
\end{equation}
and 
\begin{equation}  \label{a/b-2}
\frac{A_{0}}{B_{0}}=-\;\frac{\kappa\sqrt{p^{2}_{2}+m^{2}_{2}}+2K_{0}-2K_{-}%
} {\kappa\sqrt{p^{2}_{2}+m^{2}_{2}}-2K_{0}-2K_{-}}e^{-K_{0}z_{2}}\;.
\end{equation}
{}From (\ref{a/b-1}) and (\ref{a/b-2}) we obtain 
\begin{eqnarray}  \label{X}
&&\left(\kappa\sqrt{p^{2}_{1}+m^{2}_{1}}-2K_{0}-2K_{+}\right) \left(\kappa%
\sqrt{p^{2}_{2}+m^{2}_{2}}-2K_{0}-2K_{-}\right)  \nonumber \\
&&\makebox[5em]{}=\left(\kappa\sqrt{p^{2}_{1}+m^{2}_{1}}+2K_{0}-2K_{+}%
\right) \left(\kappa\sqrt{p^{2}_{2}+m^{2}_{2}}+2K_{0}-2K_{-}\right)
e^{K_{0}(z_{1}-z_{2})}\;,  \nonumber \\
\end{eqnarray}
which we shall refer to as the determining equation for $X$. The $\Psi$
fields in $(\pm)$ regions are 
\begin{eqnarray}  \label{psi+-}
\Psi_{+}(x)&=&-4\mbox{log}|A_{+}|-2K_{+}x \;,  \nonumber \\
\\
\Psi_{-}(x)&=&-4\mbox{log}|B_{-}|+2K_{-}x \;,  \nonumber
\end{eqnarray}
and the Hamiltonian is 
\begin{eqnarray}  \label{Ham1}
H&=&-\frac{1}{\kappa}\int dx\triangle\Psi = -\frac{1}{\kappa}\left[%
\Psi^{\prime}\right]^{\infty}_{-\infty}  \nonumber \\
&=&\frac{2(K_{+}+K_{-})}{\kappa}\;.
\end{eqnarray}
Once the solution for $X$ is obtained from (\ref{X}), the Hamiltonian is
explicitly determined from (\ref{Ham1}) in terms of the degrees of freedom
of the system (i.e. the coordinates and momenta of the particles).

>From (\ref{a+}), (\ref{b-}), (\ref{a/b-1}) and (\ref{a/b-2}), $A_{+}$ and $%
B_{-}$ are expressed in terms of $A_{0}$ as 
\begin{eqnarray}  \label{a+b-2}
A_{+}&=&\frac{4K_{0}}{2K_{0}+2K_{+}-\kappa\sqrt{p^{2}_{1}+m^{2}_{1}}} e^{%
\frac{1}{2}(K_{0}-K_{+})z_{1}}\;A_{0}\;,  \nonumber \\
\\
B_{-}&=&\frac{4K_{0}}{2K_{0}-2K_{-}+\kappa\sqrt{p^{2}_{2}+m^{2}_{2}}} e^{%
\frac{1}{2}(K_{0}+K_{-})z_{2}}\;A_{0}\;,  \nonumber
\end{eqnarray}
and from (\ref{a+b-1}) and (\ref{a+b-2}) the coefficients $A_{0}$ and $%
C_{\chi}$ (and hence $A_{+}$, $B_{-}$ and $B_{0}$) are also determined 
\begin{eqnarray}
C_{\chi}&=&\frac{1}{2\kappa}\mbox{log} \frac{(2K_{0}+2K_{+}-\kappa\sqrt{%
p^{2}_{1}+m_{1}^{2}}) (\kappa\sqrt{p^{2}_{1}+m_{1}^{2}}+2K_{0}-2K_{+})} {%
(2K_{0}+2K_{-}-\kappa\sqrt{p^{2}_{2}+m_{2}^{2}}) (\kappa\sqrt{%
p^{2}_{2}+m_{2}^{2}}+2K_{0}-2K_{-})} +\frac{1}{2\kappa}%
(K_{+}z_{1}+K_{-}z_{2})\;,  \nonumber  \label{c-chi} \\
\\
\mbox{log}|A_{+}|&=&\frac{1}{2}\mbox{log} \frac{2K_{0}+2K_{-}-\kappa\sqrt{%
p^{2}_{2}+m_{2}^{2}}} {\kappa\sqrt{p^{2}_{1}+m_{1}^{2}}+2K_{0}-2K_{+}} -%
\frac{1}{4}(K_{0}+K_{+}+\frac{\kappa\epsilon}{2}p_{1})z_{1} +\frac{1}{4}%
(K_{0}-K_{-}-\frac{\kappa\epsilon}{2}p_{2})z_{2}\;,  \nonumber  \label{a+sol}
\\
\\
\mbox{log}|B_{-}|&=&\frac{1}{2}\mbox{log} \frac{2K_{0}+2K_{+}-\kappa\sqrt{%
p^{2}_{1}+m_{1}^{2}}} {\kappa\sqrt{p^{2}_{2}+m_{2}^{2}}+2K_{0}-2K_{-}} -%
\frac{1}{4}(K_{0}-K_{+}+\frac{\kappa\epsilon}{2}p_{1})z_{1} +\frac{1}{4}%
(K_{0}+K_{-}-\frac{\kappa\epsilon}{2}p_{2})z_{2}\;,  \nonumber  \label{b-sol}
\\
\\
\mbox{log}|A_{0}|&=&\frac{1}{2}\mbox{log}\frac{ (2K_{0}+2K_{+}-\kappa\sqrt{%
p^{2}_{1}+m_{1}^{2}}) (\kappa\sqrt{p^{2}_{2}+m_{2}^{2}}+2K_{0}-2K_{-})}{%
(4K_{0})^{2}} +\frac{1}{4}(K_{+}-K_{0}-\frac{\kappa\epsilon}{2}p_{1})z_{1} 
\nonumber \\
&&-\frac{1}{4}(K_{0}+K_{-}+\frac{\kappa\epsilon}{2}p_{2})z_{2}\;,
\label{a0sol} \\
\mbox{log}|B_{0}|&=&\frac{1}{2}\mbox{log}\frac{ (2K_{0}+2K_{-}-\kappa\sqrt{%
p^{2}_{2}+m_{2}^{2}}) (\kappa\sqrt{p^{2}_{1}+m_{1}^{2}}+2K_{0}-2K_{+})}{%
(4K_{0})^{2}} +\frac{1}{4}(K_{0}+K_{+}-\frac{\kappa\epsilon}{2}p_{1})z_{1} 
\nonumber \\
&&-\frac{1}{4}(K_{-}-K_{0}+\frac{\kappa\epsilon}{2}p_{2})z_{2}\;.
\label{b0sol}
\end{eqnarray}
The parameters $C_{\pm}$ are determined from (\ref{cond4}) and (\ref{cond5}%
). From (\ref{a/b-1}), (\ref{a/b-2}) and (\ref{a+b-2}) it is evident that an
overall common sign of $A_{+}, B_{-}, A_{0}$ and $B_{0}$ has no physical
meaning, and so we can choose all these coefficients to be positive.

The previous expressions are somewhat cumbersome. We can express them more
compactly by making use of the following notation: 
\begin{eqnarray}  \label{nota}
K_{1}&\equiv& 2K_{0}+2K_{-}-\kappa\sqrt{p_{2}^{2}+m_{2}^{2}}\;,  \nonumber \\
K_{2}&\equiv& 2K_{0}+2K_{+}-\kappa\sqrt{p_{1}^{2}+m_{1}^{2}}\;,  \nonumber \\
K_{01}&\equiv& K_{0}-K_{+}+\frac{\kappa\epsilon}{2}p_{1}\;,  \nonumber \\
K_{02}&\equiv& K_{0}-K_{-}-\frac{\kappa\epsilon}{2}p_{2}\;,  \nonumber \\
{\cal M}_{1}&\equiv& \kappa\sqrt{p_{1}^{2}+m_{1}^{2}}+2K_{0}-2K_{+}\;, 
\nonumber \\
{\cal M}_{2}&\equiv& \kappa\sqrt{p_{2}^{2}+m_{2}^{2}}+2K_{0}-2K_{-}\;, \\
Y_{+}&\equiv& \kappa\left[X+\frac{\epsilon}{4}(p_{1}+p_{2})\right]\;, 
\nonumber \\
Y_{0}&\equiv& \kappa\left[X-\frac{\epsilon}{4}(p_{1}-p_{2})\right]\;, 
\nonumber \\
Y_{-}&\equiv& \kappa\left[X-\frac{\epsilon}{4}(p_{1}+p_{2})\right]\;\;. 
\nonumber
\end{eqnarray}
The coefficients $A_{+}, B_{-}, A_{0}$ and $B_{0}$ can then be rewritten as 
\begin{eqnarray}  \label{ab+-0}
A_{+}&=&\left(\frac{K_{1}}{{\cal M}_{1}}\right)^{1/2}e^{-\frac{1}{4}
(K_{01}z_{1}-K_{02}z_{2})-\frac{1}{2}K_{+}z_{1}}\;,  \nonumber \\
B_{-}&=&\left(\frac{K_{2}}{{\cal M}_{2}}\right)^{1/2}e^{-\frac{1}{4}
(K_{01}z_{1}-K_{02}z_{2})+\frac{1}{2}K_{-}z_{2}}\;,  \nonumber \\
\\
A_{0}&=&\frac{(K_{2}{\cal M}_{2})^{1/2}}{4K_{0}}e^{-\frac{1}{4}
(K_{01}z_{1}-K_{02}z_{2})-\frac{1}{2}K_{0}z_{2}}\;,  \nonumber \\
B_{0}&=&\frac{(K_{1}{\cal M}_{1})^{1/2}}{4K_{0}}e^{-\frac{1}{4}
(K_{01}z_{1}-K_{02}z_{2})+\frac{1}{2}K_{0}z_{1}}\;,  \nonumber
\end{eqnarray}
and the solution for $\phi$ is then 
\begin{eqnarray}  \label{phi-sol2}
\phi_{+}&=&\left(\frac{K_{1}}{{\cal M}_{1}}\right)^{\frac{1}{2}}\; e^{-\frac{%
1}{4}(K_{01}z_{1}-K_{02}z_{2})+\frac{1}{2}K_{+}(x-z_{1})}\;,  \nonumber \\
\phi_{0}&=&\frac{1}{4K_{0}}\;e^{-\frac{1}{4}(K_{01}z_{1}-K_{02}z_{2})}
\left\{(K_{1}{\cal M}_{1})^{1/2}e^{-\frac{1}{2}K_{0}(x-z_{1})} +(K_{2}{\cal M%
}_{2})^{1/2}e^{\frac{1}{2}K_{0}(x-z_{2})}\right\}\;, \\
\phi_{-}&=&\left(\frac{K_{2}}{{\cal M}_{2}}\right)^{\frac{1}{2}}\; e^{-\frac{%
1}{4}(K_{01}z_{1}-K_{02}z_{2})-\frac{1}{2}K_{-}(x-z_{2})}\;.  \nonumber
\end{eqnarray}

Repeating the analysis for $z_{1}<z_{2}$ yields a similar solution with $%
p_{i} \rightarrow -p_{i}$. Hence the full solution is obtained from the
preceding expressions by replacing $p_{i}$ and $z_{1}-z_{2}$ by $\tilde{p}%
_{i}=p_{i}\;\makebox{sgn}(z_{1}-z_{2})$ and $|z_{1}-z_{2}|$, respectively.
The determining equation (\ref{X}) of the Hamiltonian is then expressed as 
\begin{equation}  \label{H1}
K_{1}K_{2}={\cal M}_{1}{\cal M}_{2}\;e^{K_{0}|z_{1}-z_{2}|}\;.
\end{equation}
or 
\begin{eqnarray}  \label{H2}
&&\left(4K_{0}^{2}+[\kappa\sqrt{p_{1}^{2}+m^{2}_{1}}-2K_{+}] [\kappa\sqrt{%
p_{2}^{2}+m^{2}_{2}}-2K_{-}]\right) \mbox{tanh}\left(\frac{1}{2}%
K_{0}|z_{1}-z_{2}|\right)  \nonumber \\
&&\makebox[5em]{}=-2K_{0} \left([\kappa\sqrt{p_{1}^{2}+m^{2}_{1}}-2K_{+}]
+[\kappa\sqrt{p_{2}^{2}+m^{2}_{2}}-2K_{-}]\right)\;,
\end{eqnarray}
where the momentum $p_{i}$ is replaced by $\tilde{p}_{i}$.

For the expression (\ref{ham2}) to have a definite meaning as the
Hamiltonian, $K_{\pm}$ should be real. This imposes the restriction $H^2 +
8\Lambda/\kappa^2 > 16(p_1+p_2)^2 $. However $K_{0}$ need not be real. If $%
\Lambda$ takes a sufficiently large positive value $K_{0}$ will be imaginary
and the above analysis must be repeated. In the (0) region the soluton to
the $\phi$ equation (\ref{phi-eq}) becomes 
\begin{equation}  \label{phi-0}
\phi_{0}(x)= A_{s}\;\mbox{sin}\frac{1}{2}\tilde{K}_{0}x + A_{c}\;\mbox{cos}%
\frac{1}{2}\tilde{K}_{0}x \;,
\end{equation}
where 
\begin{eqnarray}
\tilde{K}_{0}&=& -i K_{0}  \nonumber \\
&=&\sqrt{\frac{1}{2}\Lambda -\kappa^{2}\left(X-\frac{\epsilon}{4}(\tilde{p}%
_{1}-\tilde{p}_{2})\right)^{2}} \;.
\end{eqnarray}
Under the same matching conditions (\ref{match1}-\ref{match4}) and the
boundary condition (\ref{bound}) we get, instead of (\ref{H2}), a new
determining equation for the Hamiltonian 
\begin{eqnarray}  \label{H2new}
&&\left(4\tilde{K}_{0}^{2}-[\kappa\sqrt{p_{1}^{2}+m^{2}_{1}}-2K_{+}] [\kappa%
\sqrt{p_{2}^{2}+m^{2}_{2}}-2K_{-}]\right) \mbox{tan}\left(\frac{1}{2}\tilde{K%
}_{0}|z_{1}-z_{2}|\right)  \nonumber \\
&&\makebox[5em]{}=2\tilde{K}_{0} \left([\kappa\sqrt{p_{1}^{2}+m^{2}_{1}}%
-2K_{+}] +[\kappa\sqrt{p_{2}^{2}+m^{2}_{2}}-2K_{-}]\right)\;,
\end{eqnarray}
which is just the equation derived from (\ref{H2}) by formally replacing $%
K_{0}$ with $i\tilde{K}_{0}$. Similarly, the solution for $\phi$ for
imaginary $K_{0}$ is also identical with that obtained from (\ref{phi-sol2})
by the same replacement. Hence equation (\ref{H2}) is valid for all values
of $K_{0}$, and may be regarded as a transcendental equation which
determines $H$ as a function of the independent coordinates and momenta of
the system.

We have previously shown that in the case of zero cosmological constant the
solution for $H$ can be expressed in terms of the Lambert $W$ function. In
this more general case with $\Lambda\neq 0$ the solution for $H$ from (\ref
{H2}) cannot expressed in terms of known functions. Rather we must regard $H$
as being implicitly determined in terms of the coordinates and momenta via (%
\ref{H2}).

Finally, the components of the metric may be computed from the equations (%
\ref{e-pi}), (\ref{e-gamma}), (\ref{e-Pi}) and (\ref{e-Psi}) under the
coordinate conditions (\ref{cc}). The derivation and the explicit solutions
of the metric are given in Appendix A.

The canonical equations for the 2-body system can be derived by
differentiating the determining equation (\ref{H1}) with respect to the
variables $z_{i}$ and $p_{i}$ . For the variables $p_{1}$ and $z_{1}$ this
yields 
\begin{eqnarray}  \label{p1}
\dot{p}_{1}&=&-\frac{\partial H}{\partial z_{1}} =-\frac{2}{\kappa}\left(%
\frac{\partial K_{+}}{\partial z_{1}} +\frac{\partial K_{-}}{\partial z_{1}}%
\right) =-2\left(\frac{Y_{+}}{K_{+}}+\frac{Y_{-}}{K_{-}}\right) \frac{%
\partial X}{\partial z_{1}}  \nonumber \\
&=&-\frac{2}{\kappa}\left(\frac{Y_{+}}{K_{+}}+\frac{Y_{-}}{K_{-}}\right) 
\frac{K_{0}K_{1}K_{2}}{J}\;,
\end{eqnarray}
\begin{eqnarray}  \label{z1}
\dot{z}_{1}&=&\frac{\partial H}{\partial p_{1}} =\frac{2}{\kappa}\left(\frac{%
\partial K_{+}}{\partial p_{1}} +\frac{\partial K_{-}}{\partial p_{1}}\right)
\nonumber \\
&=&\frac{\epsilon}{2}\left(\frac{Y_{+}}{K_{+}}-\frac{Y_{-}}{K_{-}}\right)
+2\left(\frac{Y_{+}}{K_{+}}+\frac{Y_{-}}{K_{-}}\right) \frac{\partial X}{%
\partial p_{1}}  \nonumber \\
&=&\epsilon\frac{Y_{+}}{K_{+}}+\frac{8}{J}\left(\frac{Y_{+}}{K_{+}} +\frac{%
Y_{-}}{K_{-}}\right)\frac{K_{0}K_{1}}{{\cal M}_{1}} \left\{\frac{p_{1}}{%
\sqrt{p_{1}^{2}+m_{1}^{2}}}-\epsilon\frac{Y_{+}}{K_{+}} \right\}\;,
\end{eqnarray}
where 
\begin{eqnarray}  \label{J}
J&=&2\left\{\left(\frac{Y_{0}}{K_{0}}+\frac{Y_{+}}{K_{+}}\right)K_{1} +\left(%
\frac{Y_{0}}{K_{0}}+\frac{Y_{-}}{K_{-}}\right)K_{2}\right\}  \nonumber \\
&&-2\left\{\left(\frac{Y_{0}}{K_{0}}-\frac{Y_{+}}{K_{+}}\right) \frac{1}{%
{\cal M}_{1}}+\left(\frac{Y_{0}}{K_{0}}-\frac{Y_{-}}{K_{-}}\right) \frac{1}{%
{\cal M}_{2}}\right\}K_{1}K_{2} -\frac{Y_{0}}{K_{0}}K_{1}K_{2}(z_{1}-z_{2})%
\;.
\end{eqnarray}

Similarly, for particle 2 the equations are 
\begin{eqnarray}
\dot{p}_{2}&=&\frac{2}{\kappa}\left(\frac{Y_{+}}{K_{+}}+\frac{Y_{-}}{K_{-}}
\right)\frac{K_{0}K_{1}K_{2}}{J}\;,  \label{p2} \\
\dot{z}_{2}&=&-\epsilon\frac{Y_{-}}{K_{-}}+\frac{8}{J}\left(\frac{Y_{+}}{%
K_{+}} +\frac{Y_{-}}{K_{-}}\right)\frac{K_{0}K_{2}}{{\cal M}_{2}} \left\{%
\frac{p_{2}}{\sqrt{p_{2}^{2}+m_{2}^{2}}}+\epsilon\frac{Y_{-}}{K_{-}}
\right\}\;.  \label{z2}
\end{eqnarray}
It is straightforward to show that these canonical equations guarantee the
conservation of the Hamiltonian (i.e. $\dot{H}=0$) and the total momentum $%
p_{1}+p_{2}$ (i.e. $\dot{p}_1 + \dot{p}_2 = 0$).

Alternatively, the equations of motion (\ref{e-p}) and (\ref{e-z}) derived
from the action (\ref{act2}) become 
\begin{eqnarray}
\dot{p}_{a}&=&-\frac{\partial N_{0}}{\partial z_{a}}\sqrt{p^{2}_{a}
+m^{2}_{a}}+\frac{\partial N_{1}}{\partial z_{a}}p_{a}\;,  \label{pa} \\
\dot{z_{a}}&=&N_{0}\frac{p_{a}}{\sqrt{p^{2}_{a}+m^{2}_{a}}}-N_{1}\;,
\label{za}
\end{eqnarray}
under the coordinate conditions (\ref{cc}). Insertion of the solutions of
the metric components given in the Appendix A into (\ref{pa}) and (\ref{za})
reproduces the canonical equations of motion (\ref{p1}), (\ref{z1}), (\ref
{p2}) and (\ref{z2}) when the partial derivatives at $z{1}, z{2}$ are
defined by 
\begin{equation}
\frac{\partial N_{0,1}}{\partial z_{i}}\equiv \frac{1}{2}\left\{\left. \frac{%
\partial N_{0,1}}{\partial x}\right|_{x=z_{i}+0} +\left.\frac{\partial
N_{0,1}}{\partial x}\right|_{x=z_{i}-0}\right\}\;.
\end{equation}
Thus consistency between the geodesic equations and the canonical equations
of motion is explicitly verified, while the formal proof of the consistency
in the case of $\Lambda=0$ \cite{OR} can be easily generalized to the $%
\Lambda \neq 0$ case.

\section{EXACT SOLUTIONS OF THE TRAJECTORIES IN THE $\Lambda=0$ CASE}

\label{exactL0}

In a previous paper \cite{2bd} we showed that in the $\Lambda=0$ case the
determining equation (\ref{H1}) can be solved explicitly and the Hamiltonian
for the equal mass is expressed in the center of inertia (C.I.) system $p_1
= -p_2 = p$ as 
\begin{equation}  \label{Ham-0}
H=\sqrt{p^2 + m^2} + \epsilon p \;\mbox{sgn}(r) - 8\;\frac{W\left[-\frac{%
\kappa}{8}(|r|\sqrt{p^2 + m^2} - \epsilon p r) \mbox{exp}\left(\frac{\kappa}{%
8}(|r|\sqrt{p^2 + m^2} - \epsilon p r\right)\right]}{\kappa |r|}\;,
\end{equation}
where $W(x)$ is the Lambert $W$ function defined via 
\begin{equation}  \label{lambertW}
y\cdot e^{y}= x \quad \Rightarrow \quad y=W(x)\;.
\end{equation}
The $W(x)$ has two real branches $W_0$ and $W_{-1}$ for real $x$ \cite
{Corless}.

The Hamiltonian (\ref{Ham-0}) is exact to infinite order in the
gravitational constant and for arbitrary values of $m$ and $p$. We can view
the whole structure of the theory from the weak field to the strong field
limits. By setting $H=H_0$ we can draw a phase space trajectory in $(r, p)$
space. This phase space trajectory should be, as a matter of course,
obtainable directly from the solution $r(t), p(t)$ to the canonical
equations by eliminating the time variable $t$. Indeed, this can be verified
by numerically solving the equations 
\begin{eqnarray}
\dot{p}&=& -\;\frac{\frac{\kappa}{4}(H - 2\epsilon \tilde{p}) (H - \epsilon%
\tilde{p} - \sqrt{p^2 + m^2})} {2 - \frac{\kappa r}{4}(H - \epsilon \tilde{p}
- \sqrt{p^2 + m^2})} \;\mbox{sgn}(r)\;,  \label{p-eq0} \\
\dot{r}&=& 2\epsilon\left\{1 - \frac{H - 2\epsilon\tilde{p}} {2 - \frac{%
\kappa r}{4}(H - \epsilon \tilde{p} - \sqrt{p^2 + m^2})} \cdot\frac{1}{\sqrt{%
p^2 + m^2}}\right\}\;\mbox{sgn}(r)\;,  \label{r-eq0}
\end{eqnarray}
which are obtained in the case of $\Lambda=0$ from (\ref{p1}), (\ref{z1}), (%
\ref{p2}) and (\ref{z2}). However for certain values of the parameters,
superficial singularities appear in $r(t)$ and $p(t)$ due to the zero points
of the denominator $\left\{2 - \kappa r(H - \epsilon \tilde{p} - \sqrt{p^2 +
m^2})/4\right\}$. These singularities correspond to $W(x)=-1$ representing
the transit point between two branches $W_0$ and $W_{-1}$. In a spacetime
description the singularities are coordinate singularities and are a
consequence of $t$ being a coordinate time.

We can deal with this problem by describing the trajectories of the
particles in terms of some invariant parameter. The natural candidate is the
proper time $\tau_a$ of each particle. From the metric components given in
Appendix A and the canonical equations (\ref{za}), the proper time is 
\begin{eqnarray}
d\tau_{a}^{2}&=&dt^{2}\left\{N_{0}(z_{a})^{2}-(N_{1}(z_{a})+\dot{z}_{a})^{2}
\right\}  \nonumber \\
&=&dt^{2}N_{0}(z_{a})^{2}\frac{m_{a}^{2}}{p_{a}^{2}+m_{a}^{2}} \qquad\qquad
(a=1,2)\;\;.  \label{tau-0}
\end{eqnarray}
For the equal mass case it is common for both particles 
\begin{equation}
d\tau=d\tau_1=d\tau_2=\frac{(H - 2\epsilon\tilde{p})m} {\left\{2 - \frac{%
\kappa r}{4}(H - 2\epsilon \tilde{p})\right\}(\sqrt{p^2 + m^2} - \epsilon%
\tilde{p})\sqrt{p^2 + m^2}}\;dt \;,
\end{equation}
and the canonical equations (\ref{p-eq0}) and (\ref{r-eq0}) become 
\begin{eqnarray}
\frac{dp}{d\tau}&=&-\frac{\kappa}{4m}\sqrt{p^2 + m^2}\left\{ H(\sqrt{p^2 +
m^2} - \epsilon\tilde{p}) - m^2\right\}\mbox{sgn}(r)\;,  \label{p-eqtau0} \\
\frac{dr}{d\tau}&=&\frac{2\epsilon}{m}(\sqrt{p^2 + m^2} - \epsilon\tilde{p})
\left\{\frac{\left[2 - \frac{\kappa r}{4}(H - \epsilon\tilde{p} - \sqrt{p^2
+ m^2})\right]\sqrt{p^2 + m^2}}{H - 2\epsilon\tilde{p}} - 1\right\}\mbox{sgn}%
(r) \;.  \label{r-eqtau0}
\end{eqnarray}
Remarkably, the equations (\ref{p-eqtau0}) and (\ref{r-eqtau0}) have an
exact solution. We can obtain it in the following way. First we solve Eq.(%
\ref{p-eqtau0}) for $p(\tau)$. From this $r(\tau)$ can be extracted by
directly solving (\ref{r-eqtau0}) after substituting the solution for $p$ or
by solving (\ref{H1}) for $r$. This yields an exact expression for the
proper separation $r$ of the two bodies as a function of their mutual proper
time. Note that since $\gamma=1$, at a fixed time $dt=0$ (and hence a fixed $%
d\tau=0$), the separation $r=z_1 - z_2$ is the proper distance between the
two particles.

In the $r > 0$ region the solution is 
\begin{eqnarray}
p(\tau)&=&\frac{\epsilon m}{2}\left(f_{0}(\tau) -\frac{1}{f_{0}(\tau)}%
\right)\;,  \label{p-exact00} \\
r(\tau)&=&\frac{16}{\kappa\left\{H - m\left(f_{0} - \frac{1}{f_{0}}\right)
\right\}}\mbox{tanh}^{-1}\left[\frac{H - m\left(f_{0} + \frac{1}{f_{0}}%
\right)} {H - m\left(f_{0} - \frac{1}{f_{0}}\right)}\right]\;,
\label{r-exact00}
\end{eqnarray}
with 
\begin{equation}
f_{0}(\tau)=\frac{H}{m}\left[1 - \frac{\sqrt{p_{0}^{2} + m^2} - \epsilon
p_{0} - \frac{m^2}{H}}{\sqrt{p_{0}^{2} + m^2} - \epsilon p_{0}}\; e^{\frac{%
\epsilon\kappa m}{4}(\tau - \tau_{0})}\right]\;,
\end{equation}
where $p_{0}$ is the initial momentum at $\tau=\tau_{0}$. In the $r < 0$
region the solution is 
\begin{eqnarray}
p(\tau)&=&-\frac{\epsilon m}{2}\left(\bar{f}_{0}(\tau) -\frac{1}{\bar{f}%
_{0}(\tau)}\right)\;,  \label{p-exact01} \\
r(\tau)&=&\frac{-16}{\kappa\left\{H - m\left(\bar{f}_{0} - \frac{1}{\bar{f}%
_{0}}\right)\right\}} \mbox{tanh}^{-1}\left[\frac{H - m\left(\bar{f}_{0} + 
\frac{1}{\bar{f}_{0}}\right)} {H - m\left(\bar{f}_{0} - \frac{1}{\bar{f}_{0}}%
\right)}\right]\;,  \label{r-exact01}
\end{eqnarray}
with 
\begin{equation}
\bar{f}_{0}(\tau)=\frac{H}{m}\left[1 - \frac{\sqrt{p_{0}^{2} + m^2} +
\epsilon p_{0} - \frac{m^2}{H}} {\sqrt{p_{0}^{2} + m^2} + \epsilon p_{0}}\;
e^{\frac{\epsilon\kappa m}{4}(\tau - \tau_{0})}\right]\;.
\end{equation}

In Fig.1 and 2 we show the typical plots of $r(\tau)$ and phase space
trajectories for various values of $H_0$ with $m=0.5$. (In all plots in this
paper we choose $\kappa=1$.) In Fig.1 it is seen that as $H_0$ increases,
not only the amplitude and the period of the bounded motion become large,
but also the shape of $r(\tau)$ deforms from the parabolic (Newtonian)
shape. This deformation is also present in the phase space trajectories
shown in Fig.2. At higher energy the trajectories become more and more $S$
shaped. This is due to the fact that the trajectory smoothly moves over the
two branches $W_0$ and $W_{-1}$ \cite{2bd}.

We see in Fig.3 the striking distinction between the separations of the
particles in the non-relativistic and relativistic cases once the value of
the energy becomes large relative to the mass (here $H_0=25$ and $m=0.5$).
The maximal separation of the particles is much smaller than in its
Newtonian counterpart (the dashed curve), and is achieved far more quickly.
After maximal separation, the particles move toward each other at a slower
velocity until they are very close together. At this point (less than 10\%
of their maximal separation), they accelerate toward the same point, after
which the motion repeats with the particles interchanged.

\hspace*{5cm}--------------------------

\vspace{5mm}

\hspace*{7cm}Fig.1

\vspace{5mm}

\begin{center}
{\it The exact $r$ vs $\tau$ curves in the case of $\Lambda=0$ for $m=0.5$
and four different values of $H_0$. }
\end{center}

\hspace*{5cm}--------------------------

\hspace*{5cm}--------------------------

\vspace{5mm}

\hspace*{7cm}Fig.2

\vspace{5mm}

\begin{center}
{\it Phase space trajectories corresponding to the $r(\tau)$ curves in
Fig.1. }
\end{center}

\hspace*{5cm}--------------------------

\hspace*{5cm}--------------------------

\vspace{5mm}

\hspace*{7cm}Fig.3

\vspace{5mm}

\begin{center}
{\it The exact $r$ vs $\tau$ curve for $H_0=25$ with $m=0.5$ and the
Newtonian curve for the same $H_0$ . }
\end{center}

\hspace*{5cm}--------------------------

\section{EXACT SOLUTIONS OF THE TRAJECTORIES FOR THE EQUAL MASS IN THE $%
\Lambda \neq 0$ CASE}

\subsection{General Discussion}

In this section we consider a system of two particles with equal mass for
the $\Lambda \neq 0$ case. In the C.I. system, depending upon the sign of $(%
\sqrt{H^2+8\Lambda/\kappa^2} -2\epsilon \tilde{p})^{2}-8\Lambda/\kappa^2$
the determining equations (\ref{H2}) and (\ref{H2new}) become 
\begin{equation}  \label{deteq-1}
({\cal J}_{\Lambda}^{\;2}+B^{2})\; \mbox{tanh}\left(\frac{\kappa}{8}{\cal J}%
_{\Lambda}\;|r|\right) =2{\cal J}_{\Lambda}B\;,
\end{equation}
and 
\begin{equation}  \label{deteq-2}
(\tilde{{\cal J}}_{\Lambda}^{\;2}-B^{2})\; \mbox{tan}\left(\frac{\kappa}{8}%
\tilde{{\cal J}}_{\Lambda}\;|r|\right) =-2\tilde{{\cal J}}_{\Lambda}B \;,
\end{equation}
respectively, where 
\begin{eqnarray}
&&{\cal J}_{\Lambda}=\sqrt{\left(\sqrt{H^2+\frac{8\Lambda}{\kappa^2}}
-2\epsilon \tilde{p}\right)^{2}-\frac{8\Lambda}{\kappa^2}}\;,  \nonumber \\
&&\tilde{{\cal J}}_{\Lambda}=\sqrt{\frac{8\Lambda}{\kappa^2}-\left(\sqrt{H^2+%
\frac{8\Lambda}{\kappa^2}} -2\epsilon \tilde{p}\right)^{2}}\;,  \nonumber \\
&&B=H-2\sqrt{p^2+m^2}\;.
\end{eqnarray}
Equation (\ref{deteq-1}) may be further divided into two types: 
\begin{equation}  \label{TanhA}
\mbox{tanh}\left(\frac{\kappa}{16}{\cal J}_{\Lambda}|r|\right) =\frac{B}{%
{\cal J}_{\Lambda}} \qquad \mbox{(tanh-type A)}\;,
\end{equation}
or 
\begin{equation}  \label{TanhB}
\mbox{tanh}\left(\frac{\kappa}{16}{\cal J}_{\Lambda}|r|\right) =\frac{{\cal J%
}_{\Lambda}}{B} \qquad \mbox{(tanh-type B)}\;.
\end{equation}
When $\Lambda=0$ the tanh-type B equation is excluded because ${\cal J}%
_{\Lambda}/B$ exceeds 1. When a cosmological constant is introduced, this
equation has solutions in some restricted range of the parameters.

Likewise, eq.(\ref{deteq-2}) may also be divided into 
\begin{equation}  \label{TanA}
\mbox{tan}\left(\frac{\kappa}{16}\tilde{{\cal J}}_{\Lambda}|r|\right)= -\;%
\frac{B}{\tilde{{\cal J}}_{\Lambda}} \qquad \mbox{(tan-type A)}\;,
\end{equation}
or 
\begin{equation}  \label{TanB}
\mbox{tan}\left(\frac{\kappa}{16}\tilde{{\cal J}}_{\Lambda}|r|\right)= \frac{%
\tilde{{\cal J}}_{\Lambda}}{B} \qquad \mbox{(tan-type B)}\;.
\end{equation}

For all of these four types of determining equations the canonical equations
of motion are identical: 
\begin{eqnarray}
\dot{p}&=&-\;\frac{\kappa{\cal J}_{\Lambda}^{2}({\cal J}_{\Lambda}^{2}-B^2)%
} {16C}\;\mbox{sgn(r)}\;,  \label{can-p} \\
\dot{r}&=&2\epsilon \;\sqrt{1+\frac{8\Lambda}{\kappa^2 H^2}} \left(1-\frac{%
{\cal J}_{\Lambda}^{2}}{C}\right)\mbox{sgn(r)} +\frac{2{\cal J}_{\Lambda}^{2}%
}{C}\frac{p}{\sqrt{p^{2}+m^{2}}}\;.  \label{can-z}
\end{eqnarray}
where 
\begin{equation}  \label{C}
C=\frac{1}{\sqrt{1+\frac{8\Lambda}{\kappa^2 H^2}}} \left\{\sqrt{1+\frac{%
8\Lambda}{\kappa^2 H^2}}\;{\cal J}_{\Lambda}^{2} -\left(\sqrt{H^2+\frac{%
8\Lambda}{\kappa^2}}-2\epsilon\tilde{p}\right) \left(B+\frac{\kappa}{16}(%
{\cal J}_{\Lambda}^{2}-B^2)\;r \right)\right\}\;.
\end{equation}

\subsection{Phase-space Trajectories}

For a given value of $\Lambda$ the equations (\ref{deteq-1}) or (\ref
{deteq-2}) describe the surface in $(r, p, H)$ space of all allowed
phase-space trajectories, from which the $(r, p)$ trajectory is obtained by
setting $H=H_0$.

Fig.4 shows phase-space plots for a small $H_0=2.2m$ and three different
values of $\Lambda$ under identical initial conditions. First we note that
the trajectories of the relativistic motion (solid curve) are are slightly
distorted compared to the Newtonian motion (dashed curve). Second, a
trajectory with $\Lambda > 0$ (dash-dot curve) is expanded, reflecting the
repulsive effect of the cosmological constant, while a trajectory of $%
\Lambda < 0$ (dotted curve) shrinks, due to the additional attractive
effect. The relativistic plots in Fig.4 correspond to the choice of $%
\epsilon =1$. Plots for $\epsilon =-1$ (the time-reversed solutions) are
obtained by reflection in the $p=0$ axis. The phase space plots for a large $%
H_0=8m$ are shown in Fig.5. The trajectories for $\Lambda\geq 0$ become
extremely $S$-shaped, while for $\Lambda < 0$ the trajectory is still a
distorted oval due to the attractive effect of $\Lambda$. The effects of the
cosmological constant ($\Lambda > 0$ repulsive; $\Lambda < 0$ attractive)
are precisely analyzed in terms of the exact $r(\tau)$ plots in the next
subsection.

\hspace*{5cm}--------------------------

\vspace{5mm}

\hspace*{7cm}Fig.4

\vspace{5mm}

\begin{center}
{\it Non-relativistic (Newtonian) and relativistic trajectories for $%
m=0.5,H_0=1.1$ and three different values of $\Lambda$. The undistorted oval
(which is really a pair of parabolas intersecting at the $r=0$ axis) is the
Newtonian trajectory. }
\end{center}

\hspace*{5cm}--------------------------

\hspace*{5cm}--------------------------

\vspace{5mm}

\hspace*{7cm}Fig.5

\vspace{5mm}

\begin{center}
{\it Relativistic trajectories for $m=0.5,H_0=4$ and three different values
of $\Lambda$. }
\end{center}

\hspace*{5cm}--------------------------

\subsection{Explicit Solutions}

The phase-space trajectories discussed in the previous subsection can be
obtained from the solution to the canonical equations (\ref{can-p}) and (\ref
{can-z}). For the equal mass case there is a common proper time for both
particles 
\begin{equation}  \label{Tau}
d\tau=d\tau_{1}=d\tau_{2}=\frac{m}{\sqrt{p^2+m^2}} \frac{{\cal J}_{\Lambda}^2%
}{C}dt\;,
\end{equation}
via which the canonical equations (\ref{can-p}) and (\ref{can-z}) may be
expressed in the form 
\begin{eqnarray}
&&\frac{dp}{d\tau}=-\;\frac{\kappa\sqrt{p^2+m^2}({\cal J}_{\Lambda}^2-B^2)} {%
16m}\;\mbox{sgn(r)}\;,  \label{p-Tau} \\
&&\frac{dr}{d\tau}=\frac{2\epsilon}{m} \left\{\sqrt{1+\frac{8\Lambda}{%
\kappa^2 H^2}} \left(\frac{C}{{\cal J}_{\Lambda}^2}-1\right)\sqrt{p^2+m^2}
+\epsilon\tilde{p}\right\}\mbox{sgn(r)} \;.  \label{r-Tau}
\end{eqnarray}

First we solve Eq.(\ref{p-Tau}) for $p$ and then (\ref{TanhA})-(\ref{TanB})
for $r$. In the $r>0$ region Eq.(\ref{p-Tau}) leads to 
\begin{eqnarray}  \label{p-int}
\int_{p_{0}}^{p}\;\frac{dp}{\left\{\sqrt{p^2+m^2} -\epsilon\sqrt{1+\frac{%
8\Lambda}{\kappa^2 H^2}}\;p -\frac{m^2}{H}\right\} \sqrt{p^2+m^2}}&=&-\;%
\frac{\kappa H}{4m}\int_{\tau_{0}}^{\tau}\;d\tau  \nonumber \\
&=&-\;\frac{\kappa H}{4m}(\tau-\tau_{0})\;,
\end{eqnarray}
provided the condition 
\begin{equation}
1+\frac{8\Lambda}{\kappa^2 H^2} \geq 0
\end{equation}
is satified. Hence for $\Lambda<0$ the motion is allowed as long as $H$
satisfies 
\begin{equation}  \label{H-bound}
H \geq \sqrt{-\;\frac{8\Lambda}{\kappa^2}}\;.
\end{equation}
We perfom the integration of the LHS of (\ref{p-int}) for three separate
cases which depend on the value of $\Lambda$ relative to $m$ and $H$. The
solution $p(\tau)$ is 
\begin{equation}  \label{p-exact1}
p(\tau)=\frac{\epsilon m}{2}\left(f(\tau) -\frac{1}{f(\tau)}\right)\;,
\end{equation}
with 
\begin{equation}
f(\tau) = \left\{ 
\begin{array}{ll}
\frac{\frac{H}{m}\left(1+\sqrt{\gamma_H}\right) \left\{1-\eta\;e^{\frac{%
\epsilon\kappa m}{4}\sqrt{\gamma_m}(\tau-\tau_{0})}\right\}} {1+\sqrt{%
\gamma_m} +\left(\sqrt{\gamma_m}-1\right) \;\eta\;e^{\frac{\epsilon\kappa m}{%
4}\sqrt{\gamma_m}(\tau-\tau_{0})}} & \qquad \gamma_{m}> 0 \;, \\ 
&  \\ 
\frac{1+\sqrt{\gamma_H}} {\frac{m}{H}+\frac{\sigma}{m-\sigma\frac{%
\epsilon\kappa H}{8}(\tau-\tau_{0})}} & \qquad \gamma_{m}=0 \;, \\ 
&  \\ 
\frac{\frac{H}{m}(1+ \sqrt{\gamma_H})} {1+\sqrt{-\gamma_m} \frac{\sigma+%
\frac{m^2}{H}\sqrt{-\gamma_m} \tan\left[\frac{\epsilon\kappa m}{8}\sqrt{%
-\gamma_m}(\tau-\tau_{0})\right]} {\frac{m^2}{H}\sqrt{-\gamma_m} -\sigma \tan%
\left[\frac{\epsilon\kappa m}{8}\sqrt{-\gamma_m}(\tau-\tau_{0})\right]}} & 
\qquad \gamma_{m} < 0 \;,
\end{array}
\right.
\end{equation}
where 
\begin{equation}
\begin{array}{ll}
\gamma_{H}= 1+\frac{8\Lambda}{\kappa^2 H^2}\;, & \qquad \gamma_{m}= 1+\frac{%
8\Lambda}{\kappa^2 m^2}\;, \\ 
\eta=\frac{\sigma -\frac{m^2}{H}\sqrt{\gamma_m}} {\sigma + \frac{m^2}{H}%
\sqrt{\gamma_m}}\;, & \qquad \sigma = (1+\sqrt{\gamma_H}) (\sqrt{p_{0}^2+m^2}%
-\epsilon p_{0})-\frac{m^2}{H}\;,
\end{array}
\end{equation}
with $p_0$ being the initial momentum at $\tau=\tau_0$.

Similarly the solution in $r<0$ region is 
\begin{equation}  \label{p-exact2}
p(\tau)=-\frac{\epsilon m}{2}\left(\bar{f}(\tau)-\frac{1}{\bar{f}(\tau)}
\right)\;,
\end{equation}
with 
\begin{equation}
\bar{f}(\tau) = \left\{ 
\begin{array}{ll}
\frac{\frac{H}{m}\left(1+\sqrt{\gamma_H}\right) \left\{1-\bar{\eta}\;e^{%
\frac{\epsilon\kappa m}{4}\sqrt{\gamma_m}(\tau-\tau_{0})}\right\}} {1+\sqrt{%
\gamma_m} +\left(\sqrt{\gamma_m}-1\right) \;\bar{\eta}\;e^{\frac{%
\epsilon\kappa m}{4}\sqrt{\gamma_m}(\tau-\tau_{0})}} & \qquad \gamma_{m}> 0
\;, \\ 
&  \\ 
\frac{1+\sqrt{\gamma_H}} {\frac{m}{H}+\frac{\bar{\sigma}} {m-\bar{\sigma}%
\frac{\epsilon\kappa H}{8}(\tau-\tau_{0})}} & \qquad \gamma_{m}=0 \;, \\ 
&  \\ 
\frac{\frac{H}{m}(1+ \sqrt{\gamma_H})} {1+\sqrt{-\gamma_m} \frac{\bar{\sigma}%
+\frac{m^2}{H}\sqrt{-\gamma_m} \tan\left[\frac{\epsilon\kappa m}{8}\sqrt{%
-\gamma_m}(\tau-\tau_{0})\right]} {\frac{m^2}{H}\sqrt{-\gamma_m} -\bar{\sigma%
} \tan\left[\frac{\epsilon\kappa m}{8}\sqrt{-\gamma_m}(\tau-\tau_{0})\right]}%
} & \qquad \gamma_{m} < 0 \;,
\end{array}
\right.
\end{equation}
where 
\begin{equation}
\bar{\sigma} = (1+\sqrt{\gamma_H}) (\sqrt{p_{0}^2+m^2}+\epsilon p_{0})-\frac{%
m^2}{H} \;, \qquad \bar{\eta}=\frac{\bar{\sigma} -\frac{m^2}{H}\sqrt{\gamma_m%
}} {\bar{\sigma} + \frac{m^2}{H}\sqrt{\gamma_m}}\;\;.
\end{equation}

The solution for $r(\tau)$ for each of the determining equations (\ref{TanhA}%
)- (\ref{TanB}) is obtained as follows

\noindent tanh-type A: 
\begin{equation}  \label{sol-tanhA}
r(\tau)=\left\{ 
\begin{array}{ll}
\frac{16}{\sqrt{\left(\sqrt{\kappa^2 H^2+8\Lambda} -m\kappa(f(\tau)-\frac{1}{%
f(\tau)})\right)^{2} -8\Lambda }}\; \mbox{tanh}^{-1}\left[\frac{%
\kappa\left(H- m\left|f(\tau)+\frac{1}{f(\tau)} \right|\right)} {\sqrt{\left(%
\sqrt{\kappa^2 H^2+8\Lambda} -m\kappa(f(\tau)-\frac{1}{f(\tau)}%
)\right)^{2}-8\Lambda }} \right] & \qquad r>0\;, \\ 
&  \\ 
\frac{-16}{\sqrt{\left(\sqrt{\kappa^2 H^2+8\Lambda} -m\kappa({\bar f}(\tau)-%
\frac{1}{{\bar f}(\tau)})\right)^{2} -8\Lambda }}\; \mbox{tanh}^{-1}\left[%
\frac{\kappa\left(H- m\left|{\bar f}(\tau)+\frac{1} {{\bar f}(\tau)}%
\right|\right)} {\sqrt{\left(\sqrt{\kappa^2 H^2+8\Lambda} -m\kappa({\bar f}%
(\tau)-\frac{1}{{\bar f}(\tau)})\right)^{2} -8\Lambda }}\right] & \qquad r<0
\;,
\end{array}
\right.
\end{equation}
tanh-type B: 
\begin{equation}  \label{sol-tanhB}
r(\tau)=\left\{ 
\begin{array}{ll}
\frac{16}{\sqrt{\left(\sqrt{\kappa^2 H^2+8\Lambda} -m\kappa(f(\tau)-\frac{1}{%
f(\tau)})\right)^{2} -8\Lambda }}\; \mbox{tanh}^{-1}\left[\frac{\sqrt{\left(%
\sqrt{\kappa^2 H^2+8\Lambda} -m\kappa(f(\tau)-\frac{1}{f(\tau)})\right)^{2}
-8\Lambda }} {\kappa\left(H-m\left|f(\tau)+\frac{1}{f(\tau)}\right|\right)}%
\right] & \qquad r>0\;, \\ 
&  \\ 
\frac{-16}{\sqrt{\left(\sqrt{\kappa^2 H^2+8\Lambda} -m\kappa({\bar f}(\tau)-%
\frac{1}{{\bar f}(\tau)})\right)^{2} -8\Lambda }}\; \mbox{tanh}^{-1}\left[%
\frac{\sqrt{\left(\sqrt{\kappa^2 H^2+8\Lambda} -m\kappa({\bar f}(\tau)-\frac{%
1}{{\bar f}(\tau)})\right)^{2} -8\Lambda }} {\kappa\left(H-m\left|{\bar f}%
(\tau)+\frac{1}{{\bar f}(\tau)}\right|\right)} \right] & \qquad r<0 \;,
\end{array}
\right.
\end{equation}
tan-type A: 
\begin{equation}  \label{sol-tanA}
r(\tau)=\left\{ 
\begin{array}{ll}
\frac{16}{\sqrt{8\Lambda - \left(\sqrt{\kappa^2 H^2+8\Lambda}
-m\kappa(f(\tau)-\frac{1}{f(\tau)})\right)^{2} }}\; \left(\mbox{tan}^{-1}%
\left[\frac{\kappa\left(m \left|f(\tau)+\frac{1}{f(\tau)} \right|-H\right)} {%
\sqrt{8\Lambda - \left(\sqrt{\kappa^2 H^2+8\Lambda} -m\kappa(f(\tau)-\frac{1%
}{f(\tau)})\right)^{2} }} \right]+n\pi\right) & \quad r>0\;, \\ 
&  \\ 
\frac{-16}{\sqrt{8\Lambda - \left(\sqrt{\kappa^2 H^2+8\Lambda} -m\kappa({%
\bar f}(\tau)-\frac{1}{{\bar f}(\tau)})\right)^{2} }}\; \left(\mbox{tan}^{-1}%
\left[\frac{\kappa\left(m \left|{\bar f}(\tau) +\frac{1}{{\bar f}(\tau)}%
\right|-H\right)} {\sqrt{8\Lambda - \left(\sqrt{\kappa^2 H^2+8\Lambda}
-m\kappa({\bar f}(\tau)-\frac{1}{{\bar f}(\tau)})\right)^{2} }}\right]%
+n\pi\right) & \quad r<0 \;,
\end{array}
\right.
\end{equation}
tan-type B: 
\begin{equation}  \label{sol-tanB}
r(\tau)=\left\{ 
\begin{array}{ll}
\frac{16}{\sqrt{8\Lambda - \left(\sqrt{\kappa^2 H^2+8\Lambda}
-m\kappa(f(\tau)-\frac{1}{f(\tau)})\right)^{2} }}\; \left(\mbox{tan}^{-1}%
\left[\frac{\sqrt{8\Lambda - \left(\sqrt{\kappa^2 H^2+8\Lambda}
-m\kappa(f(\tau)-\frac{1}{f(\tau)})\right)^{2} }}{\kappa\left(H-m
\left|f(\tau)+\frac{1}{f(\tau)}\right|\right)}\right]+n\pi\right) & \quad
r>0\;, \\ 
&  \\ 
\frac{-16}{\sqrt{8\Lambda - \left(\sqrt{\kappa^2 H^2+8\Lambda} -m\kappa({%
\bar f}(\tau)-\frac{1}{{\bar f}(\tau)})\right)^{2} }}\; \left(\mbox{tan}^{-1}%
\left[\frac{\sqrt{8\Lambda - \left(\sqrt{\kappa^2 H^2+8\Lambda} -m\kappa({%
\bar f}(\tau)-\frac{1}{{\bar f}(\tau)})\right)^{2} }} {\kappa\left(H-m \left|%
{\bar f}(\tau)+\frac{1}{{\bar f}(\tau)}\right|\right)} \right]+n\pi\right) & 
\quad r<0 \;\;.
\end{array}
\right.
\end{equation}

The exact $r(\tau)$ solutions corresponding to the phase space trajectories
in Fig.4 are given by tanh-type A solution (\ref{sol-tanhA}) and are plotted
in Fig.6. The motions are bounded and periodic. Comparison of three curves
in Fig.6 indicates that a negative cosmological constant $\Lambda <0$ acts
effectively as an attractive force: for the same value of $H_0$, the
particles do not achieve as wide a proper separation, and the frequency of
oscilliation is more rapid. As well, a positive $\Lambda >0$ acts as a
repulsive force: the frequency of oscillation decreases and the particles
achieve a wider proper separation.

\hspace*{5cm}--------------------------

\vspace{5mm}

\hspace*{7cm}Fig.6

\vspace{5mm}

\begin{center}
{\it The exact $r$ vs $\tau$ curves corresponding to the phase space
trajectories in Fig.4. }
\end{center}

\hspace*{5cm}--------------------------

The period $T$ for the bounded motion is obtained from tanh-type A solution
with the condition $r=0$ and $p=\pm\;p0$: 
\begin{equation}
T=\cases{\frac{16}{\sqrt{\kappa^2m^2+8\Lambda}}\tanh^{-1}\left(\frac{
\sqrt{\kappa^2m^2+8\Lambda}\sqrt{H^2-4m^2}}{\kappa Hm}\right)
&$\gamma_m>0$,\vspace{2mm}\cr \frac{16\sqrt{H^2-4m^2}}{\kappa
Hm}&$\gamma_m=0$,\vspace{2mm}\cr
\frac{16}{\sqrt{-\kappa^2m^2-8\Lambda}}\tan^{-1}\left(\frac{
\sqrt{-\kappa^2m^2-8\Lambda}\sqrt{H^2-4m^2}}{\kappa Hm}\right)
&$\gamma_m<0$.}
\end{equation}

In figure 7 we plot $r(\tau)$ for fixed $\Lambda=-1.5$ and $H_0=16$ for
several different values of $m$. Though the attractive effect of a negative $%
\Lambda$ is common in all cases, a special (and rather surprising) situation
arises. As the motion becomes more relativistic (i.e. $m$ gets smaller) we
find that a second maximum develops in the curve (see $m=0.05$ curve). The
description of the motion is as follows. The two particles start at $r=0$
depart in opposite directions, reaching a maximum separation. They then go
back toward one another for a certain period of proper time. However at some
point they each reverse direction, reaching a second maximal separation.
They then reverse direction again, finally returning to their starting point
where the motion then repeats itself.

As the mass becomes very small, the second maximum prevails. This peculiar
behavior \cite{2bdcossh} is due to a subtle interplay between the
gravitational attraction, cosmological constant and relativistic motion of
the particles. To our knowledge it has never been previously observed. The
changes of the peaks are clearly grasped in the phas$\frac{{}}{{}}$e space
trajectories in Fig.8. (In Fig.7 the first maximum of $m=0.001$ curve could
not be drawn due to plotting precision.)

\hspace*{5cm}--------------------------

\vspace{5mm}

\hspace*{7cm}Fig.7

\vspace{5mm}

\begin{center}
{\it A sequence of equal mass curves for $\Lambda=-1.5$ and $H_0=16$. \\[0pt]
Note the presence of the second maximum for $m=0.05$. }
\end{center}

\hspace*{5cm}--------------------------

\hspace*{5cm}--------------------------

\vspace{5mm}

\hspace*{7cm}Fig.8

\vspace{5mm}

\begin{center}
{\it Change of peaks for $\Lambda=-1.5$ and $H_0=16$ as $m$ gets smaller. }
\end{center}

\hspace*{5cm}--------------------------

As the negative value of $\Lambda$ approaches its lower bound of $- \kappa^2
H^2 /8$, the form of the phase space trajectories transforms from an $S$%
-shaped curve to a double peaked one and then to a diamond shape. Figure 9
illustrates these characteristics for the case of $m=0.5$ and $H_0=100$, in
which a double peak structure appears for $\Lambda=-70$. For each trajectory
in Fig.9 the corresponding $r(\tau)$ plot is shown in Fig.10.

\hspace*{5cm}--------------------------

\vspace{5mm}

\hspace*{7cm}Fig.9

\vspace{5mm}

\begin{center}
{\it Phase space trajectories for $m=0.5, H_0=100$ and different values of
negative $\Lambda$. }
\end{center}

\hspace*{5cm}--------------------------

\hspace*{5cm}--------------------------

\vspace{5mm}

\hspace*{7cm}Fig.10

\vspace{5mm}

\begin{center}
{\it The $r$ vs $\tau$ curves corresponding the phase space trajectories in
Fig.9. }
\end{center}

\hspace*{5cm}--------------------------

\vspace{5mm}

The double-peak structure shown in Figs.7-10 is a consequence of having a
momentum-dependent potential. We can gain some insight into this behaviour
by computing a perturbative Hamiltonian. The structure of the determining
equations (\ref{deteq-1}) and (\ref{deteq-2}) suggests that we can carry out
a 2-parameter expansion of the Hamiltonian in terms of $\kappa$ and $%
\Lambda/\kappa^2$. To the third order the result is 
\begin{eqnarray}  \label{Happrox}
H&=&H_0(p,r)+H_\Lambda(p,r)  \nonumber \\
&=&2\sqrt{p^{2}+m^{2}} +\frac{\kappa}{4}(\sqrt{p^{2}+m^{2}}-\epsilon\tilde{p}%
)^{2}\;|r| +\frac{\kappa^{2}}{4^{2}}(\sqrt{p^{2}+m^{2}}-\epsilon\tilde{p}%
)^{3}\;r^{2}  \nonumber \\
&&+\frac{7\kappa^{3}}{6\times 4^{3}}(\sqrt{p^{2}+m^{2}} -\epsilon\tilde{p}%
)^{4}\;|r|^{3} -\frac{\Lambda}{2\kappa}\cdot\frac{\epsilon\tilde{p}}{\sqrt{%
p^{2}+m^{2}}} \;|r| -\frac{\Lambda}{16}\cdot\frac{\epsilon\tilde{p}\;m^{2}}{%
p^{2}+m^{2}}\;r^{2}  \nonumber \\
&&+\frac{\Lambda^{2}}{4\kappa^{3}}\cdot\frac{\epsilon\tilde{p}}{%
(p^{2}+m^{2})^{3/2}}\;|r| +\cdot\cdot\cdot\;\;.
\end{eqnarray}
It is straightforward to show that the terms in $H_\Lambda$ have the form 
\begin{equation}  \label{general}
\frac{p^{s_1}}{(\sqrt{p^2 + m^2})^{s_2}}|r|^{s_3} \;,
\end{equation}
to arbitrary order in $\Lambda/\kappa^2$, where the $s_{i}$ are positive
integers and $s_2\geq s_1$. One of the characteristics common to all such
terms is that they vanish as $p \rightarrow 0$, since in this case the
determining equation of the Hamiltonian becomes 
\begin{equation}  \label{p=0determine}
\mbox{tanh}\left(\frac{\kappa}{16}H|r|\right) = \frac{H - 2m}{H}\;.
\end{equation}
which is $\Lambda$-independent. Another important characteristic is that
they have a single maximum at $p^2=\frac{s_1 m^2}{s_2-s_1}$ if $s_2>s_1$ in
the case of $\Lambda<0$.

Consider the situation depicted in Figs.7-10, that of two particles
initially at the origin, each having initial momentum $p_{0}$, with $\Lambda
<0$. In this case $H_{0}=2\sqrt{p^{2}+m^{2}}$, $H_{\Lambda }=0$, and the
particles initially move apart as though they were free. From (\ref{p-exact1}%
) the momentum is a monotonically decreasing function of time. For $p_{0}/m$
sufficiently small, the particles will execute a motion which is a
perturbation of that described in section \ref{exactL0} since the momentum
never becomes large enough to cross the maximum in $H_{\Lambda }$. However
if $p_{0}/m$ is sufficiently large, the terms in $H_{\Lambda }$ will grow as 
$p$ decreases, and the $\Lambda $-dependent part of the potential will
continue to increase. The terms in $H_{0}$ will decrease as $p$ decreases,
even though $r$ is increasing. Eventually a maximum in $r$ is reached, after
which both $r$ and $p$ are decreasing. In the generic case $r$ will continue
to decrease toward zero. However a second extremum will appear if the $%
H_{\Lambda }$ terms get too small too rapidly before $p=0$, which can happen
for $\Lambda $ within a certain range. Since $H=H_{0}+H_{\Lambda }$ is a
constant of the motion, the only way to preserve the constancy of $H$ will
be for $r$ to increase again. Essentially the particles are repelled due to
their kinetic energy within this range of $\Lambda $. Of course $r$ cannot
increase too much, because $p$ continues to get small -- eventually $r$ must
reach a 2nd maximum, and then turns around again until $p=0$, after which
the motion continues to $r=0$ whereupon the particles interchange roles.

We can see this effect in a simple non-relativistic model with 
\begin{equation}
\hat{H}=\hat{p}^{2}/m+Km\frac{\hat{p}}{1+\hat{p}^{2}/m^{2}}|\hat{r}|=m\left( 
{p}^{2}+K\frac{{p}}{1+{p}^{2}}|{r}|\right) \;,  \label{nonrelmod}
\end{equation}
where in the latter equation $p$ and $r$ have been rescaled in units of $m$.
The potential has been chosen so that $s_{2}=1+s_{1}=2$ in terms of (\ref
{general}) above. Since $\hat{H}=mh$ is a constant of the motion, we can
write 
\begin{equation}
|r|={\displaystyle\frac{(-p^{2}+h)\,(1+p^{2})}{K\,p}}\;,  \label{nonrelmod2}
\end{equation}
which has extrema at $p=p_{\pm }\equiv \frac{1}{6}\sqrt{-6+6\,h\pm 6\,\sqrt{%
1-14\,h+h^{2}}}$. For $p_{0}^{2}=h<7+4\sqrt{3}$, the extreme values of $|r|$
are not real, and so there will be no double peak structure in the
phase-space trajectory. However for $p_{0}^{2}=h>7+4\sqrt{3}$ two extrema
appear, and the particle gets a bounce. This motion is viewed qualitatively
in a potential diagram shown in Fig.11. As the momentum decreases from a
initial value $p_{0}$ to zero, the potential curve (which is linear in $|r|$%
) changes from $A$ to $B\rightarrow \cdot \cdot \cdot \rightarrow
E\rightarrow F\rightarrow E\rightarrow \cdot \cdot \cdot \rightarrow
B\rightarrow A\rightarrow G\rightarrow I\rightarrow \cdot \cdot \cdot $.
Accordingly the particle moves from points $0$ to $10$ in numerical order.

The equations of motion for the Hamiltonian (\ref{nonrelmod}) are easily
solved, yielding 
\begin{equation}
p(t)=\sqrt{{W}(h\,e^{h-2Kt})}\;,\qquad r(t)={\displaystyle\frac{\left\{ h-{W}%
(h\,\,e^{h-2Kt})\right\} \,\left\{ 1+W(he^{h-2Kt})\right\} }{K\,\sqrt{%
W(he^{h-2Kt})}}}\;,  \label{nonrelmod3}
\end{equation}
where $W(x)$ is the Lambert-W function (\ref{lambertW}). The function $p(t)$
is monotonically decreasing. Provided $h>7+4\sqrt{3}$ the particle gets a
bounce at $r=r(p_{-})$ before moving out to infinity, as illustrated in
Figs. 12 and 13 for the case of{\bf \ $h=(7+4\sqrt{3})+20$. }In these
figures the numbers on the curves denote the corresponding numbers in the
potential diagram of Fig.11.

\hspace*{5cm}--------------------------

\vspace{5mm}

\hspace*{7cm}Fig.11

\vspace{5mm}

\begin{center}
{\it A schematic view of the motion with a large $p_0$ in a potential
diagram. }
\end{center}

\hspace*{5cm}--------------------------

\hspace*{5cm}--------------------------

\vspace{5mm}

\hspace*{7cm}Fig.12

\vspace{5mm}

\begin{center}
{\it The $p(t)$ and $r(t)$ curves in the non-relativistic model for $h=(7+4%
\sqrt{3})+20$. \\[0pt]
A bounce occurs at $t=t_{-}=18.08$.}
\end{center}

\hspace*{5cm}--------------------------

\hspace*{5cm}--------------------------

\vspace{5mm}

\hspace*{7cm}Fig.13

\vspace{5mm}

\begin{center}
{\it The phase space trajectory in the non-relativistic model for $h=(7+4%
\sqrt{3})+20$.}
\end{center}

\hspace*{5cm}--------------------------

The expansion (\ref{Happrox}) has similar features, except that there is an
additional gravitational and cosmological attraction which prevents the
separation from diverging. To order $\Lambda$ the potential is a sum of two
terms of the form in (\ref{general}), with $s_2=s_1=1$ in the first term and 
$s_2=1+s_1=2$ in the second term. The latter provides the bounce effect
described above, but is always overwhelmed by the first term for small $|r|$%
. However the $\Lambda^2$ term has $s_2=2+s_1=3$ and is a pure bounce term.
Hence there exists a range of $\Lambda$ which can provide a bounce. Fig.14
shows the phase space trajectories in $r>0$ region for $\Lambda=-0.3, H_0 =5$
and two different masses ($m=0.2, 0.7$), in which the dotted curves
represent the motions to the first order of $\Lambda$ in the Hamiltonian (%
\ref{Happrox}) and the solid curves do the motions corresponding to the
second order of $\Lambda$. We see that to order $\Lambda$, as the motion
becomes relativistic ($m$ gets smaller) the trajectory simply expands and
changes from $S$-shape to a diamond shape. When the $\Lambda^2$ term is
included, the double peak structure (a solid curve with $m=0.2$) appears.

\hspace*{5cm}--------------------------

\vspace{5mm}

\hspace*{7cm}Fig.14

\vspace{5mm}

\begin{center}
{\it Phase space trajectories for the perturbative Hamiltonian \\[0pt]
for $\Lambda=-0.3, H_0 =5$ and $m=0.2,0.7$. }
\end{center}

\hspace*{5cm}--------------------------

This perturbative analysis indicates that the bounce effect is a result of
the negative cosmological constant inducing a momentum dependent potential
with positive coefficients. For $\Lambda>0$, odd powers of $\Lambda/\kappa^2$
are strongly repulsive, and suppress the attractive effects of even powers
of $\Lambda/\kappa^2$, eliminating the double peak structure.

More generally, a positive cosmological constant acts effectively as a
repulsive force. Figure 15 shows $r(\tau)$ plots for fixed $\Lambda=1.5,
H_0=16$ and several different values of $m$. The motion becomes unbounded
between $m=4.72$ and $m=4.73$. The $r(\tau)$ plots in Fig.16 are for fixed $%
m=0.5, H_0=16$ and different values of $\Lambda$, showing also the
transition from bounded to unbounded motion. This transition occurs at $%
{\cal J}_{\Lambda}=0$ and the critical value of $\Lambda$ is given by $%
\Lambda_{c}=\frac{\kappa^2 m^4}{2(H^2-4m^2)}$.

As $\Lambda \longrightarrow \Lambda_{c}$ the particles rapidly separate,
remaining nearly stationary for an increasingly large period of proper time
before coming together again. At $\Lambda = \Lambda_{c}$ this separation
time becomes infinite, and for $\Lambda > \Lambda_{c}$, the separation
diverges at finite $\tau$.

\hspace*{5cm}--------------------------

\vspace{5mm}

\hspace*{7cm}Fig.15

\vspace{5mm}

\begin{center}
{\it A sequence of curves of equal mass for $\Lambda=1.5, H_0=16$. }
\end{center}

\hspace*{5cm}--------------------------

\hspace*{5cm}--------------------------

\vspace{5mm}

\hspace*{7cm}Fig.16

\vspace{5mm}

\begin{center}
{\it A sequence of curves near $\Lambda_c=20.008333$ for $m=7, H_0=16$. }
\end{center}

\hspace*{5cm}--------------------------

Though all the above solutions are derived from tanh-type equations (\ref
{TanhA}) and (\ref{TanhB}), for a positive cosmological constant there exist
also a countably infinite set of unbounded motions specified by tan-type A,
B equations (\ref{TanA}) and (\ref{TanB}). Then for $0< \Lambda <
\Lambda_{c} $, both bounded and unbounded motions are realized for a fixed
value of $H$, as shown in Fig.17. In the unbounded motion two particles
simply approach one another at some minimal value of $|r|$ and then reverse
direction toward infinity. In the trajectories the dotted curves come from
tan-type A solution (\ref{sol-tanA}) and the dashed curves do from tan-type
B solution (\ref{sol-tanB}). As $\Lambda$ approaches $\Lambda_{c}$, two
bulges of the solid curve (tanh-type A) and the dotted curve (tan-type A:
n=0) come close and contact. When $\Lambda$ exceeds $\Lambda_{c}$ two curves
switch to the unbounded trajectories as shown in the solid curves in Fig.18.
The particles cross one another before receding toward infinity. The upper
solid curve represents the motion in which $p$ approaches the asymptotic
values $p_{\pm}\equiv\frac{1}{2\kappa}(\pm \sqrt{\kappa^2 H^2+8\Lambda} + 
\sqrt{8\Lambda})$ as $r\longrightarrow \pm\infty$.

As noted previously, one peculiar feature of this motion is that the two
particles diverges to infinite separation at finite proper time. The time $%
\tau_{\infty}$ for $r\longrightarrow \infty$ is 
\begin{equation}
\tau_{\infty}=\frac{4}{\kappa m \sqrt{\gamma_{m}}} \mbox{log} \left( \frac{%
H(1+\sqrt{\gamma_H}) - (p_{+} + \sqrt{p_{+}^{2}+m^2}) (1+\sqrt{\gamma_m})}{%
\eta\left[H(1+\sqrt{\gamma_H}) - (p_{+} + \sqrt{p_{+}^{2}+m^2})(\sqrt{%
\gamma_m}-1)\right]}\right)\;\;.
\end{equation}
The lower solid curve represents the motion in reversed direction. For $%
\Lambda > \Lambda_{c}$ only unbounded motions are realized.

\hspace*{5cm}--------------------------

\vspace{5mm}

\hspace*{7cm}Fig.17

\vspace{5mm}

\begin{center}
{\it Phase space trajectories of the bouned and the unbounded motions \\[0pt]
for $\Lambda=1, m=1$ and $H_0 =2.1$. }
\end{center}

\hspace*{5cm}--------------------------

\hspace*{5cm}--------------------------

\vspace{5mm}

\hspace*{7cm}Fig.18

\vspace{5mm}

\begin{center}
{\it Phase space trajectories of the unbounded motions \\[0pt]
for $\Lambda=1.5, m=1$ and $H_0 =2.1$. }
\end{center}

\hspace*{5cm}--------------------------

We discuss in Appendix C the causal relationship between the two particles
in the unbounded case.

\section{THE UNEQUAL MASS CASE}

For the unequal masses the proper time (\ref{tau-0}) of each particle is 
\begin{eqnarray}  \label{tau-2}
d\tau_1&=&dt\;\frac{16YK_{0}K_{1}m_{1}}{JKM_{1}\sqrt{p^2+m_{1}^{2}}}\;, 
\nonumber \\
\\
d\tau_2&=&dt\;\frac{16YK_{0}K_{2}m_{2}}{JKM_{2}\sqrt{p^2+m_{2}^{2}}}\;, 
\nonumber
\end{eqnarray}
where $K\equiv K_{+}=K_{-}$ and $Y\equiv Y_{+}=Y_{-}$. In this situation,
choosing the time coordinate to be the proper time of one particle
introduces an asymmetry into the description of the motion. Instead we seek
a time variable which is symmetric with respect to $1 \leftrightarrow 2$ and
reduces to the proper time (\ref{Tau}) when $m_1=m_2$. From (\ref{tau-2}) we
choose 
\begin{equation}  \label{tau-3}
d\tilde{\tau}\equiv dt\;\frac{16YK_{0}}{JK} \left(\frac{K_{1}K_{2}m_{1}m_{2}%
}{M_{1}M_{2}\sqrt{p^2+m_{1}^{2}} \sqrt{p^2+m_{1}^{2}}}\right)^{1/2}\;.
\end{equation}
In terms of this variable the canonical equations are expressed as 
\begin{eqnarray}
\frac{dp}{d\tilde{\tau}}&=&-\frac{1}{4\kappa}\left( \frac{%
K_{1}K_{2}M_{1}M_{2}\sqrt{p^2+m_{1}^{2}}\sqrt{p^2+m_{2}^{2}}} {m_{1}m_{2}}%
\right)^{1/2}\;,  \label{unequal-p} \\
\frac{dz_{i}}{d\tilde{\tau}}&=&(-1)^{i+1} \left( \frac{M_{1}M_{2}\sqrt{%
p^2+m_{1}^{2}}\sqrt{p^2+m_{2}^{2}}} {K_{1}K_{2}m_{1}m_{2}}\right)^{1/2}
\left\{\frac{\epsilon J}{16K_{0}} +\frac{K_{i}}{M_{i}}\left(\frac{p}{\sqrt{%
p^2+m_{i}^{2}}} -\epsilon\frac{Y}{K}\right)\right\}\;,  \nonumber
\label{unequal-z} \\
\\
\frac{dr}{d\tilde{\tau}}&=&\left( \frac{M_{1}M_{2}\sqrt{p^2+m_{1}^{2}}\sqrt{%
p^2+m_{2}^{2}}} {K_{1}K_{2}m_{1}m_{2}}\right)^{1/2}  \nonumber \\
&&\makebox[3em]{}\times\left\{\frac{\epsilon J}{8K_{0}} +\frac{K_{1}}{M_{1}}%
\left(\frac{p}{\sqrt{p^2+m_{1}^{2}}} -\epsilon\frac{Y}{K}\right) +\frac{K_{2}%
}{M_{2}}\left(\frac{p}{\sqrt{p^2+m_{2}^{2}}} -\epsilon\frac{Y}{K}%
\right)\right\}\;.  \label{unequal-r}
\end{eqnarray}
Note that $r$ still describes the proper distance between the particles at
any fixed instant.

Unlike the equal mass case, the integration $\int dp (K_{1}K_{2}M_{1}M_{2}%
\sqrt{p^2+m_{1}^{2}}\sqrt{p^2+m_{2}^{2}} \;)^{-1/2}$ can not be performed
within the framework of elementary calculus. Hence we solve (\ref{unequal-p}%
) numerically.

In the case of a negative cosmological constant the $r(\tau)$ plots in
Fig.19 show the trajectories for various mass ratios $m_{1}/m_{2}$ in the
fixed $\Lambda=-1, m_2=1$ and $H_0=10$. Compared with the equal mass case $%
m1=m2=1$, as the mass ratio gets larger, the gravitational attraction is
stronger and the proper distance between two particles as well as the period
become shorter. When the mass ratio gets a small value than unity, the
gravity becomes weak. However, for quite a small mass ratio a strong
attractive effect of the cosmological constant prevails and the period
changes to become shorter. At the same time the double peak structure (the
second maximum) appears and finally the first maximum fades out. These
characteristics are very clear in the unequal mass case.

\hspace*{5cm}--------------------------

\vspace{5mm}

\hspace*{7cm}Fig.19

\vspace{5mm}

\begin{center}
{\it $r(\tau)$ plots for the different values of the mass ratio $m_{1}/m_{2}$
\\[0pt]
for $\Lambda=-1, m_{2}=1$ and $H_0 =10$. }
\end{center}

\hspace*{5cm}--------------------------

\vspace{5mm} For a positive cosmological constant the situation is simple.
As shown in Fig.20, as the mass ratio becomes small, the particles separate
with an increasingly larger period of bounded motion. This is due to a
repulsive effect of the cosmological constant and beyond the critical value
the motion bocomes unbounded.

\hspace*{5cm}--------------------------

\vspace{5mm}

\hspace*{7cm}Fig.20

\vspace{5mm}

\begin{center}
{\it $r(\tau)$ plots for the different values of the mass ratio $m_{1}/m_{2}$
\\[0pt]
for $\Lambda=0.003, m_{2}=1$ and $H_0 =10$. }
\end{center}

\hspace*{5cm}--------------------------

\section{CONCLUSIONS}

In general relativity the relationship between the motion of a set of $N$
bodies and the structure of space-time is non-linear and quite complicated,
even for $N=2$. Expanding upon the solution presented in \cite{2bdcossh}, we
have obtained an exact solution to the 2-body problem in $(1+1)$ dimensions
with a cosmological constant. To our knowledge this is the first
non-perturbative relativistic curved-spacetime treatment of this problem,
providing new avenues for investigation of one-dimensional self-gravitating
systems.

We recapitulate the main results of our paper: \newline
(1) We formulated the canonical formalism for a system of $N$ bodies in a
lineal theory of gravity with a cosmological constant $\Lambda $. The system
is described by a conservative Hamiltonian. The effect of $\Lambda $ is
incorporated into the potential. \newline
(2) For $N=2$ the determining equation of the Hamiltonian is a
transcendental equation derived from the matching conditions and appropriate
boundary conditions at infinity. From these the canonical equations of
motion may be derived. The metric components are also completely determined. 
\newline
(3) For the equal mass case we obtained explicitly the exact solutions to
the canonical equations in terms of the mutual proper time of the particles.
Using the solutions we analyzed the motion in both $r(\tau )$ plots and
phase-space trajectories. \newline
(4) As expected, a positive cosmological constant yields a repulsive effect
on the motion relative to their mutual gravitational attraction. For $%
0<\Lambda <\Lambda _{c}$ both bounded and unbounded motions are realized,
while for $\Lambda _{c}<\Lambda $ only the unbounded motions are allowed. As 
$\Lambda \rightarrow \Lambda _{c}$ the particles separate to an infinite
proper distance in infinite proper time. For $\Lambda >\Lambda _{c}$ this
infinite separation occurs in finite proper time.\newline
(5) A negative cosmological constant has an additional attractive effect,
and the motion of the particles is bounded. However for a certain range of
the parameters, a repulsive effect sets in, resulting the double-peaked
structures of Figs.7-10. This effect is due to a subtle interplay between
the momentum-dependent $\Lambda $ potential and the gravitational
attraction. \newline
(6) In the unequal mass case the same basic features also occur; indeed the
double peak behavior shows up more clearly than in the equal mass case.
Although eq. (\ref{unequal-p}) cannot be integrated in terms of elementary
functions, it is straightforward to numerically integrate. An exact solution
in the small mass limit of the particle 1 was also obtained. \newline

Several interesting features of the motion remain to be explored. The
divergent separation of the bodies at finite proper time needs to be better
understood. Another issue concerns the condition (\ref{H-bound}) which means
that for a given value of $\Lambda =-|\Lambda |$ the motion is allowed for
the total energy larger than $\sqrt{8|\Lambda |/\kappa ^{2}}$. What is the
physical meaning of this condition? It seems to suggest that as the
attractive effect of $\Lambda <0$ exceeds a critical value the two particle
system is no longer stable and transforms into some other system (probably
making a black hole). To formulate the canonical formalism to treat this
problem is our next subject.

\section*{APPENDIX A: SOLUTION OF THE METRIC TENSOR}

Under the coordinate conditions (\ref{cc}) the field equations (\ref{e-pi}),
(\ref{e-gamma}), (\ref{e-Pi}) and (\ref{e-Psi}) become 
\begin{eqnarray}
&&\dot{\pi}+N_{0}\left\{\frac{3\kappa}{2}\pi^{2} +\frac{1}{8\kappa}%
(\Psi^{\prime})^{2} -\frac{1}{4}\left(\frac{\Lambda}{\kappa}\right) -\frac{%
p_{1}^{2}}{2\sqrt{p_{1}^{2}+m_{1}^{2}}}\delta(x-z_{1}(t)) -\frac{p_{2}^{2}}{2%
\sqrt{p_{2}^{2}+m_{2}^{2}}}\delta(x-z_{2}(t))\right\}  \nonumber \\
&&+N_{1}\left\{\pi^{\prime}+p_{1}\delta(x-z_{1}(t))+p_{2}\delta(x-z_{2}(t))
\right\}+\frac{1}{2\kappa}N^{\prime}_{0}\Psi^{\prime} +N^{\prime}_{1}\pi=0
\;,  \label{eq-pi1} \\
&&\kappa \pi N_{0}+N^{\prime}_{1}=0\;,  \label{eq-N1} \\
&&\partial_{1}(\frac{1}{2}N_{0}\Psi^{\prime}+N^{\prime}_{0})=0\;,
\label{eq-N0} \\
&&\dot{\Psi}+2\kappa N_{0}\pi-N_{1}\Psi^{\prime}=0\;.  \label{eq-Psi1}
\end{eqnarray}
The solution to (\ref{eq-N0}) is 
\begin{equation}  \label{N0}
N_{0}=A\;e^{-\frac{1}{2}\Psi}=A\phi^{2}= \left\{ 
\begin{array}{ll}
A\phi_{+}^{2} & \qquad \mbox{(+) region} \\ 
A\phi_{0}^{2} & \qquad \mbox{(0) region} \\ 
A\phi_{-}^{2} & \qquad \mbox{(-) region}
\end{array}
\right. \;,
\end{equation}
$A$ being an integration constant. \newline
\noindent Eq.(\ref{eq-N1}) is 
\begin{equation}
N_{1}^{\prime}=-\kappa A \chi^{\prime}\phi^{2}\;.
\end{equation}
The solution in each region is 
\begin{equation}  \label{N1-1}
\left\{ 
\begin{array}{lll}
N_{1(+)}=\epsilon\left\{A\;\frac{Y_{+}}{K_{+}}\phi_{+}^{2}-D_{+}\right\} & 
\qquad \mbox{(+) region}\;, &  \\ 
N_{1(0)}=\epsilon\left\{A\frac{Y_{0}}{K_{0}}\left[A_{0}^{2}\;e^{K_{0}x}
-B_{0}^{2}\;e^{-K_{0}x}\right]+2AY_{0}A_{0}B_{0}x+D_{0}\right\} & \qquad %
\mbox{(+) region}\;, &  \\ 
N_{1(-)}=-\epsilon\left\{A\;\frac{Y_{-}}{K_{-}}\phi_{-}^{2}-D_{-}\right\} & 
\qquad \mbox{(+) region}\;, & 
\end{array}
\right.
\end{equation}
where $D_{+}, D_{-}$ and $D_{0}$ are integration constants.

The matching conditions $N_{1(+)}(z_{1})=N_{1(0)}(z_{1})$ and $%
N_{1(-)}(z_{2})=N_{1(0)}(z_{2})$ lead to 
\begin{eqnarray}
A&=&\frac{8K_{0}(D_{+}+D_{-})}{J}\;e^{\frac{1}{2}(K_{01}z_{1}-K_{02}z_{2})}%
\;, \\
D_{0}&=&-\frac{D_{+}-D_{-}}{2}+\frac{D_{+}+D_{-}}{2J}\left\{ 2\left[\left(%
\frac{Y_{0}}{K_{0}}+\frac{Y_{+}}{K_{+}}\right)K_{1} -\left(\frac{Y_{0}}{K_{0}%
}+\frac{Y_{-}}{K_{-}}\right)K_{2}\right] \right.  \nonumber \\
&&\left.-2\left[\left(\frac{Y_{0}}{K_{0}}-\frac{Y_{+}}{K_{+}}\right) \frac{1%
}{{\cal M}_{1}}-\left(\frac{Y_{0}}{K_{0}}-\frac{Y_{-}}{K_{-}}\right) \frac{1%
}{{\cal M}_{2}}\right]K_{1}K_{2} -\frac{Y_{0}}{K_{0}}K_{1}K_{2}(z_{1}+z_{2})%
\right\}\;.
\end{eqnarray}
In deriving these relations the expressions (\ref{ab+-0}) for $A_{+,0},
B_{-,0}$ and the determining equation (\ref{H1}) were used. As for the
equation (\ref{eq-pi1}), first take the $\delta$ function at $x=z_{1}$: 
\begin{eqnarray}
\lefteqn{\left\{\delta\;\mbox{function part of LHS}\; (\ref{eq-pi1})\; \mbox{%
at} \;x=z_{1}\right\}}  \nonumber \\
&=&\frac{1}{2}p_{1}\;\delta(x-z_{1})\left\{\dot{z}_{1} -N_{0}(z_{1})\frac{%
p_{1}}{\sqrt{p_{1}^{2}+m_{1}^{2}}}+N_{1}(z_{1})\right\}  \nonumber \\
&=&\frac{1}{2}p_{1}\;\delta(x-z_{1})\epsilon \left(\frac{Y_{+}}{K_{+}}-D_{+}
\right)\;,
\end{eqnarray}
where $N_{0}(z_{1}), N_{1}(z_{1})$ and the canonical equation were inserted.
Then the integration constant $D_{+}$ should be 
\begin{equation}  \label{D+}
D_{+}=\frac{Y_{+}}{K_{+}}\;,
\end{equation}
and similary 
\begin{equation}  \label{D-}
D_{-}=\frac{Y_{-}}{K_{-}}\;.
\end{equation}
Now the metric tensor is completely determined: 
\begin{eqnarray}
N_{0(+)}(x)&=&\frac{8}{J}\left(\frac{Y_{+}}{K_{+}}+\frac{Y_{-}}{K_{-}}%
\right) \frac{K_{0}K_{1}}{{\cal M}_{1}}\;e^{K_{+}(x-z_{1})}\;,  \nonumber \\
N_{0(0)}(x)&=&\frac{1}{2K_{0}J}\left(\frac{Y_{+}}{K_{+}}+\frac{Y_{-}}{K_{-}}
\right)\left[(K_{1}{\cal M}_{1})^{\frac{1}{2}} e^{-\frac{1}{2}%
K_{0}(x-z_{1})}+(K_{2}{\cal M}_{2})^{\frac{1}{2}} e^{\frac{1}{2}%
K_{0}(x-z_{2})}\right]^{2}\;,  \nonumber \\
N_{0(-)}(x)&=&\frac{8}{J}\left(\frac{Y_{+}}{K_{+}}+\frac{Y_{-}}{K_{-}}%
\right) \frac{K_{0}K_{2}}{{\cal M}_{2}}\;e^{-K_{-}(x-z_{2})}\;,  \nonumber \\
\\
N_{1(+)}&=&\epsilon\frac{Y_{+}}{K_{+}} \left\{\frac{8}{J}\left(\frac{Y_{+}}{%
K_{+}}+\frac{Y_{-}}{K_{-}}\right) \frac{K_{0}K_{1}}{{\cal M}_{1}}%
\;e^{K_{+}(x-z_{1})}-1\right\}\;,  \nonumber \\
N_{1(0)}&=&\epsilon\left\{\frac{Y_{0}}{2JK_{0}^{2}} \left(\frac{Y_{+}}{K_{+}}%
+\frac{Y_{-}}{K_{-}}\right) \left[K_{2}M_{2}\;e^{K_{0}(x-z_{2})}-K_{1}M_{1}%
\;e^{-K_{0}(x-z_{1})} \right.\right.  \nonumber \\
&&\makebox[5em]{}\left.\left.+2K_{0}(K_{1}K_{2}M_{1}M_{2})^{1/2} \;e^{\frac{1%
}{2}K_{0}(z_{1}-z_{2})}\;x\right]+D_{0}\right\}\;,  \nonumber \\
N_{1(-)}&=&-\epsilon\frac{Y_{-}}{K_{-}} \left\{\frac{8}{J}\left(\frac{Y_{+}}{%
K_{+}}+\frac{Y_{-}}{K_{-}}\right) \frac{K_{0}K_{2}}{{\cal M}_{2}}%
\;e^{-K_{-}(x-z_{2})}-1\right\}\;.  \nonumber
\end{eqnarray}
With this solution and the canonical equations, the field equation (\ref
{eq-pi1}) can be proved to hold in a whole $x$ space.

As we showed in the previous paper, to satisfy (\ref{eq-Psi1}) the dilaton
field $\Psi$ needs an extra function $f(t)$, which has no effect on the
dynamics of particles. After lengthy calculation Eq.(\ref{eq-Psi1}) leads to 
\begin{eqnarray}  \label{dot-f}
\dot{f}(t)&=&-\frac{d}{dt}(K_{01}z_{1}-K_{02}z_{2}) +\frac{2}{J}\left(\frac{%
Y_{+}}{K_{+}}+\frac{Y_{-}}{K_{-}}\right)\left\{ 2K_{0}K_{1}\frac{p_{1}}{%
\sqrt{p_{1}^{2}+m_{1}^{2}}} -2K_{0}K_{2}\frac{p_{2}}{\sqrt{%
p_{2}^{2}+m_{2}^{2}}}\right.  \nonumber \\
&&\left.+\epsilon\frac{Y_{0}}{K_{0}}\left(K_{0}K_{1}+K_{0}K_{2} -\frac{%
K_{0}K_{1}K_{2}}{{\cal M}_{1}}-\frac{K_{0}K_{1}K_{2}}{{\cal M}_{2}}
\right)+4\epsilon Y_{0}K_{0}\left(\frac{K_{1}}{{\cal M}_{1}}+\frac{K_{2}}{%
{\cal M}_{2}}\right)\right\}\;.  \nonumber \\
\end{eqnarray}
Thus $f(t)$ is uniquely determined.

\section*{APPENDIX B: A TEST PARTICLE APPROXIMATION}

For a single static source $M$ the solution to the field equations (\ref
{e-pi}) -(\ref{e-Psi}) under the coordinate conditions (\ref{cc}) is 
\begin{eqnarray}
N_{0} &=&e^{\frac{\kappa M}{4}|x|}\;,\qquad N_{1}=\epsilon \;\sqrt{1+\frac{%
8\Lambda }{\kappa ^{2}M^{2}}}\;\frac{x}{|x|}\left[ e^{\frac{\kappa M}{4}%
|x|}-1\right] \;,  \nonumber \\
&& \\
\pi  &=&-\frac{\epsilon M}{4}\sqrt{1+\frac{8\Lambda }{\kappa ^{2}M^{2}}}%
\;,\qquad \Psi =-\frac{\kappa M}{2}\;|x|+\frac{\kappa \epsilon M}{2}\sqrt{1+%
\frac{8\Lambda }{\kappa ^{2}M^{2}}}\;\;t\;.  \nonumber
\end{eqnarray}
\{\}From (\ref{e-p}) and (\ref{e-z}) the canonical equations for a test
particle (mass $\mu $) under the gravity of a static source are 
\begin{eqnarray}
\dot{p} &=&-\sqrt{p^{2}+\mu ^{2}}\;\frac{\partial N_{0}(z)}{\partial z}+p\;%
\frac{\partial N_{1}(z)}{\partial z}\;,  \label{test-can-p} \\
\dot{z} &=&\frac{p}{\sqrt{p^{2}+\mu ^{2}}}\;N_{0}(z)-N_{1}(z)\;.
\label{test-can-z}
\end{eqnarray}
The Hamiltonian leading to these equations is 
\begin{eqnarray}
H &=&M+\sqrt{p^{2}+\mu ^{2}}\;N_{0}(z)-p\;N_{1}(z)  \nonumber
\label{test-Ham} \\
&=&M+\sqrt{p^{2}+\mu ^{2}}\;e^{\frac{\kappa M}{4}|z|}-\epsilon p\sqrt{1+%
\frac{8\Lambda }{\kappa ^{2}M^{2}}}\;\frac{z}{|z|}\left[ e^{\frac{\kappa M}{4%
}|z|}-1\right] \;.
\end{eqnarray}
This Hamiltonian is also derived from the determining eq. (\ref{H1}) by
setting 
\begin{eqnarray}
&&z_{1}=z,\quad m_{1}=\mu ,\quad p_{1}=p,\quad \tilde{p}_{1}=\tilde{p}=p%
\frac{z}{|z|},  \nonumber \\
&&z_{2}=0,\quad m_{2}=M,\quad p_{2}=0,  \nonumber
\end{eqnarray}
and retaining only the linear terms of $\sqrt{p^{2}+\mu ^{2}}$ and $\tilde{p}
$. \newline

In terms of the proper time of the test particle 
\begin{equation}
d\tau^2=dt^{2}\left\{N_{0}(z)^2-(N_{1}(z)+\dot{z})^2\right\} =dt^{2}N_{0}^{2}%
\frac{\mu^2}{p^2+\mu^2}\;,
\end{equation}
the canonical equations (\ref{test-can-p}) and (\ref{test-can-z}) are
expressed as 
\begin{eqnarray}
\frac{dp}{d\tau}&=&-\frac{\kappa M}{4\mu}\sqrt{p^2+\mu^2} \left\{\sqrt{%
p^2+\mu^2}\mbox{sgn}(z) - \epsilon p\;\sqrt{1+\frac{8\Lambda}{\kappa^{2}M^{2}%
}}\right\}\;,  \label{test-p2} \\
\frac{dz}{d\tau}&=&\frac{p}{\mu} -\epsilon\sqrt{1+\frac{8\Lambda}{%
\kappa^{2}M^{2}}}\mbox{sgn}(z) \left(1-e^{-\frac{\kappa M}{4}|z|}\right)%
\frac{\sqrt{p^2+\mu^2}}{\mu}\;.  \label{test-z2}
\end{eqnarray}
Eq.(\ref{test-p2}) can be integrated and in $z > 0$ region the solution $%
p(\tau)$ is 
\begin{equation}
p(\tau)=\frac{\epsilon\mu}{2}\left(h(\tau)-\frac{1}{h(\tau)}\right)\;,
\end{equation}
with 
\begin{equation}
h(\tau) = \left\{ 
\begin{array}{ll}
\frac{\left(1+\sqrt{\gamma_M}\right) \left\{1-\rho\;e^{\epsilon\sqrt{\frac{%
\Lambda}{2}}(\tau-\tau_{0})}\right\}} {\sqrt{\gamma_{M}-1}
\left\{1+\rho\;e^{\epsilon\sqrt{\frac{\Lambda}{2}}(\tau-\tau_{0})}\right\}}
& \qquad \Lambda > 0 \;, \\ 
&  \\ 
\frac{\sqrt{p_{0}^{2}+\mu^2}+\epsilon p_{0}}{\mu} -\frac{\epsilon\kappa M}{4}%
(\tau-\tau_0) & \qquad \Lambda=0 \;, \\ 
&  \\ 
\frac{ 1-\frac{(1+\sqrt{\gamma_M})(\sqrt{p_{0}^{2}+\mu^2}-\epsilon p_{0})} {%
\mu\;\sqrt{1-\gamma_{M}}}\;\mbox{tan}\left[\epsilon\sqrt{-\frac{\Lambda}{8}}
(\tau-\tau_{0})\right] } {\frac{\sqrt{p_{0}^{2}+\mu^2}-\epsilon p_{0}}{\mu} +%
\frac{\sqrt{1-\gamma_{M}}}{1+\sqrt{\gamma_M}} \;\mbox{tan}\left[\epsilon%
\sqrt{-\frac{\Lambda}{8}} (\tau-\tau_{0})\right] } & \qquad \Lambda < 0 \;,
\end{array}
\right.
\end{equation}
where 
\begin{equation}
\gamma_{M}= 1+\frac{8\Lambda}{\kappa^2 M^2}\;, \qquad \rho=\frac{(1+\sqrt{%
\gamma_M})(\sqrt{p_{0}^{2}+\mu^2}-\epsilon p_{0}) -\mu\sqrt{\gamma_{M}-1}} {%
(1+\sqrt{\gamma_M})(\sqrt{p_{0}^{2}+\mu^2}-\epsilon p_{0}) +\mu\sqrt{%
\gamma_{M}-1}}\;,
\end{equation}
with $p_0$ being the initial momentum at $\tau=\tau_0$.

In $z < 0$ region the solution is 
\begin{equation}
p(\tau)=-\frac{\epsilon\mu}{2}\left(\bar{h}(\tau)-\frac{1}{\bar{h}(\tau)}
\right)\;,
\end{equation}
with 
\begin{equation}
\bar{h}(\tau) = \left\{ 
\begin{array}{ll}
\frac{\left(1+\sqrt{\gamma_M}\right) \left\{1-\bar{\rho}\;e^{\epsilon\sqrt{%
\frac{\Lambda}{2}}(\tau-\tau_{0})}\right\}} {\sqrt{\gamma_{M}-1} \left\{1+%
\bar{\rho}\;e^{\epsilon\sqrt{\frac{\Lambda}{2}}(\tau-\tau_{0})}\right\}} & 
\qquad \Lambda > 0 \;, \\ 
&  \\ 
\frac{\sqrt{p_{0}^{2}+\mu^2}-\epsilon p_{0}}{\mu} -\frac{\epsilon\kappa M}{4}%
(\tau-\tau_0) & \qquad \Lambda=0 \;, \\ 
&  \\ 
\frac{ 1-\frac{(1+\sqrt{\gamma_M})(\sqrt{p_{0}^{2}+\mu^2}+\epsilon p_{0})} {%
\mu\;\sqrt{1-\gamma_{M}}}\;\mbox{tan}\left[\epsilon\sqrt{-\frac{\Lambda}{8}}
(\tau-\tau_{0})\right] } {\frac{\sqrt{p_{0}^{2}+\mu^2}+\epsilon p_{0}}{\mu} +%
\frac{\sqrt{1-\gamma_{M}}}{1+\sqrt{\gamma_M}} \;\mbox{tan}\left[\epsilon%
\sqrt{-\frac{\Lambda}{8}} (\tau-\tau_{0})\right] } & \qquad \Lambda < 0 \;,
\end{array}
\right.
\end{equation}
where 
\begin{equation}
\bar{\rho}=\frac{(1+\sqrt{\gamma_M})(\sqrt{p_{0}^{2}+\mu^2}+\epsilon p_{0})
-\mu\sqrt{\gamma_{M}-1}} {(1+\sqrt{\gamma_M})(\sqrt{p_{0}^{2}+\mu^2}%
+\epsilon p_{0}) +\mu\sqrt{\gamma_{M}-1}}\;.
\end{equation}

When $p=p_0$ and $z=0$ at $\tau=\tau_0$, the total energy is $H=H_0=M+\sqrt{%
p_{0}^2+\mu^2}$. The solution for $z(\tau)$ is obtained from (\ref{test-Ham}%
) and $p(\tau)$ as 
\begin{equation}  \label{sol-testz}
z(\tau)=\left\{ 
\begin{array}{ll}
\frac{4}{\kappa M}\;\mbox{log} \frac{\frac{2\sqrt{p_{0}^{2}+\mu^2}}{\mu}%
-\left(h-\frac{1}{h}\right) \sqrt{\gamma_M}} {\left(h+\frac{1}{h}%
\right)-\left(h-\frac{1}{h}\right)\sqrt{\gamma_M}} & \qquad z>0\;, \\ 
&  \\ 
-\frac{4}{\kappa M}\;\mbox{log} \frac{\frac{2\sqrt{p_{0}^{2}+\mu^2}}{\mu}%
-\left(\bar{h}-\frac{1}{\bar{h}} \right)\sqrt{\gamma_M}} {\left(\bar{h}+%
\frac{1}{\bar{h}}\right)-\left(\bar{h}-\frac{1}{\bar{h}} \right)\sqrt{%
\gamma_M}} & \qquad z<0 \;.
\end{array}
\right.
\end{equation}
For the test particle solution the critical value of $\Lambda$ is $\Lambda_c
=\frac{\kappa^2 \mu^2 M^2}{8p_{0}^2}$. \newline

Fig.21 and 22 show typical trajectories of the test particle $\mu=0.1$ for $%
M=10$ and $\Lambda=-10, 0, 0.4$ and $2$. The characteristics of these plots
are common to those of the unequal mass case.

\hspace*{5cm}--------------------------

\vspace{5mm}

\hspace*{7cm}Fig.21

\vspace{5mm}

\begin{center}
{\it Phase space trajectories of a test particle for different values of $%
\Lambda$. }
\end{center}

\hspace*{5cm}--------------------------

\hspace*{5cm}--------------------------

\vspace{5mm}

\hspace*{7cm}Fig.22

\vspace{5mm}

\begin{center}
{\it The $r$ plots corresponding to the trajectories in Fig.21. }
\end{center}

\hspace*{5cm}--------------------------

\section*{APPENDIX C: CAUSAL RELATIONSHIPS BETWEEN PARTICLES IN UNBOUNDED
MOTION}

We can explicitly verify that the particles lose causal contact with one
another for $\tau > \tau_\infty$. Consider the unbounded motion of tan-type $%
A (n=0)$ with $m=1, H=2.1$ and $\Lambda=1.5$. The path $x(t)$ of light
emitted from particle 2 at time $T$ is governed by $d\tau=0$, which reads 
\begin{equation}
\left(\frac{dx}{dt}\right)^2 + 2N_{1}\frac{dx}{dt} -(N_{0}^{2}-N_{1}^{2})=0
\end{equation}
and so the equation of the light signal directed to particle 1 is 
\begin{equation}  \label{eq1light}
\frac{dx}{dt}=N_{0}(x(t), z_{1}(t), z_{2}(t), p(t)) - N_{1}(x(t), z_{1}(t),
z_{2}(t), p(t)) \;.
\end{equation}
The light emitted in the opposite direction is described by 
\begin{equation}  \label{eq2light}
\frac{dx}{dt}=-N_{0}(x(t), z_{1}(t), z_{2}(t), p(t)) - N_{1}(x(t), z_{1}(t),
z_{2}(t), p(t))\;.
\end{equation}
Numerically solving (\ref{eq1light}) and (\ref{eq2light}) yields the
solutions shown in Figs.23 and 24, where the trajectories of light signals
emitted from particle 2 at various times $T$ are plotted. For small $T (T<3)$%
, the particles are in causal contact (a dotted curve in (+) direction in
Fig.23), but for $T \approx 3$ the signal just barely catches up with
particle 1, which is almost in light-like motion (a dashed curve in (+)
direction in Fig.23). For $T=4$ the world line $x(t)$ in the (+) direction
is parallel to $z_{1}(t)$ at large $t$ and in the (-) direction it goes
nearly on the same trajectory with the particle 2. For large $T$ ($T > 4.82$%
) the particles are out of causal contact with each other (Fig.24): a light
ray sent from particle 2 toward particle 1 receives a strong repulsive
effect and ultimately reverses direction, following behind particle 2. In
Fig.24 the trajectories of the light signal emitted to (-) direction can not
be discriminated from those of the particle 2.

\hspace*{5cm}--------------------------

\vspace{5mm}

\hspace*{7cm}Fig.23

\vspace{5mm}

\begin{center}
{\it The trajectories of light signals emitted at $T=2, \;3$ and $4$. }
\end{center}

\hspace*{5cm}--------------------------

\hspace*{5cm}--------------------------

\vspace{5mm}

\hspace*{7cm}Fig.24

\vspace{5mm}

\begin{center}
{\it The trajectories of light signals emitted at $T=4.82, \;5$ and $6$. }
\end{center}

\hspace*{5cm}--------------------------

A flat-space model of these effects can be constructed as follows. Consider
the following expression for the 2-velocity 
\begin{equation}  \label{eq2v}
u^\mu = (f(\sigma\tau),\sqrt{f^2(\sigma\tau)-1})
\end{equation}
where $f(\sigma\tau)$ is some function and 
\begin{equation}
d\tau^2 = dt^2-dx^2
\end{equation}
is the flat metric. We have $\frac{dt}{d\tau} = f$, $\frac{dx}{d\tau} = 
\sqrt{f^2-1}$ and so 
\begin{equation}
\frac{dx}{dt} = \frac{\sqrt{f^2-1}}{f} \;.
\end{equation}

The general expression for the acceleration of a particle with 2-velocity (%
\ref{eq2v}) is 
\begin{equation}
a^\mu = \frac{d u^\mu}{d\tau} = \sigma f^\prime (1,\frac{f}{\sqrt{%
f^2(\sigma\tau)-1}})
\end{equation}
where $f^\prime = df(\tau)/d\tau$. We have $u\cdot u = 1$ and $a \cdot u = 0$
and 
\begin{equation}
a \cdot a = \frac{(\sigma f^\prime)^2 }{f^2(\sigma\tau)-1}
\end{equation}
for the magnitude of the acceleration. In general we have the following
possibilites:

1) The function $f \to f_0$ where $f_0$ is finite at $\tau\to\infty$. In
this case the particle never becomes lightlike.

2) The function $f \to \infty$ as $\tau\to\infty$. In this case the particle
becomes lightlike, but it takes an infinite proper time (and coordinate
time) for this to happen. The standard example is $f = \cosh(\sigma \tau)$,
the constant acceleration example.

3) The function $f \to \infty$ as $\tau\to \tau_0$, where $\tau_0$ is
finite. In this case the particle becomes lightlike in a finite amount of
proper time, but an infinite amount of coordinate time. An example would be $%
f = \sec(\sigma \tau)$. The acceleration is not constant, but increases as a
function of proper time, diverging at $\tau=\tau_0$. This last situation is
realized by our exact solutions (\ref{sol-tanA}--\ref{sol-tanB}) with $%
\Lambda > \Lambda_c$.

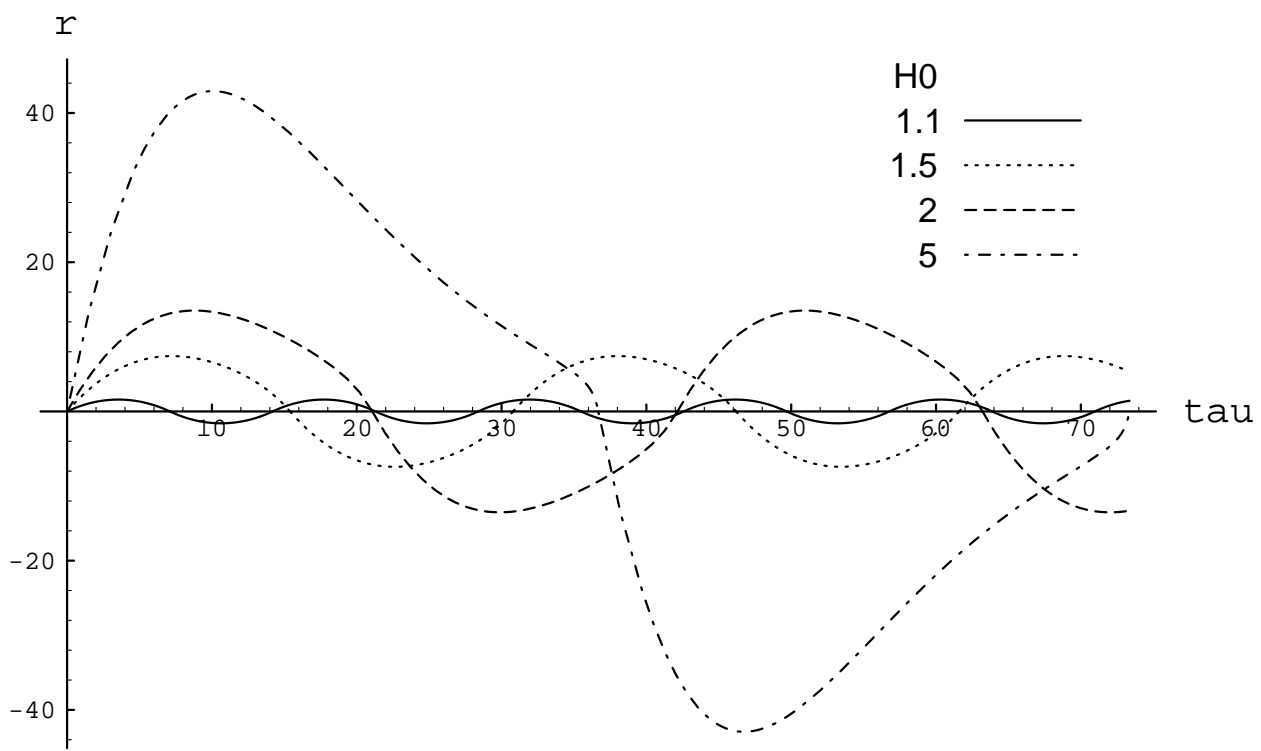
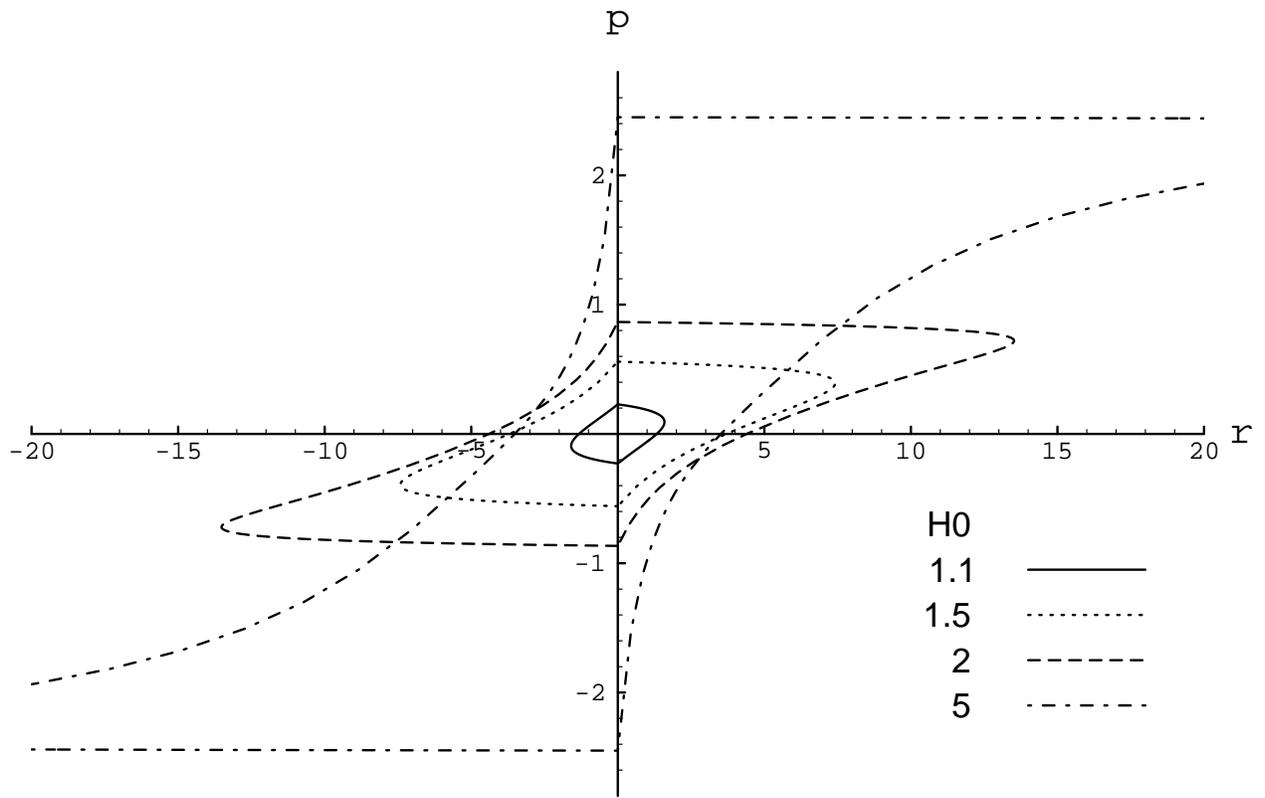
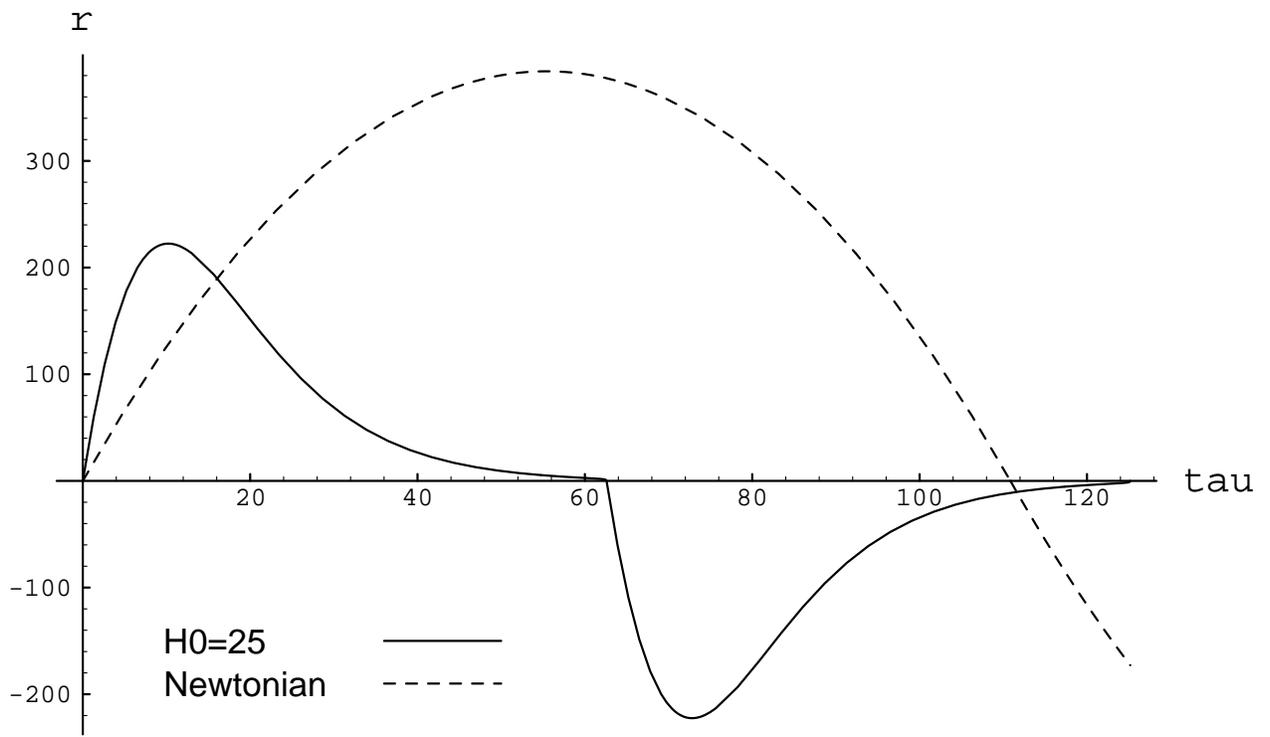
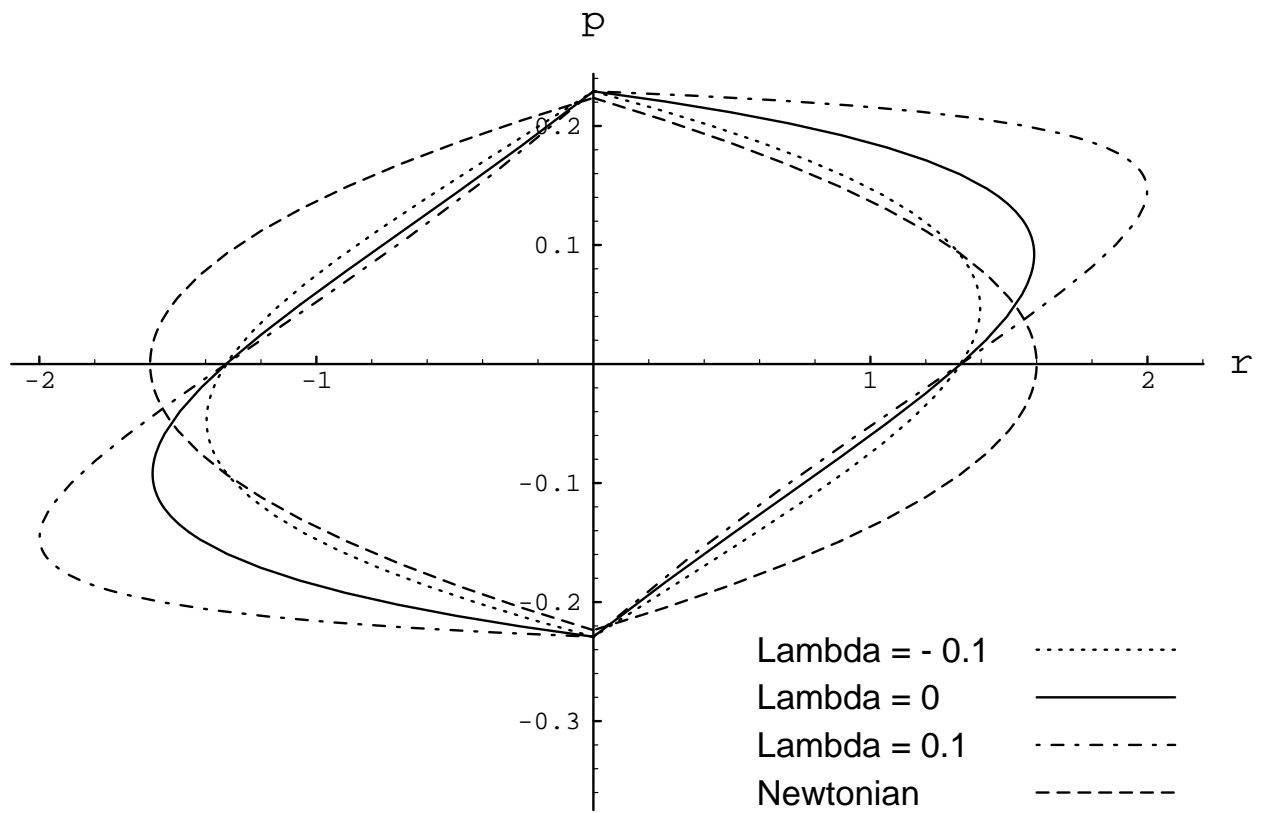
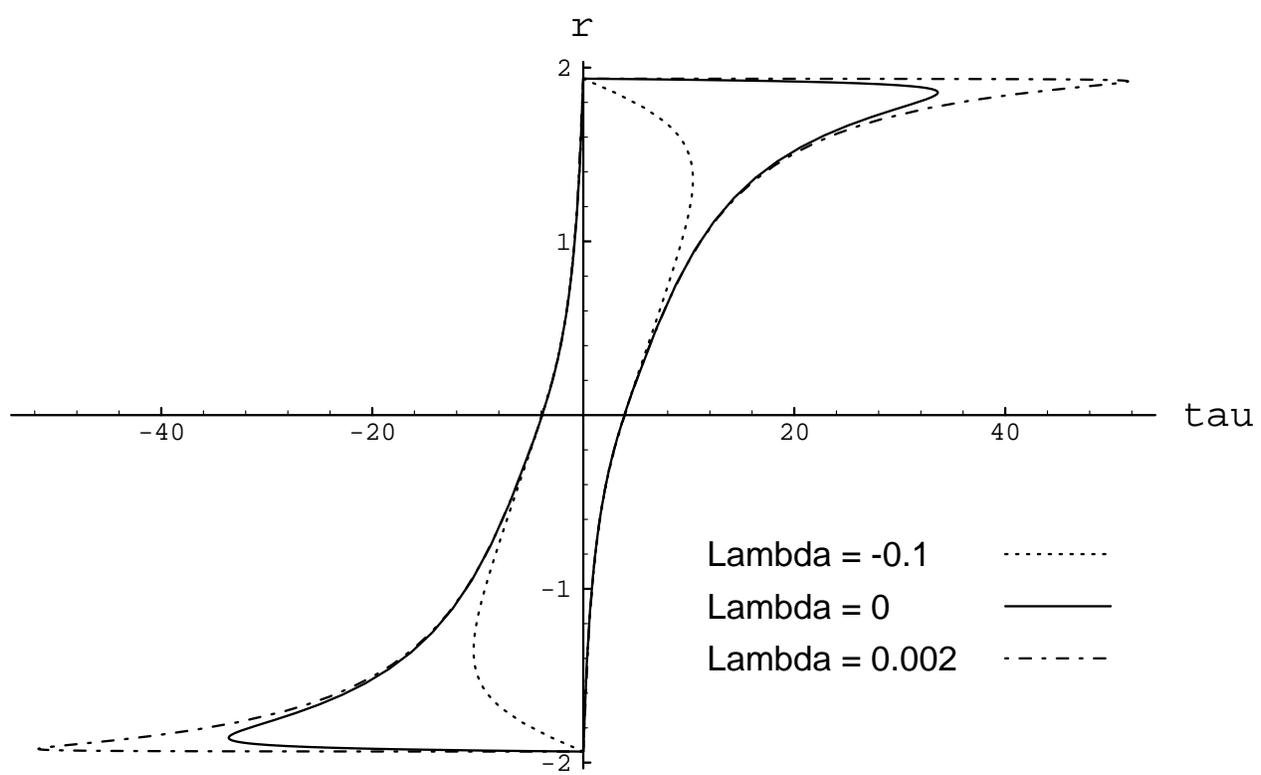
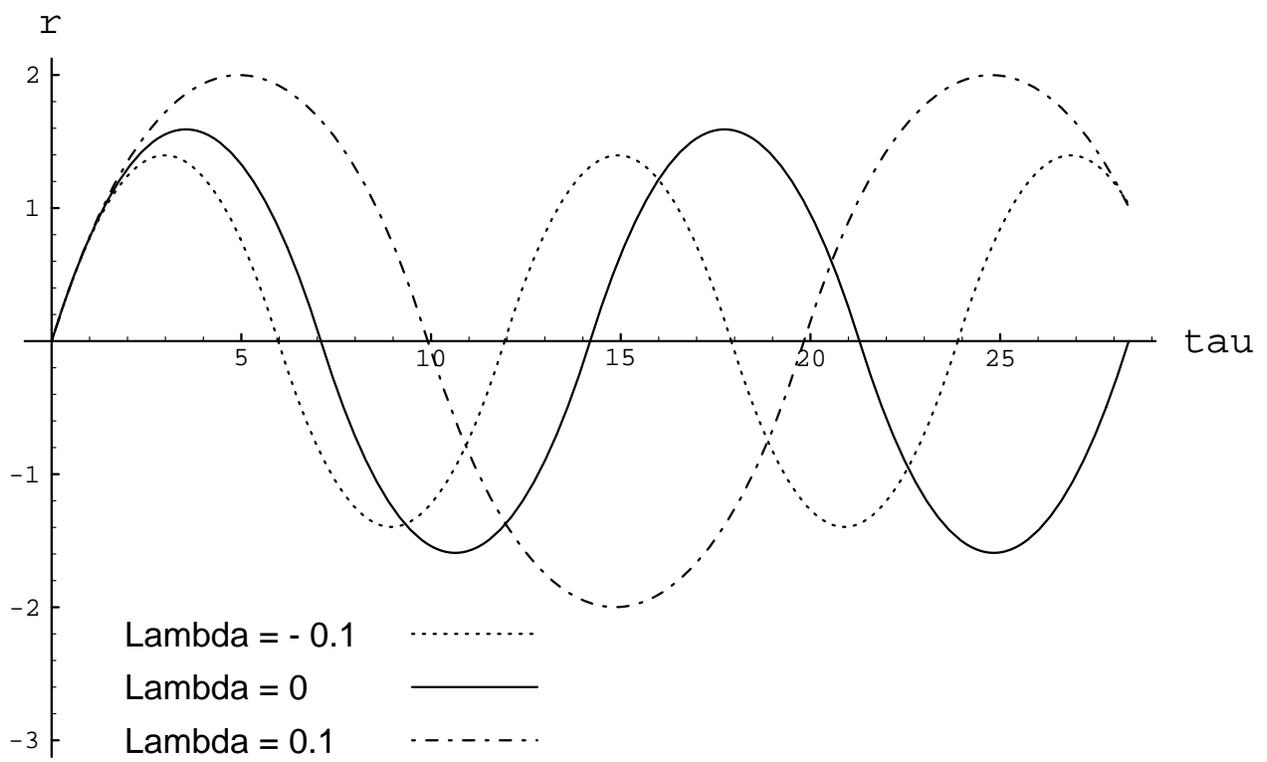
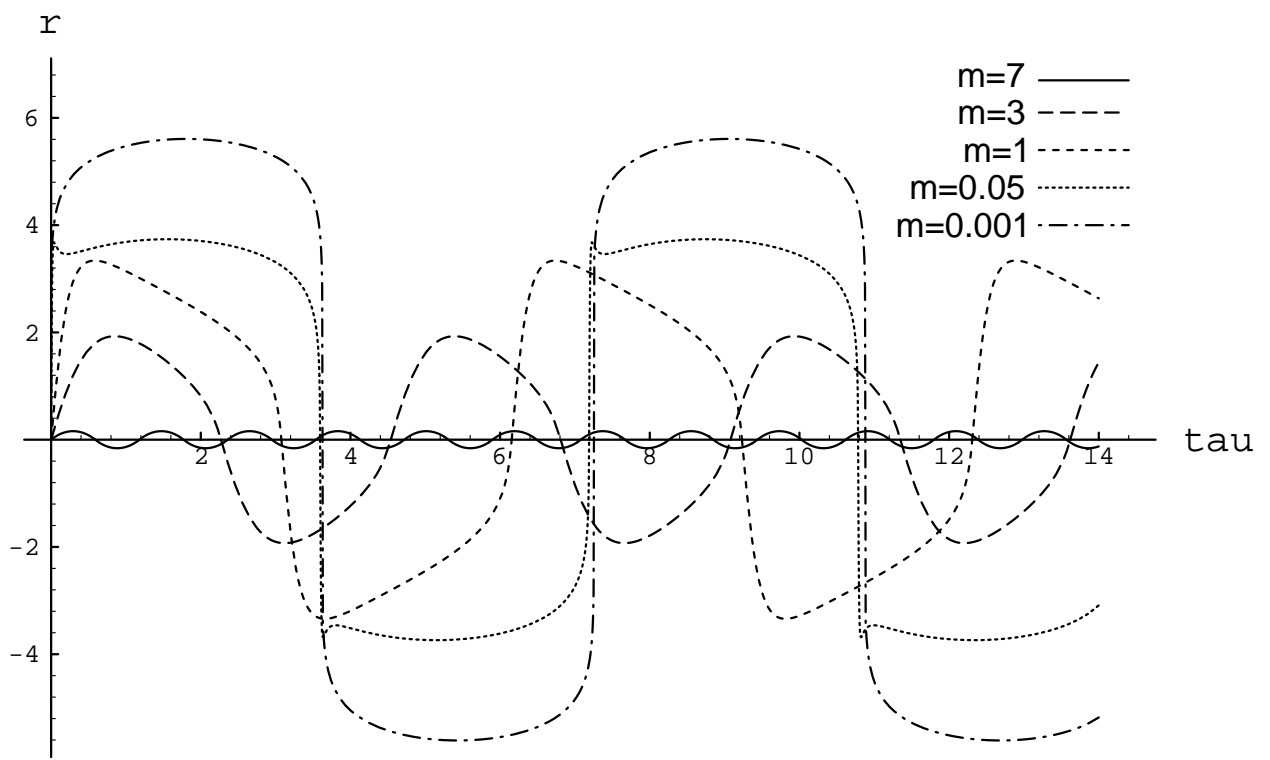
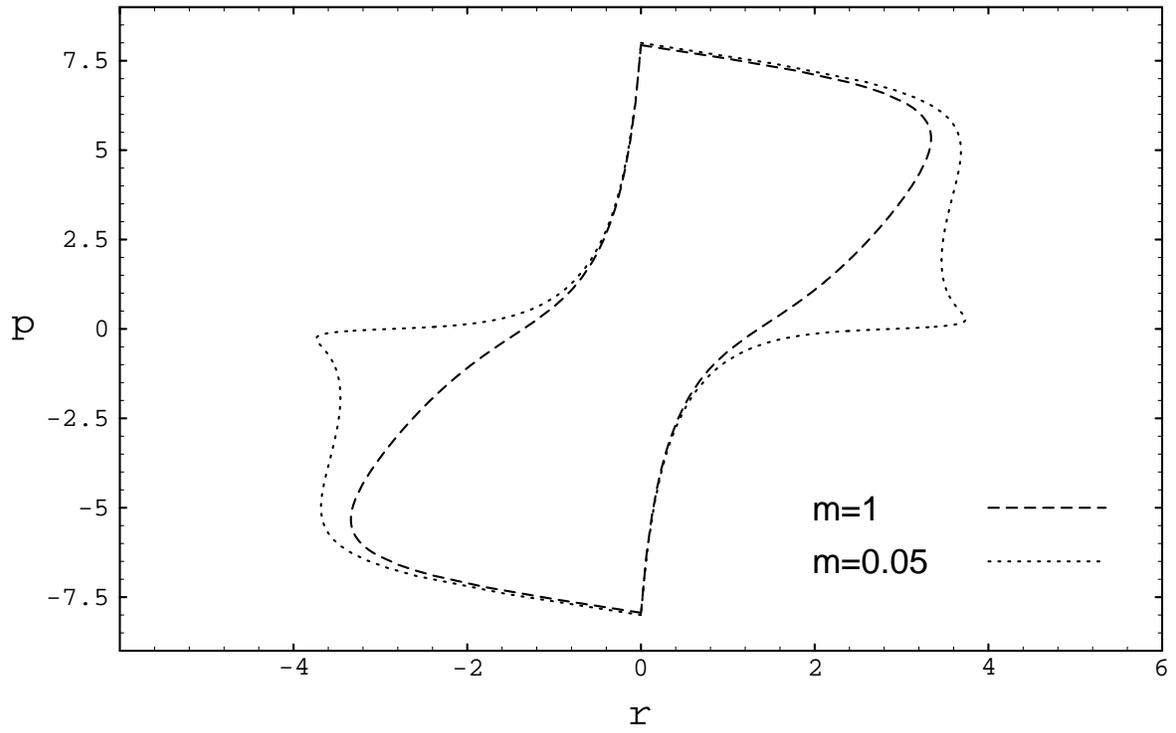
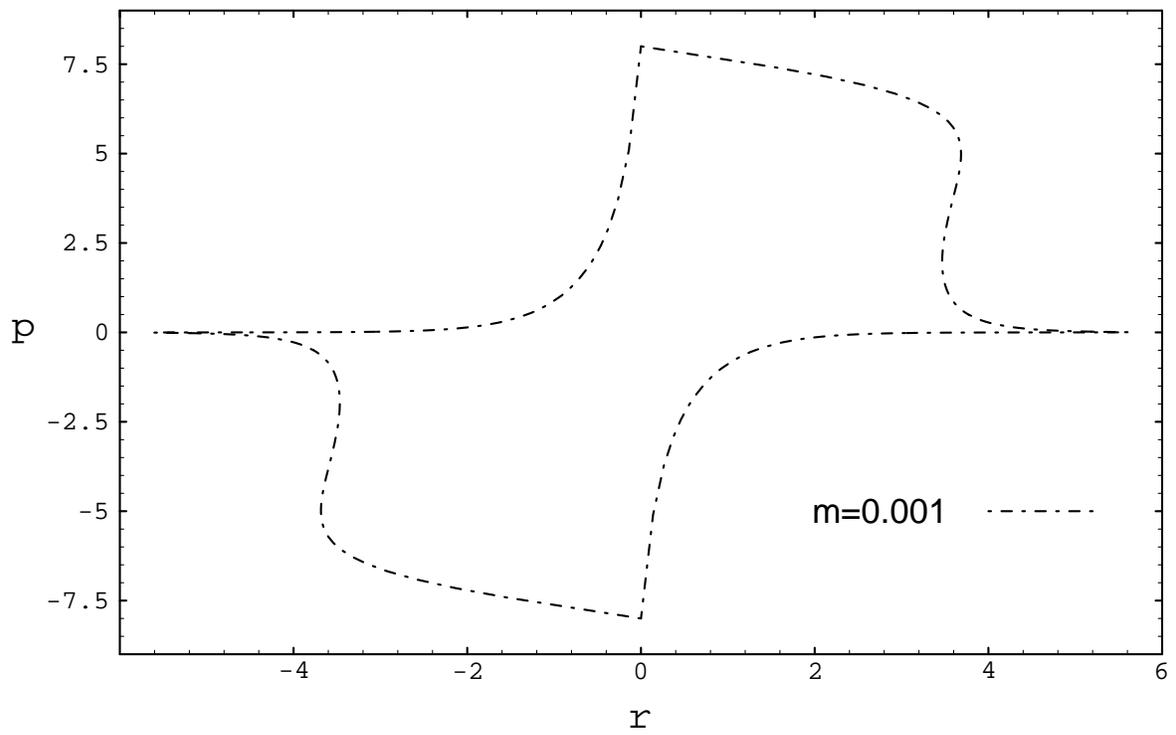
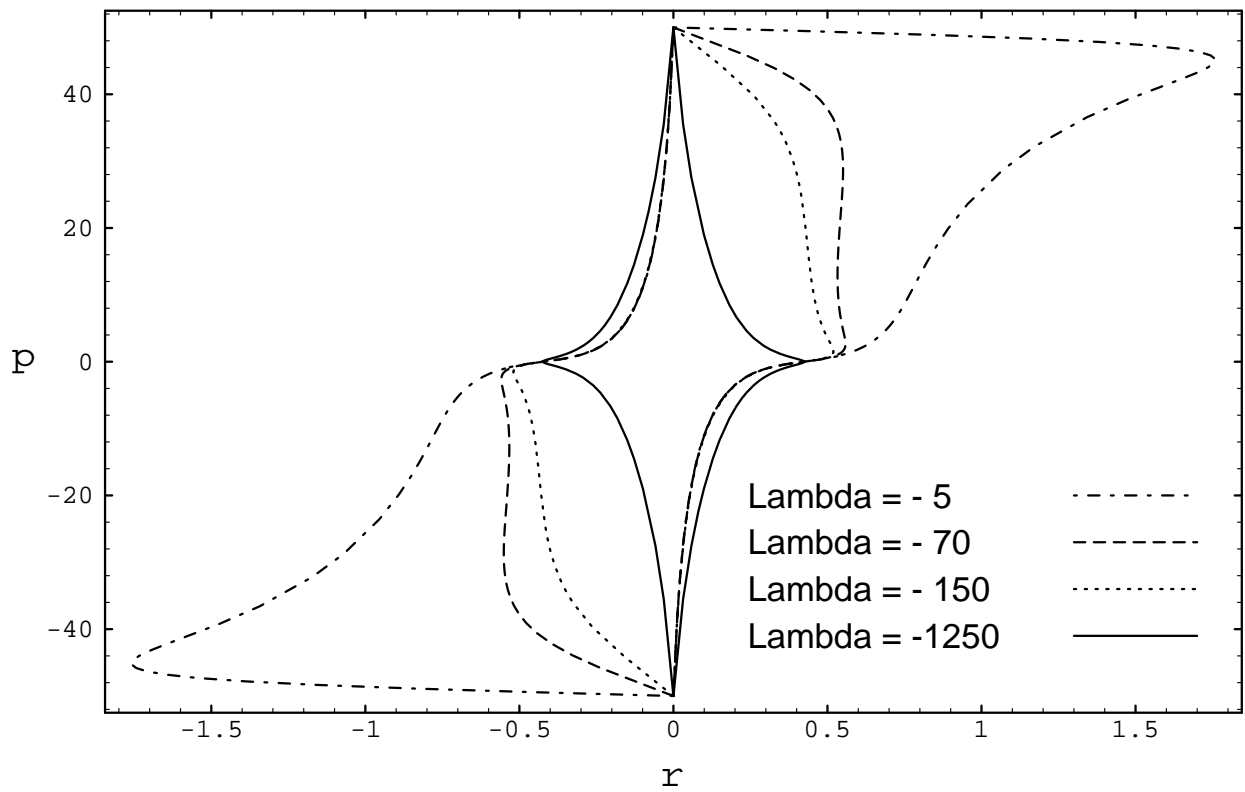
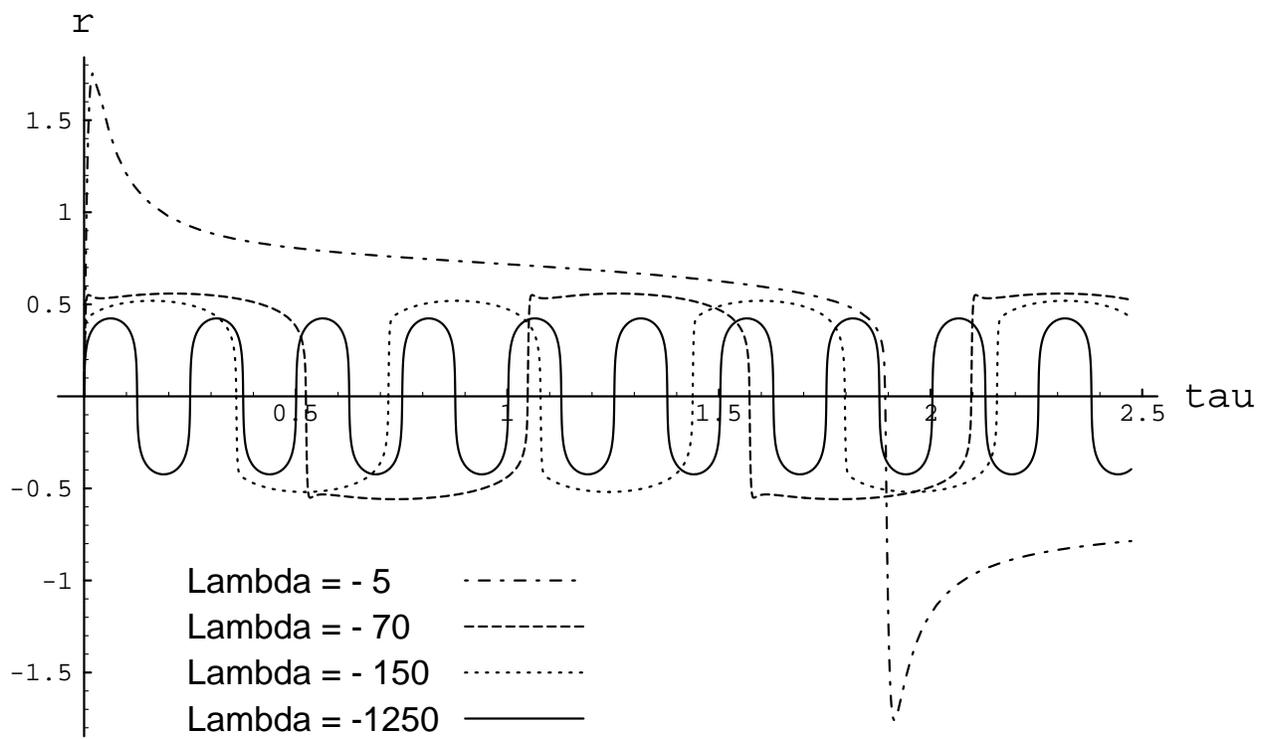
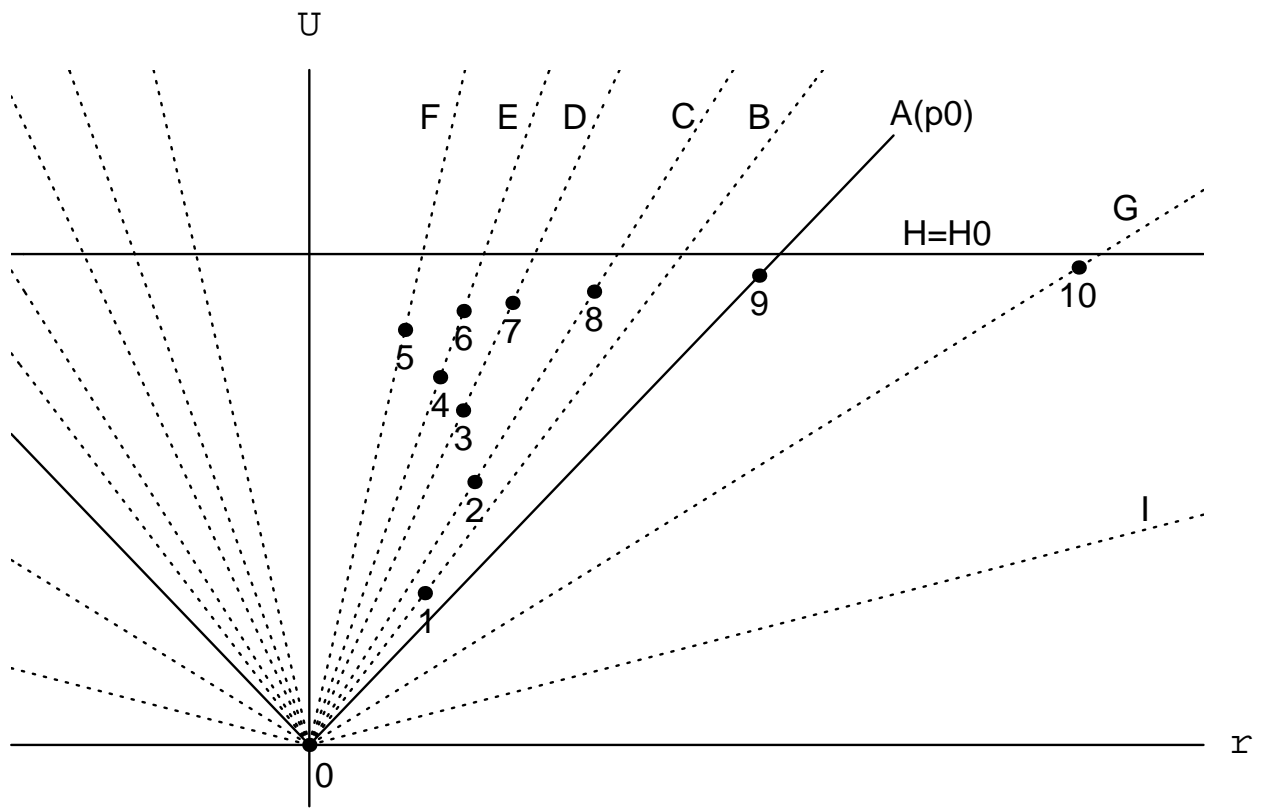
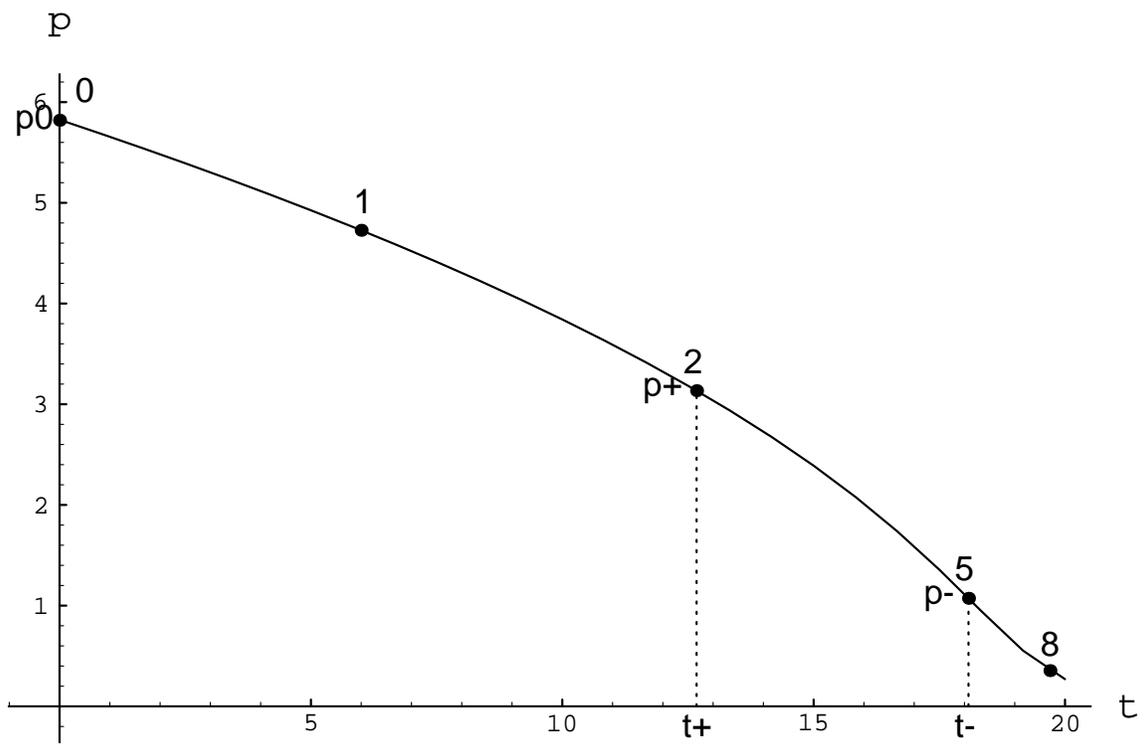
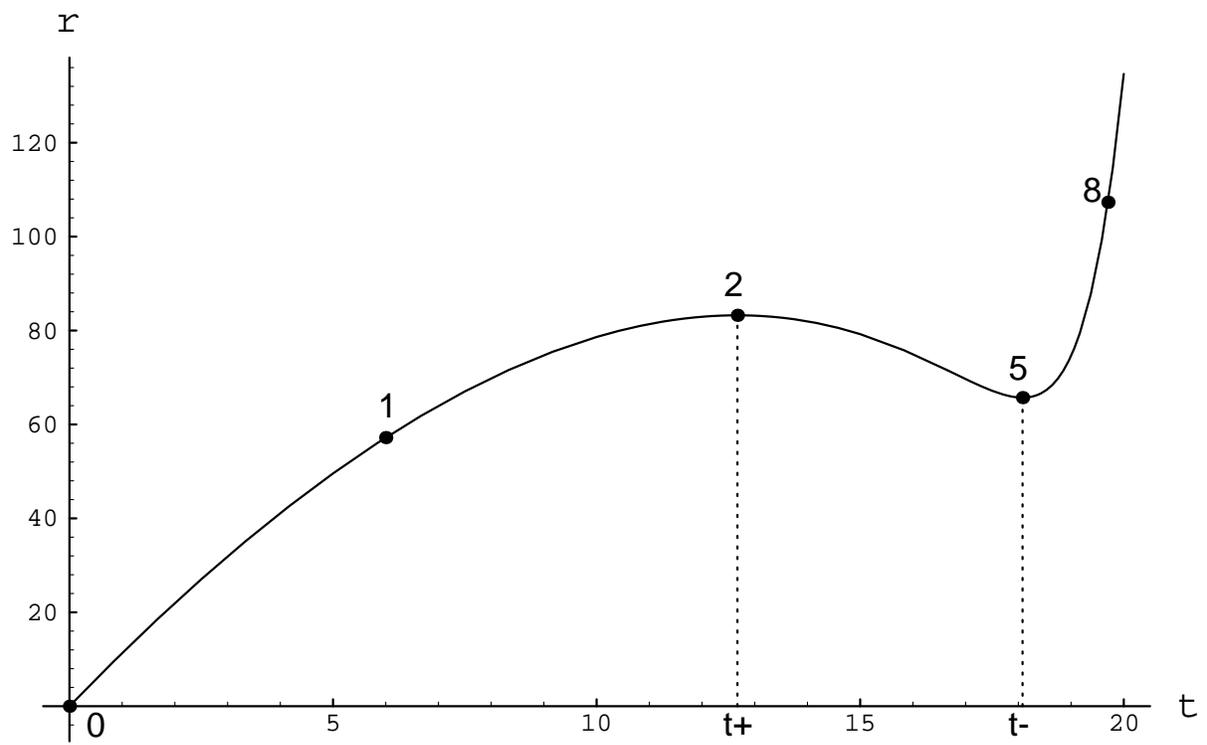
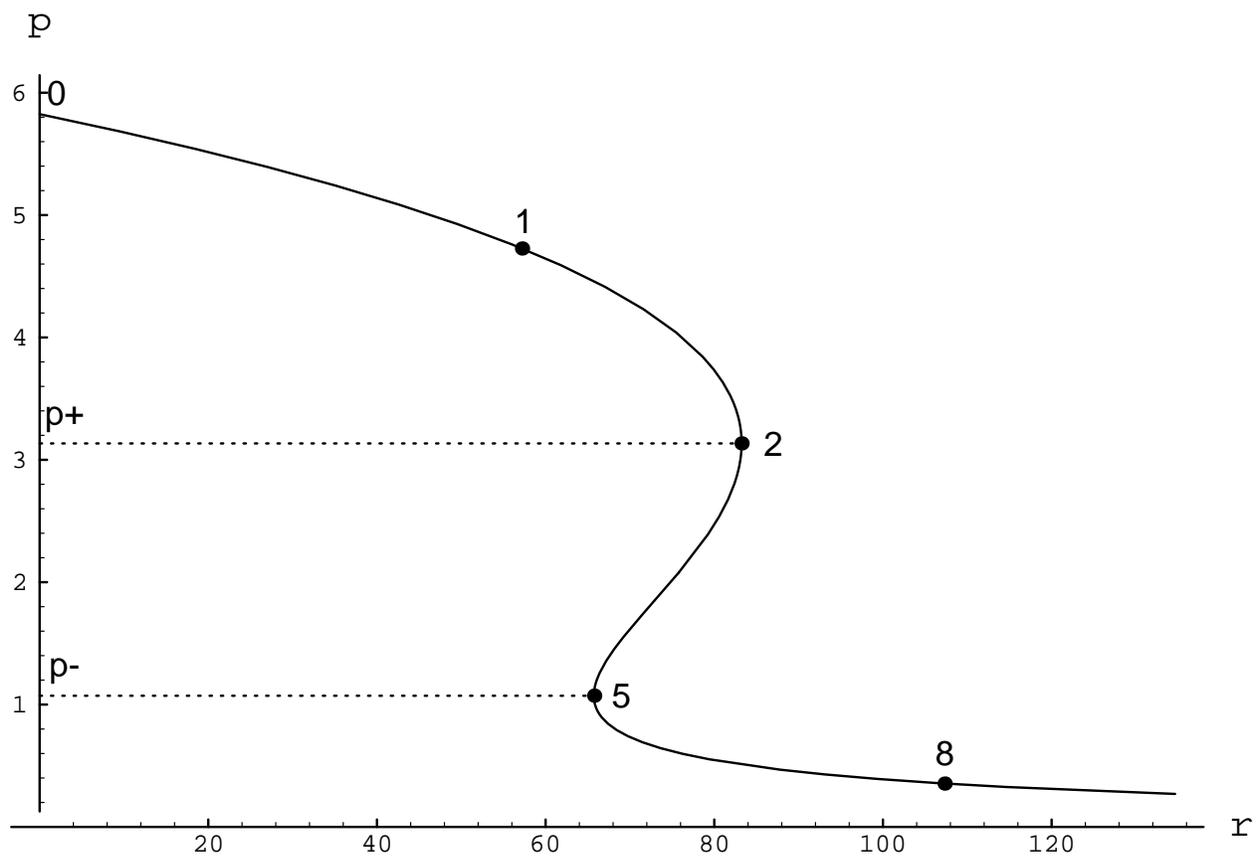
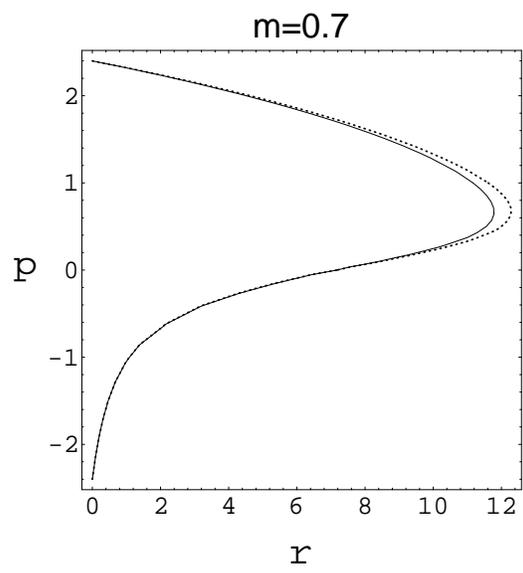
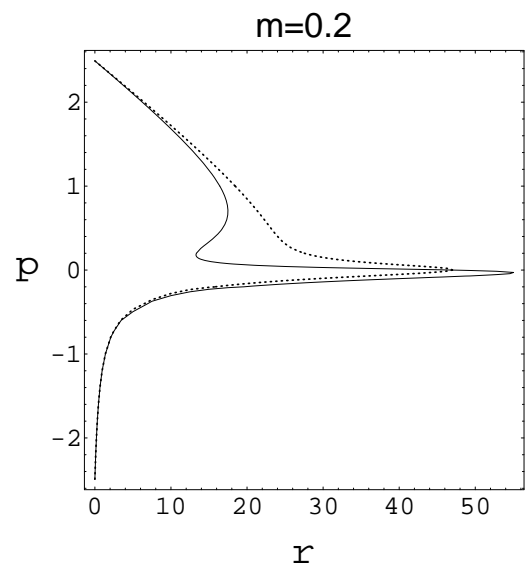
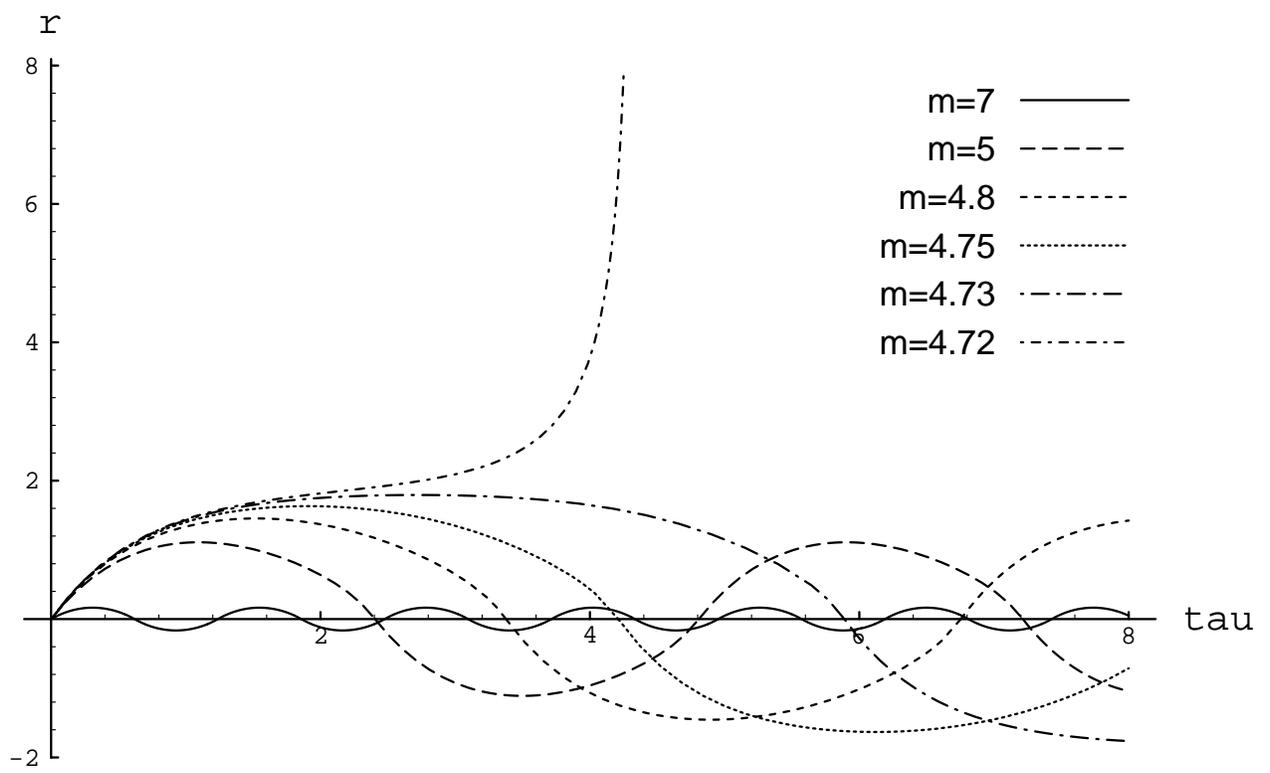
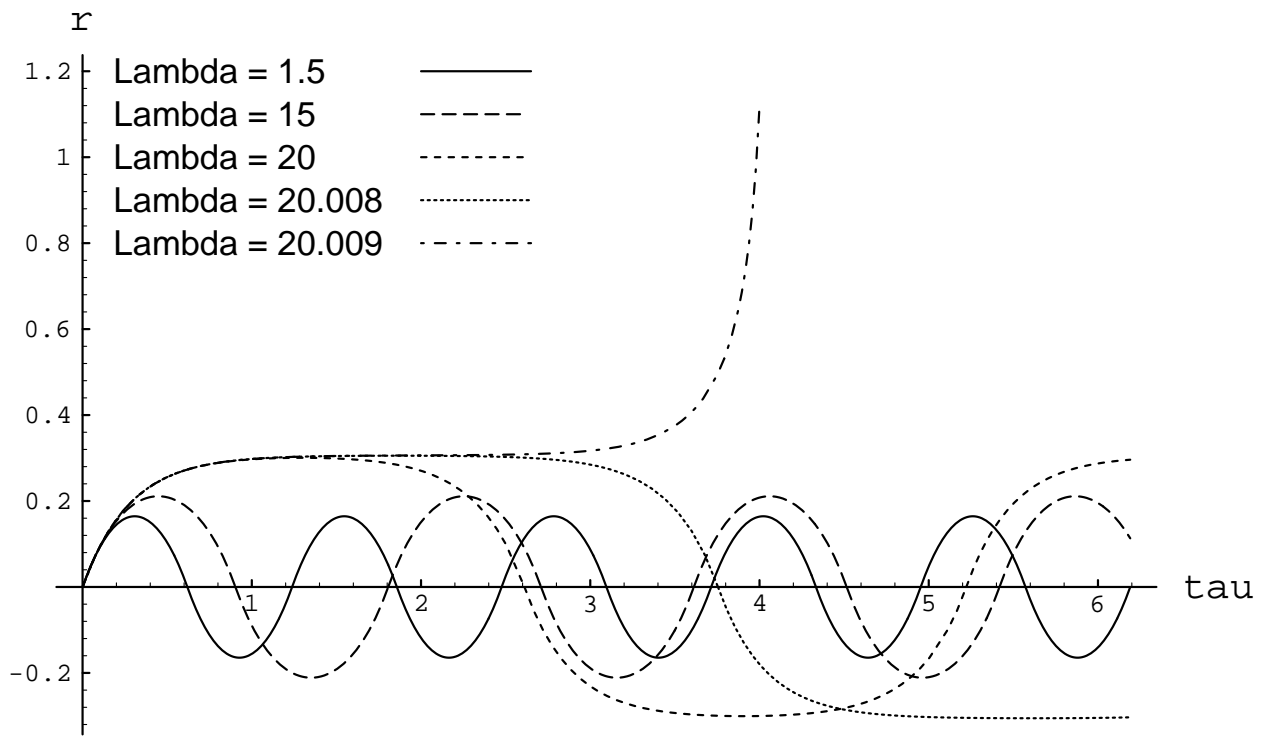
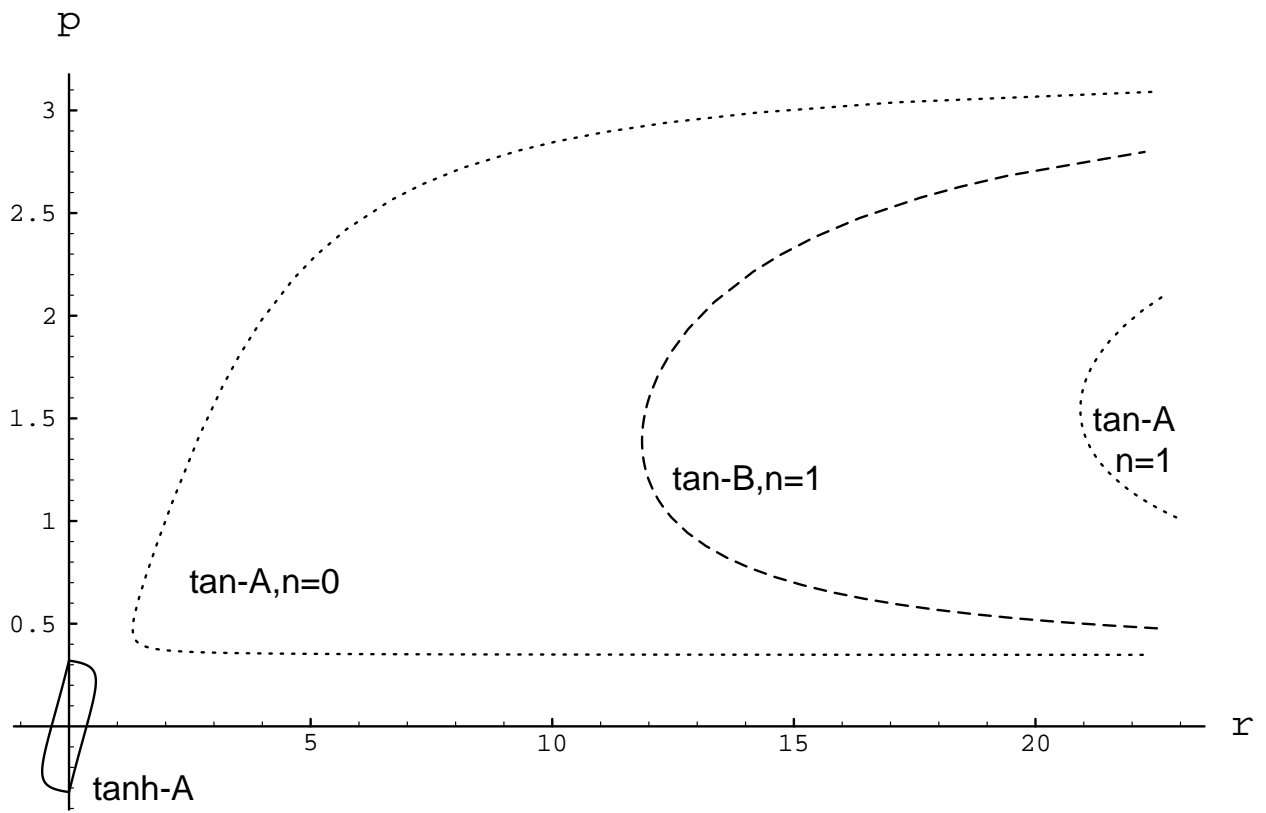
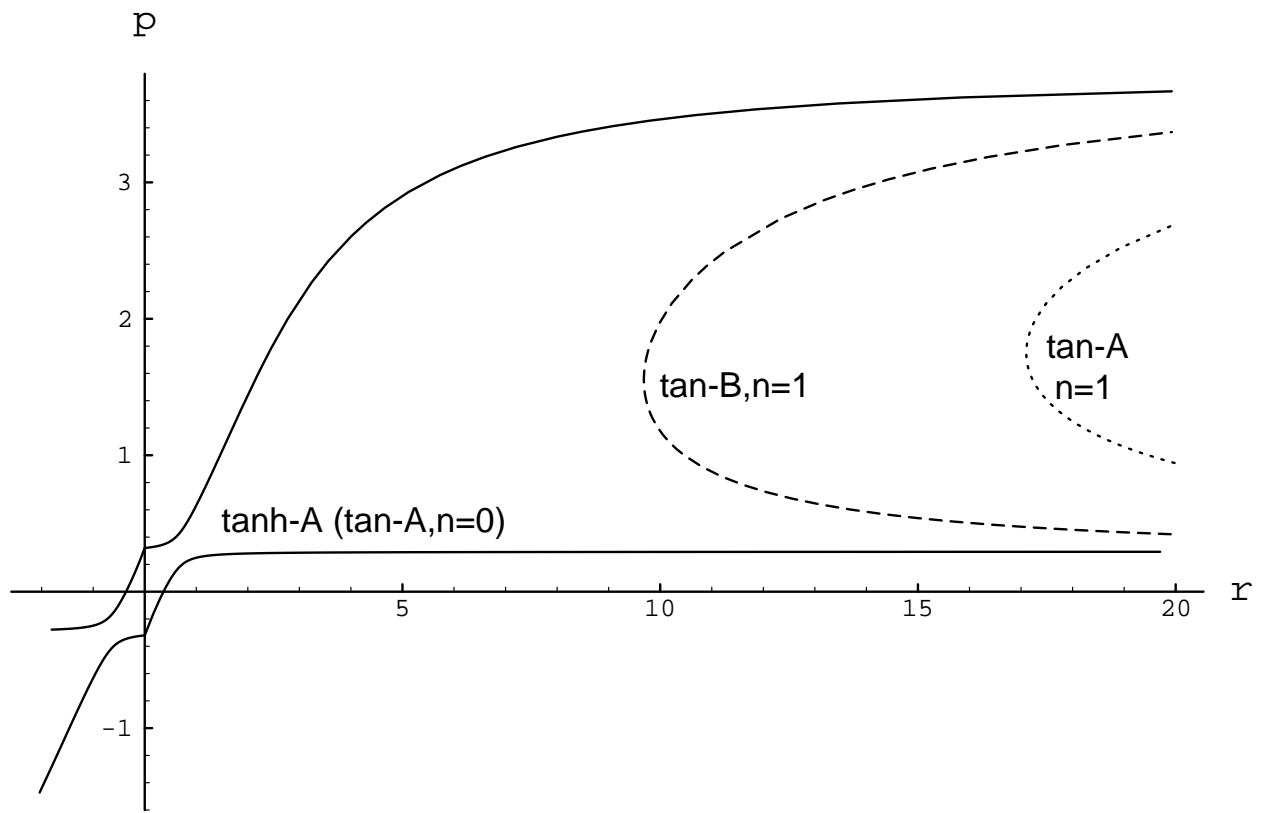
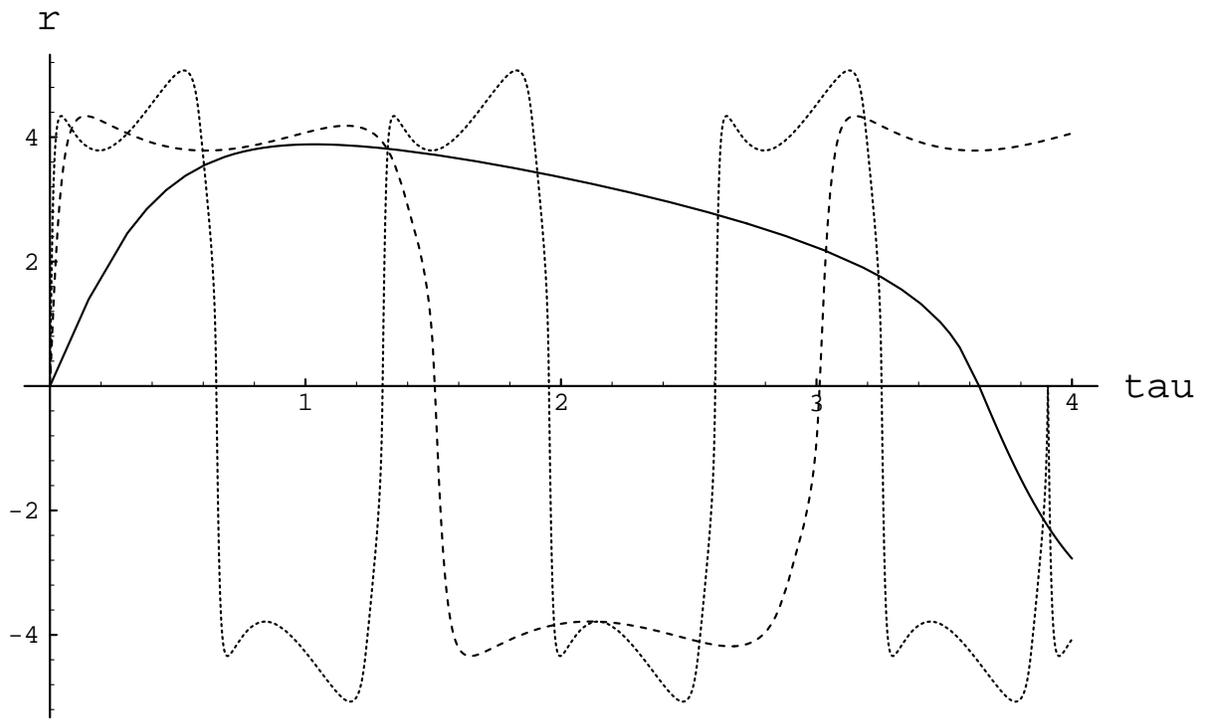
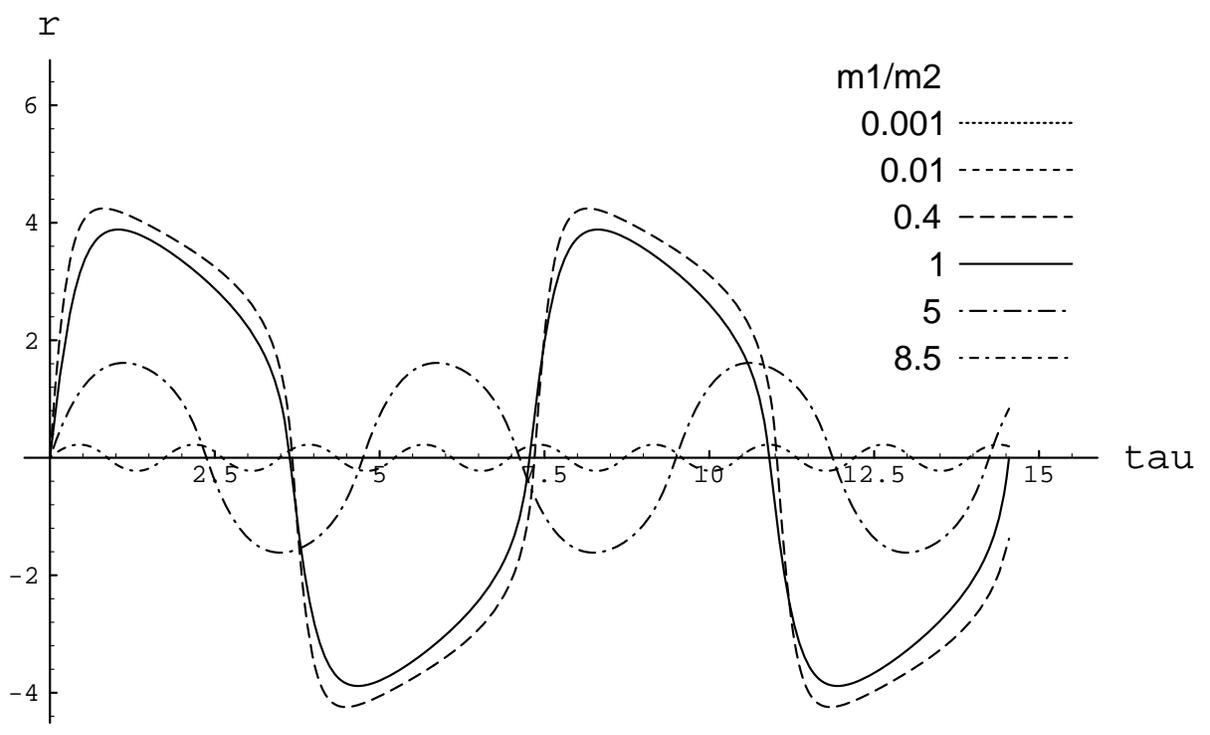
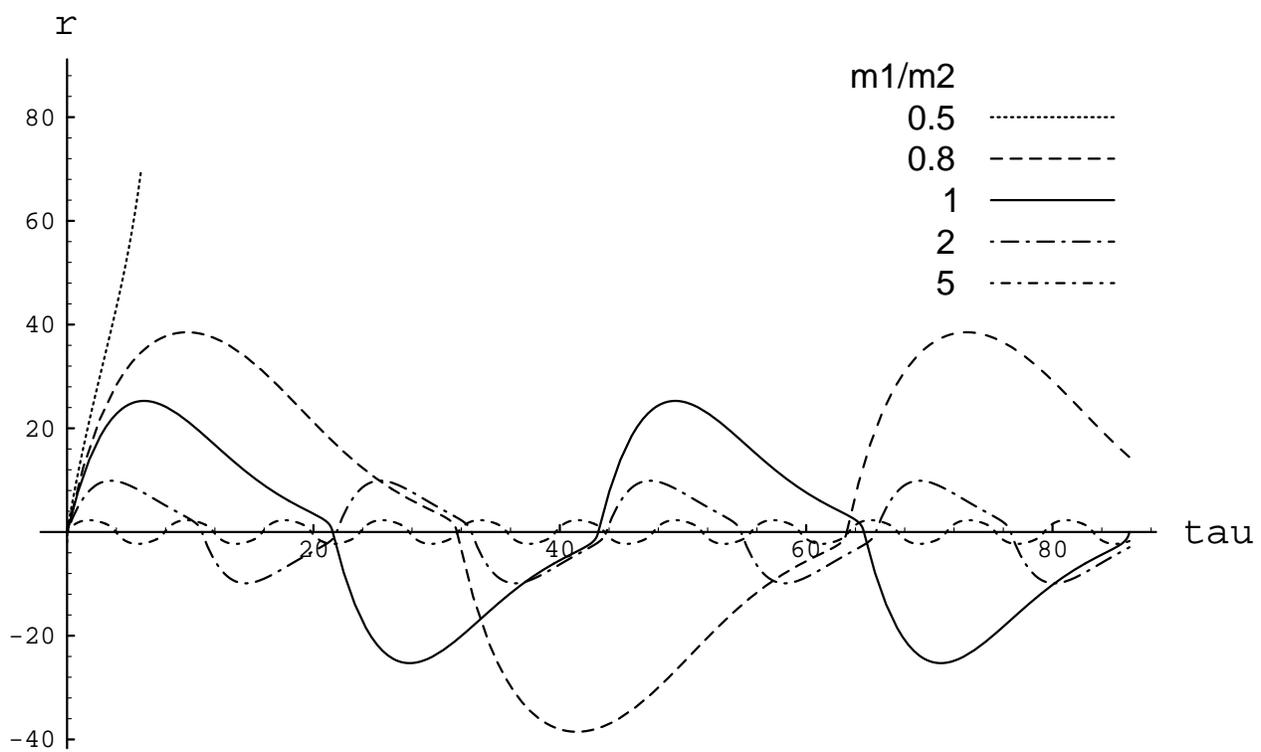
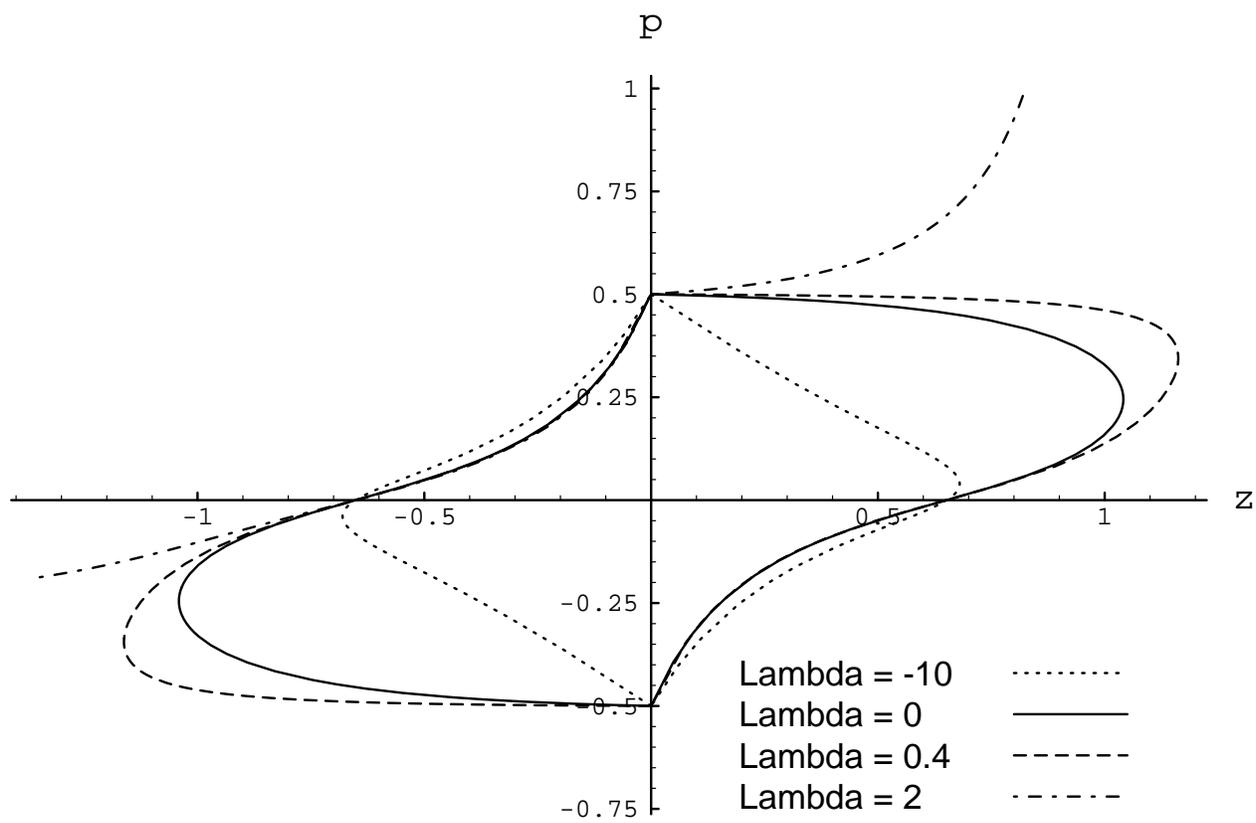
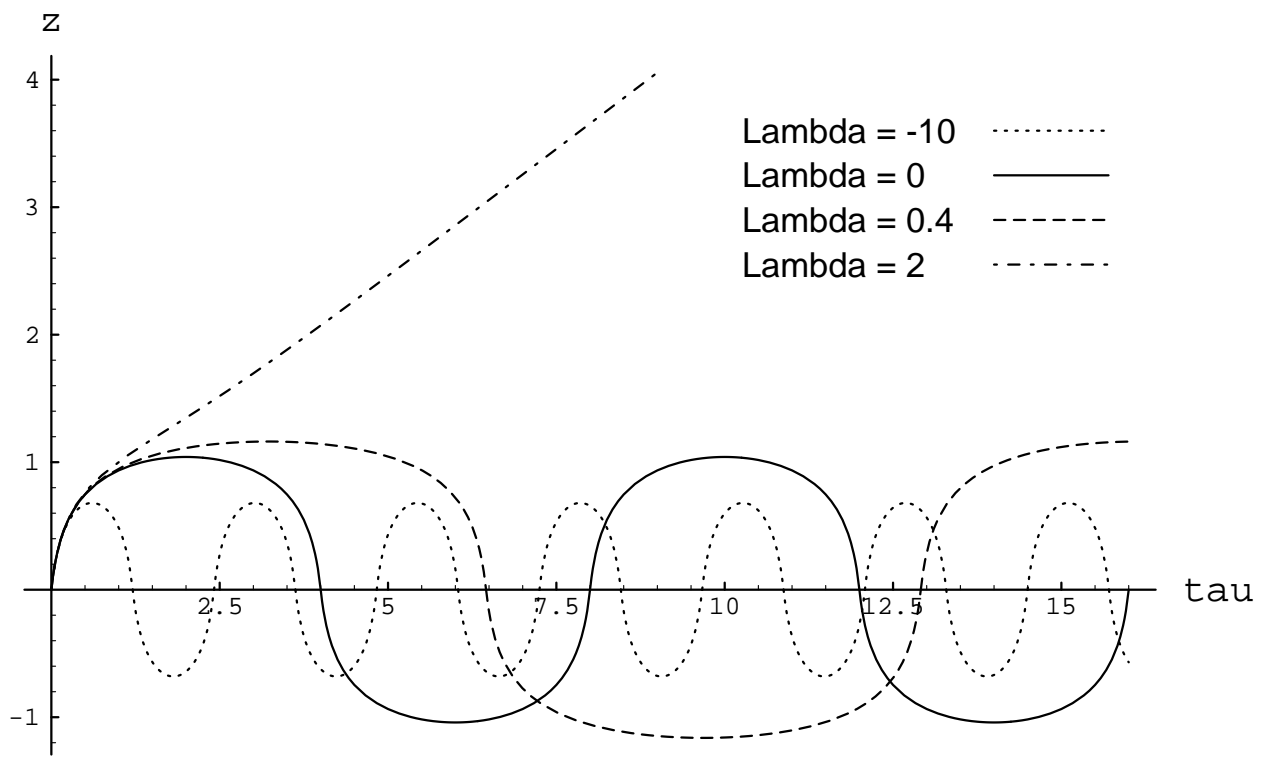
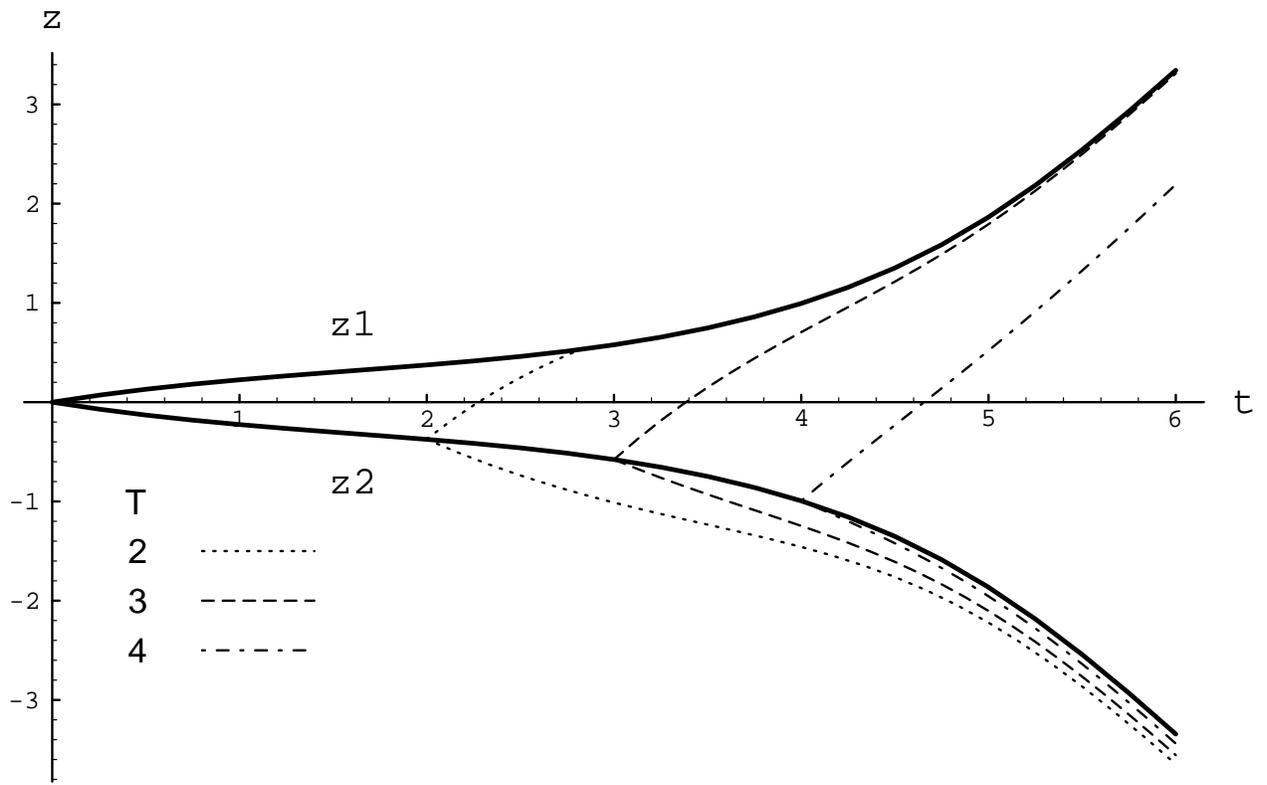
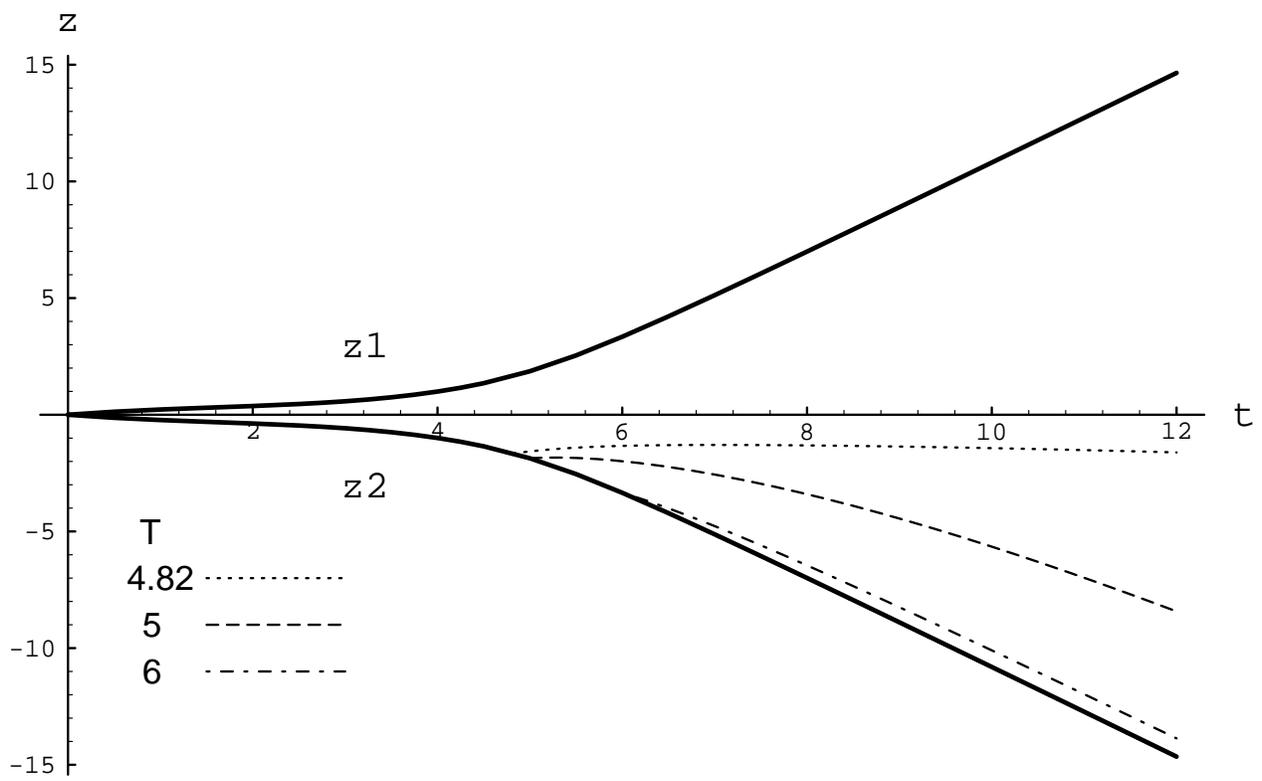
\end{document}